\documentclass[a4paper,11pt]{article}
\pdfoutput=1 
\usepackage{graphicx}
\usepackage{float}
\usepackage{mciteplus}
\usepackage{slashed}
\usepackage{dirtytalk}
\usepackage{bbold}
\usepackage{braket}
\usepackage{mathtools,amsmath,amssymb,mathbbol}
\usepackage{booktabs}
\usepackage{tabularx}
\usepackage{longtable}
\usepackage{xcolor}
\usepackage{tikz}
\usetikzlibrary{decorations.pathmorphing,calc}
\usepackage{tikz-feynman}
\tikzfeynmanset{compat=1.1.0}
\tikzfeynmanset{warn luatex=false} 

\usepackage{subcaption}

\tikzset{
  graviton/.style={decorate, decoration={snake, amplitude=0.9mm, segment length=3.2mm}},
  lbl/.style={font=\scriptsize, fill=white, inner sep=1.2pt, rounded corners=1pt}
}

\tikzset{graviton/.style={decorate, decoration={snake, amplitude=0.9mm, segment length=3.2mm}}}

\usepackage{jhep} 

\usepackage[T1]{fontenc} 

\renewcommand{\[}{\begin{equation}\begin{aligned}}
\renewcommand{\]}{\end{aligned}\end{equation}}
\def\beq{\begin{equation}}
\def\eeq{\end{equation}}

\def\({$}
\def\){$}

\renewcommand{\texttt}{{}}
\def\bs{\begin{subequations}}
\def\es{\end{subequations}}

\def\Ec{\mathcal{E}}

\def\Kc{\mathcal{K}}
\def\Lc{\mathcal{L}}

\def\Nc{\mathcal{N}}
\def\Oc{\mathcal{O}}
\def\Pc{\mathcal{P}}

\def\Rc{\mathcal{R}}

\def\Tc{\mathcal{T}}

\newcommand{\tia}[1]{}

\def\polarizationtensor{\rm e}

\newcommand{\bea}{\begin{eqnarray}}
\newcommand{\eea}{\end{eqnarray}}
\newcommand{\beas}{\begin{eqnarray*}}
\newcommand{\eeas}{\end{eqnarray*}}
\newcommand{\bal}{\begin{aligned}}
\newcommand{\eal}{\end{aligned}}

\newcommand{\LF}{\left(}
\newcommand{\RF}{\right)}
\newcommand{\LT}{\left[}
\newcommand{\RT}{\right]}

\newcommand{\pd}{\partial}

\renewcommand{\imath}{\ensuremath{\mathrm{i}}}
\renewcommand{\vec}[1]{\ensuremath{\mathbf{#1}}}

\title{Quantum (quadratic) gravity: replacing the massive tensor ghost with an inverted harmonic oscillator-like instability}

\author[a]{K. Sravan Kumar}

\author[b]{Jo\~ao Marto}

\affiliation[a~]{Institute of Cosmology \& Gravitation,
	University of Portsmouth,
	Dennis Sciama Building, Burnaby Road,
	Portsmouth, PO1 3FX, United Kingdom}
\affiliation[b~]{
	Departamento de F\'isica, Centro de Matem\'atica e Aplicações (CMA-UBI), Universidade da Beira Interior, Rua Marquês D'Ávila e Bolama, 6201-001 Covilhã, Portugal}

\emailAdd{sravan.kumar@port.ac.uk}
\emailAdd{jmarto@ubi.pt}

\abstract{The quadratic theory of gravity is the unique renormalizable theory of quantum gravity in 4 dimensions, as proved by K. S. Stelle in 1977. Over the decades, the theory has been understood to contain a massive tensor ghost, and several attempts have been made to evade its adverse effects by proposing new quantization prescriptions and interpretations. In this paper, we show that the additional spin--2 of quadratic gravity can be turned into a healthy inverted harmonic oscillator (IHO)-like instability, which can be quantized consistently with direct-sum quantum field theory (DQFT), which incorporates geometric superselection sectors. Such modes possess a well-defined quantum description yet do not admit a particle interpretation and are not part of the asymptotic spectrum, being characterized by hyperbolic evolution and spacelike momentum support. We argue that, as a consequence, the extra spin--2 degree of freedom remains off-shell and effectively decoupled from ordinary matter fields, avoiding unitarity violations in observable processes. We argue that this IHO instability is a prevalent feature of fundamental physics, whether it concerns quantum fields on curved spacetimes or the Higgs $\mathbb{Z}_2$ symmetry breaking in the Standard Model of particle physics. Thus, we demonstrate that our new understanding of quadratic gravity offers a fundamental pathway to quantum gravity and a safe beginning for the Universe. Furthermore, we derive key observational predictions of this construction in the view of primordial gravitational waves with new bounds on the tensor-to-scalar ratio and the parity asymmetric features on the large angular scales. }


\makeatother

\makeatletter
\gdef\@fpheader{}
\makeatother

\begin{document}
	
\maketitle

\section{Introduction}

Understanding gravity and quantum mechanics (QM) until the Planck scales
has been a milestone sought by physicists worldwide for generations \cite{Buoninfante:2024yth,Loll:2022ibq,deBoer:2022zka}. The proposals of quantum gravity have evolved from beautiful physical ideas to mathematical formulations, although we lack robust observational evidence, and even most importantly, concrete, unique, testable predictions from many of the approaches. Nevertheless, theories of quantum gravity have led to speculative toy model-building, which has been driving the fields of cosmology and black hole physics for decades with very minimal success in underpinning physics at the foundational level. A noteworthy fact is that popular theories of quantum gravity suffer from a lack of a definitive action with the fewest possible parameters. For example, string theory, which has been described as a promising venture to unify all known laws of physics, suffers from a multitude of possibilities involving extra dimensions, an infinitely large number of vacua, or uncertainty in the number of degrees of freedom at Planck scales, all of which impede the theory from even recovering the standard model (SM) of particle physics \cite{Woodard:2009ns}. Another popular approach, loop quantum gravity, is a nonperturbative quantization program for 3+1 dimensional general relativity formulated in connection variables. While it is grounded in a well-defined classical action, there remains no single unanimously accepted formulation of the full quantum dynamics \cite{NicolaiPeetersZamaklar2005}. The other contender, asymptotic safety, is built on a novel idea by Weinberg on the behavior of coupling constants towards the Planck scale. Over the decades, the asymptotic safety program has focused on developing an effective action for quantum gravity in Euclidean spacetime rather than Lorentzian, assuming the existence of an ultraviolet (UV) fixed point \cite{Donoghue:2019clr}. 
The other approaches, such as causal sets and causal dynamical triangulations, aim to build spacetime fundamentally from discrete and continuous approximations \cite{Henson:2006kf,Surya:2019ndm}, which also lack a definitive action. All these endeavors towards quantum gravity, over the years, kept looking for observational evidence despite their respective setbacks in providing predictions. The majority of quantum gravity investigations often partially, or fully, avoid the concern of unitarity loss in quantum field theory in curved spacetime (QFTCS). The widely accepted notion is that once we fix the Planck scale quantum gravity, QFTCS unitarity would be an emergent outcome at the low energies \cite{Banks:1983by,Hawking:1982dj,Almheiri:2020cfm,Giddings:2022jda,Bire82}. 
 Ultimately, the laws of physics can only be established through observation and experiment but at the same time, we do need theories that are robust rather than with extreme flexibility to fit any future observational outcome. The ultimate goal of theoretical physics is not only to ensure logical consistency but also to make testable predictions that ultimately push the boundaries of our knowledge and technology. The present paper explores not only the logical consistency of quantum gravity within foundational principles of QFT, but also its observational consequences which can be validated by the future gravitational wave observations.

Almost all the quantum gravity approaches do, in some way or the other, give up (or circumvent) perturbative renormalizability \cite{Buoninfante:2024yth,Buoninfante:2025dgy}. 
Sticking to the strict renormalizability that led to the success of the SM of particle physics, there is only one renormalizable theory of quantum gravity in 4 dimensions with a unique Lagrangian structure: quadratic gravity \cite{Stelle:1976gc} (also known as "Stelle gravity"). This theory is a modification of general relativity (GR), that involves quadratic curvatures (i.e., $R^2$ and $W_{\mu\nu\rho\sigma}W^{\mu\nu\rho\sigma}$). In the SM, the renormalizability criterion dictates how we write a Lagrangian with a finite set of parameters\footnote{Weinberg's QFT textbook \cite{Weinberg:1995mt} clearly states "A quantum field theory (QFT) is renormalizable if all interaction terms in the Lagrangian have total operator dimension $\leq 4$ in four spacetime dimensions (4D); equivalently, all couplings are of dimension $\geq 0$". This essentially means we should stop writing any interaction terms once we reach the coupling constants carrying zero dimension. In other words, renormalizability demands the non-existence of any operators greater than dimension 4 being part of the Lagrangian (in 4D).
}\cite{Dyson:1949bp,Sakata:1952rq}.  To be precise, renormalizability in its strict form informs us that the UV behavior of the fundamental theory is ruled by the interaction terms that carry dimensionless coupling constants. 
The formulation of quadratic gravity follows the same successful path \cite{Buoninfante:2025dgy}, but, nearly 50 years later, the puzzle of a ghost-like massive spin--2 field remains unresolved, leaving the theory incomplete. As a result, the quantum gravity community has widely perceived that quadratic gravity is not a consistent theory of quantum gravity; thus, the majority of quantum gravity frameworks go beyond the standard perturbative strict renormalizability.

It is worth checking if we exhausted all the options of understanding the ghost problem in the only unique renormalizable quantum (quadratic gravity) before dismissing this approach entirely. We are not the ones to first asking this question; already, pioneering works of D. Anselmi \cite{Anselmi:2018ibi,Anselmi:2018tmf,Anselmi:2019xac}, A. Salvio \& A. Strumia \cite{Salvio:2014soa,Salvio:2018crh}, J. Donoghue, and G. Menezes \cite{Donoghue:2021eto,Donoghue:2021cza,Donoghue:2019ecz,Donoghue:2020mdd} have approached this problem with new interpretations of the spin--2 ghost, modifying Feynman's $-i\epsilon$ prescription for the ghost propagators. The physics outcome of all these attempts is to project a ghost either as unphysical (fake) or an unstable state. However, the fundamental premise about the presence of a ghost (an effective state with a negative definite Hamiltonian) is never changed. This is exactly where the current paper proposes an alternative to the widely accepted understanding of the ghost pathology. With a change in sign of the coefficient of the Weyl square, we can turn the ghost into an inverted harmonic oscillator-like instability. At first glance, this appears pathological because most physicists view it as an unhealthy tachyonic degree of freedom. We demonstrate that this is not necessarily the case; in fact, it is a healthy instability analogous to the one that leads to ${\mathbb Z}_2$ symmetry breaking in the SM of particle physics (which in turn is inspired by the instability in the Landau-Ginzburg theory of phase transitions) \cite{Higgs:2014aqa}. The physics of this instability lies in the understanding of the quantum inverted harmonic oscillator (IHO) \cite{Gaztanaga:2025awe}. Furthermore, IHO is an interesting physical system that appears across domains such as condensed matter physics, quantum chemistry, and biology \cite{Subramanyan:2020fmx,Sundaram:2024ici}, and it was famously studied by Berry and Keating (BK) in 1999, where BK matched the (quantum) energy spectrum of IHO with the non-trivial zeros of the Riemann zeta function \cite{Berry1999}. Quantization of IHO necessitates one to consider physics with two arrows of time at the parity conjugate points of the physical space. This is exactly what the direct-sum quantum theory brings, and remarkably, it resolves two important issues, such as the unitarity of QFT in curved spacetime and the black hole information paradox \cite{Kumar:2023ctp,Kumar:2023hbj,Kumar:2024oxf,Kumar:2024ahu}, defining direct-sum quantum field theory (DQFT) in curved spacetime. This is because quantum fields in curved spacetime naturally decompose into modes whose dynamics are governed by IHOs. (See \cite{Gaztanaga:2025awe} for more details). This formulation (DQFT) not only explained the longstanding CMB anomalies with a significant statistical evidence over the standard theory (by 650 times) but also provided new predictions for large scale correlations in the primordial gravitational waves \cite{Gaztanaga:2024vtr,Gaztanaga:2024nwn,Gaztanaga:2025awe}. In this paper, we demonstrate how the universal physics of IHO (together with the criteria of renormalizability) would elegantly land us with a quantum theory of gravity. In other words, the quantum gravity we formulate here is built on the (IHO) physics that widely occurs in various contexts, and is also strongly built on the rich understanding of quantum fields in curved spacetime. 

In nutshell, the present work resolves the longstanding tension between perturbative renormalizability and unitarity in quadratic gravity that arises from an incomplete understanding of the dynamical structure of the additional spin–2 sector. We show that for the physically consistent sign of the Weyl–squared coefficient, the extra tensor mode is not a ghost excitation but is governed by the Hamiltonian of a IHO-like degree of freedom (which we named dual-IHO). Such systems describe controlled dynamical instabilities rather than negative–norm degrees of freedom, and they admit a consistent quantum description when formulated within a direct–sum Hilbert space structure. Within this framework, the spin–2 sector remains off–shell and contributes only through its dispersive, virtual effects, thereby preserving the renormalizability of quadratic gravity while avoiding violations of the optical theorem. The resulting theory retains the massless graviton and the scalaron as the only asymptotic propagating degrees of freedom, while the dual–IHO sector plays the role of a virtual UV completion.

Additionally, the purpose of quantum gravity is also to avoid gravitational singularities that GR is plagued by.  We argue that the Quadratic Quantum Gravity (QQG), being a unitary and renormalizable theory, could be a theory that doesn't support any singularity under the finite action principle \cite{Barrow1987,Barrow:2019gzc}. This point of view, recently embraced by Lenhers and Stelle \cite{Lehners:2023fud}, who argued that the inclusion of higher-order terms, such as the Weyl term, in the gravitational action can lead to a safe beginning for the Universe, characterized by the absence of singular spacetimes at the Big Bang. This conclusion is derived from equation (18) of \cite{Lehners:2023fud}, which does not explicitly depend on the sign of the Weyl term's coefficient. As such, the argument for a non-singular beginning remains valid even if the sign of the Weyl term is switched (as in the unitary QQG we obtain here). We also obtain, in this paper, the implications of the unitary QQG for inflationary cosmology as it provides a legitimate UV completion to the well-celebrated theory of Starobinsky inflation \cite{Starobinsky:1980te,Starobinsky:1981vz}. In our formulation of (unitary) QQG, the physical (asymptotic) observable degrees of freedom remain the scalaron (massive scalar responsible for inflation) and the massless graviton. The so-called massive spin--2 would be an off-shell IHO-type degree of freedom, which can never appear as an asymptotic state and thus only contributes to renormalizability and UV completion. This eliminates any possibility of extra degrees of freedom being visible in the CMB or primordial gravitational waves. However, the IHO-type spin--2 mode does leave its imprints by modifying the tensor-to-scalar ratio of the Starobinsky model in a predictive manner. One can derive further predictions of the unitary QQG in the higher-order cosmological correlations under the cosmological collider physics program \cite{Arkani-Hamed:2015bza}, which is beyond the scope of this paper. We also derive the parity asymmetric features of primordial power spectra in the unitary QQG driven inflationary scenario which results in the oscillatory features in the even-odd angular power spectra of scalar and tensor sector on large scales. This would test the DQFT construction of inflationary quantum vacuum \cite{Gaztanaga:2024vtr,Gaztanaga:2025awe} (which we call "Direct-Sum Inflation (DSI)) that breaks the degeneracy between parity conjugate geometric superselection sectors.  

This paper is organized as follows. In Sec.~\ref{sec:IHO-SM} we discuss the HO, IHO and dual-IHO physics and their role in QFT, SM, QFTCS. We highlight how quantum mechanics and QFT usually rely on the presumption of the arrow of time, but the occurrence of IHO instabilities challenges this view. In Sec.~\ref{sec:DQFT}, we review the formulation of direct-sum quantum (field) theory which splits the quantum states as direct-sum of two components with  two arrows of time in the parity conjugate geometric superselection sectors. We also detail on how this approach unveils a new understanding of quantum IHO and its connection of the spectrum with non-trivial zeros of the Riemann zeta function.  In Sec.~\ref{sec:QG-Renorm}  we present a detailed review of quadratic gravity and its spin-0 and spin-2 degrees of freedom. We especially discuss the cases of spin-2 ghost and the dual-IHO spin-2 cases exclusively. We discuss on the literature so-far on saving unitarity of quadratic gravity with spin-2 ghost and make heuristic comparisons between various approaches. In Sec.~\ref{sec:UQQG-IHOrenorm} we explore the case of dual-IHO spin--2 field naturally restoring unitarity in quadratic gravity. We present the analysis of scattering amplitudes, 1-loop effects and renormalizability with the virtual contributions from dual-IHO spin--2 field in the physical processes. We explicitly prove how the dual spin--2 field does not enter the unitarity cuts, does not go on-shell and consequently the optical theorem can be respected. We discuss the quantization of dual-IHO spin--2 sector with DQFT approach, which cannot be interpreted as a particle and the asymptotic state. In Sec.~\ref{sec:safe-begin}, we discuss the finite action principle and its role in avoiding singular solutions in the unitary quadratic gravity. We argue that Starobinsky inflationary scenario could be an emergent in QQG since the dual-IHO spin-2 sector can potentially drive the Universe towards a safe beginning. We analyze the UV behavior of 1-loop beta functions in unitary QQG and find the analogies with quantum electrodynamics (QED). In Sec.~\ref{sec:QG-early universe} we compute predictions of inflationary scenario in the unitary QQG which can be tested with future gravitational wave data. We derive,  in particular, parity asymmetry of primordial power spectra that naturally emerges with DSI in QQG. In Sec.~\ref{sec:conclusions}, we provide the highlights of our investigation with future outlook. The Appendix.~\ref{sec:ZT-derivation} derives the effective action for the massless graviton during inflation. 

We work in Minkowski space with signature $\eta_{\mu\nu}=\mathrm{diag}(-,+,+,+)$, so
$ k^2 \equiv \eta_{\mu\nu}k^\mu k^\nu = -k_0^2 + \mathbf{k}^2.
$
We use the Feynman prescription $-i\epsilon$, i.e., denominators are of the form
$
\frac{1}{k^2 + m^2 - i\epsilon}\,.
$
Throughout the paper, we use an overbar to indicate background quantities and use the natural units $\hbar = c=1, M_p^2 = \frac{1}{8\pi G}$. 

\section{The inverted harmonic oscillator  and its prevalence in the standard model \& quantum aspects of gravity}
\label{sec:IHO-SM}

The IHO occupies a distinctive position in quantum theory, lying conceptually between the stable harmonic oscillator and genuinely pathological ghost systems. While the ordinary harmonic oscillator describes fluctuations around a stable minimum and underpins much of perturbative quantum field theory, the IHO governs dynamics near unstable saddle points and encodes exponential sensitivity rather than oscillatory motion. Importantly, this instability should not be confused with the presence of negative norm states or violations of unitarity. In quantum field theory, IHOs arise naturally in physically well-understood settings, including tachyonic mass terms in the Higgs mechanism and the near-horizon dynamics of fields in curved spacetime. In both cases, the underlying Hamiltonian remains Hermitian and the Hilbert space well defined, even though the spectrum is continuous and unbounded. This sharply contrasts with ghost Hamiltonians, where negative kinetic terms lead to fundamental inconsistencies such as loss of unitarity or negative probabilities.

This distinction motivates treating IHO dynamics as a controlled and physically meaningful description of instability or horizon-induced amplification, rather than as evidence of a pathological degree of freedom. In the following, we contrast the HO, ghost, and IHO Hamiltonians explicitly, and clarify the role played by each in quantum field theory and gravity.

The common lore in learning and teaching physics majorly involves understanding the HO, whose Hamiltonian is bounded from below as 
\begin{equation}
      H_{ho} = \frac{p^2}{2m}+ \frac{1}{2}m\omega^2 q^2 = \frac{\omega}{2}\LF \tilde p^2+ \tilde q^2 \RF,\quad \omega>0
\end{equation}
where $\Tilde{p} = \frac{p}{\sqrt{ m\omega}},\quad \Tilde{q}= \sqrt{m\omega} q$. This means the position and momenta are the oscillatory functions and the phase space contains the compact trajectories. 
\begin{equation}
    H_{ho} = \frac{\omega}{2}\LF \tilde{p}^2+\tilde{q}^2 \RF,\quad \tilde{p}= \sqrt{\vert E\vert }\cos\LF \omega t \RF,\quad \tilde{q}= \sqrt{\vert E\vert }\sin\LF \omega t \RF,\quad E>0
\end{equation}
When we go to field theory we basically collect infinitely many HOs coupled in space. To illustrate this, let us consider an example of a Klein-Gordon (KG) field in 1+1 dimensions 
\begin{equation}
\begin{aligned}
S^{1+1}_{KG} & = \frac{1}{2}\int dt_p \Bigg\{\int dx\Bigg[ \LF \frac{\pd\phi}{\pd t_p} \RF^2- \LF \frac{\pd\phi}{\pd x} \RF^2 -m^2\phi^2 \Bigg]\Bigg\}  \\ 
 & = \frac{1}{2}\int dt_p \Bigg\{\int dx\, \phi\Bigg[ \LF -\frac{\pd^2}{\pd t_p^2} \RF +\LF \frac{\pd^2}{\pd x^2} \RF -m^2 \Bigg]\phi \Bigg\}
\end{aligned}
\label{11KGaction}
\end{equation}
which is a continuous approximation of $n$ HOs separated by an infinitesimal distance $\delta x$, 
\begin{equation}
S^{1+1}_{KG} = \frac{1}{2}\int dt_p \Bigg\{\sum_n \delta x\Bigg[ \LF \frac{\pd\phi_n}{\pd t_p} \RF^2-\frac{1}{\LF \delta x \RF^2}\LF \phi_{n+1}-\phi_n \RF^2 -  m^2\phi_n^2 \Bigg]\Bigg\}
\label{lattice-action}
\end{equation}
which is known as a lattice model. We can easily check that in the limit $n\to \infty$ and $\delta x\to 0$ the action \eqref{lattice-action} becomes \eqref{11KGaction}. 

The entire (quantum) field theory is built on an understanding of (quantum) HO physics. However, one essential ingredient of the SM, the Higgs field's $\mathbb Z_2$ symmetry breaking, is, in fact, remarkably associated with IHO physics instead. Before elucidating this fact, we recall here the quick understanding of the IHO phase space. The Hamiltonian of IHO is 
\begin{equation}
	H_{iho} =
\frac{ \omega}{2} \LF \Tilde{p}^2-\Tilde{q}^2 \RF 
	\label{IHOhamil}
\end{equation}
which is indefinite and unbounded. If we develop a field theory out of IHO, we get a classical field with a negative mass square term 
\begin{equation}
    S^{1+1}_{ihoKG}  = \frac{1}{2}\int dt_p \Bigg\{\int dx\Bigg[ \LF \frac{\pd\phi}{\pd t_p} \RF^2- \LF \frac{\pd\phi}{\pd x} \RF^2 -(-\mu^2)\phi^2 \Bigg]\Bigg\}
\end{equation}
which is often called the tachyonic field, but here we stick to the terminology of "IHO field". In 1+3D, the IHO field is 
\begin{equation}
    S^{1+3}_{ihoKG}  = \frac{1}{2}\int dt_p \Bigg\{\int d^3x\Bigg[ \LF \frac{\pd\phi}{\pd t_p} \RF^2- \LF \nabla\phi \RF^2 -(-\mu^2)\phi^2 \Bigg]\Bigg\},\quad \mu^2>0
    \label{IHOfield}
\end{equation}
IHO Hamiltonian is different from the negative definite Hamiltonian (associated with a ghost degree of freedom, which we discuss in the later sections) 
\begin{equation}
    H_{\rm ghost} = -\frac{\omega}{2}\LF \tilde p^2+ \tilde q^2 \RF
    \label{ghostho}
\end{equation}
which is not bounded from below. A Hilbert space associated with a negative definite Hamiltonian leads to violation of unitarity because of negative probabilities or a negative metric. In quantum mechanics, probabilities are defined through the inner product.
Unitarity requires that time evolution preserve a positive definite norm,
\begin{equation}
\langle \psi(t)\,|\,\psi(t)\rangle
=
\langle \psi(0)\,|\,\psi(0)\rangle
>
0
\qquad \forall\, t .
\end{equation}
If the inner product is indefinite, negative-norm states arise $ \langle \psi | \psi \rangle < 0$, and the probabilistic interpretation breaks down \cite{Salvio:2018crh}. Consequently, a Hilbert space with an indefinite metric does not admit a unitary quantum theory. Thus, if a theory contains a degree of freedom with a negative definite Hamiltonian, unitarity is certainly violated. We later discuss this in the context of non-unitary quadratic gravity. 

The crux of this paper is to obtain a unitary QQG, which involves IHO physics. The seemingly simple IHO system \eqref{IHOhamil} challenges the way we normally understand QM. The Hamiltonian equations of motion for IHO give us the following classical solution 
\begin{equation}
    \begin{aligned}
        \tilde p & = \sqrt{\vert E\vert }\sinh\LF \omega t \RF,\,\tilde q = \sqrt{\vert E\vert }\cosh\LF \omega t \RF,\,t: -\infty\to \infty\,(\rm Region\,I,\,E<0)  \\
          \tilde p & = \sqrt{\vert E\vert }\sinh\LF \omega t \RF,\,\tilde q = -\sqrt{\vert E\vert }\cosh\LF \omega t \RF,\,t: \infty\to -\infty\,(\rm Region\,II,\,E<0) \\ 
        \tilde p & = \sqrt{\vert E\vert }\cosh\LF \omega t \RF,\,\tilde q = -\sqrt{\vert E\vert }\sinh\LF \omega t \RF,\,t: -\infty\to \infty\,(\rm Region\,III,\,E>0)\\
          \tilde p & = -\sqrt{\vert E\vert }\cosh\LF \omega t \RF,\,\tilde q = -\sqrt{\vert E\vert }\sinh\LF \omega t \RF,\,t: \infty\to -\infty\,(\rm Region\,IV,\,E>0) \,,
    \end{aligned}
    \label{phihoeq}
\end{equation}
We can write the IHO Hamiltonian \eqref{IHOhamil} in the canonically rotated coordinates as 
\begin{equation}
	H_{iho} = \frac{ \omega}{2}\LF Q\cdot P+P\cdot Q  \RF,\quad Q= \frac{\Tilde{p}+\Tilde{q}}{\sqrt{2}},\, \quad P= \frac{\Tilde{p}-\Tilde{q}}{\sqrt{2}}
 \label{BKHamilt}
\end{equation}
The form of \eqref{BKHamilt} is famously known as Berry-Keating's IHO, whose energy spectrum is matched with non-trivial zeros of the Riemann zeta function under certain quantization conditions that can naturally emerge through our direct-sum quantum theory \cite{Gaztanaga:2025awe}. 
\begin{figure}
    \centering
    \includegraphics[width=0.9\linewidth]{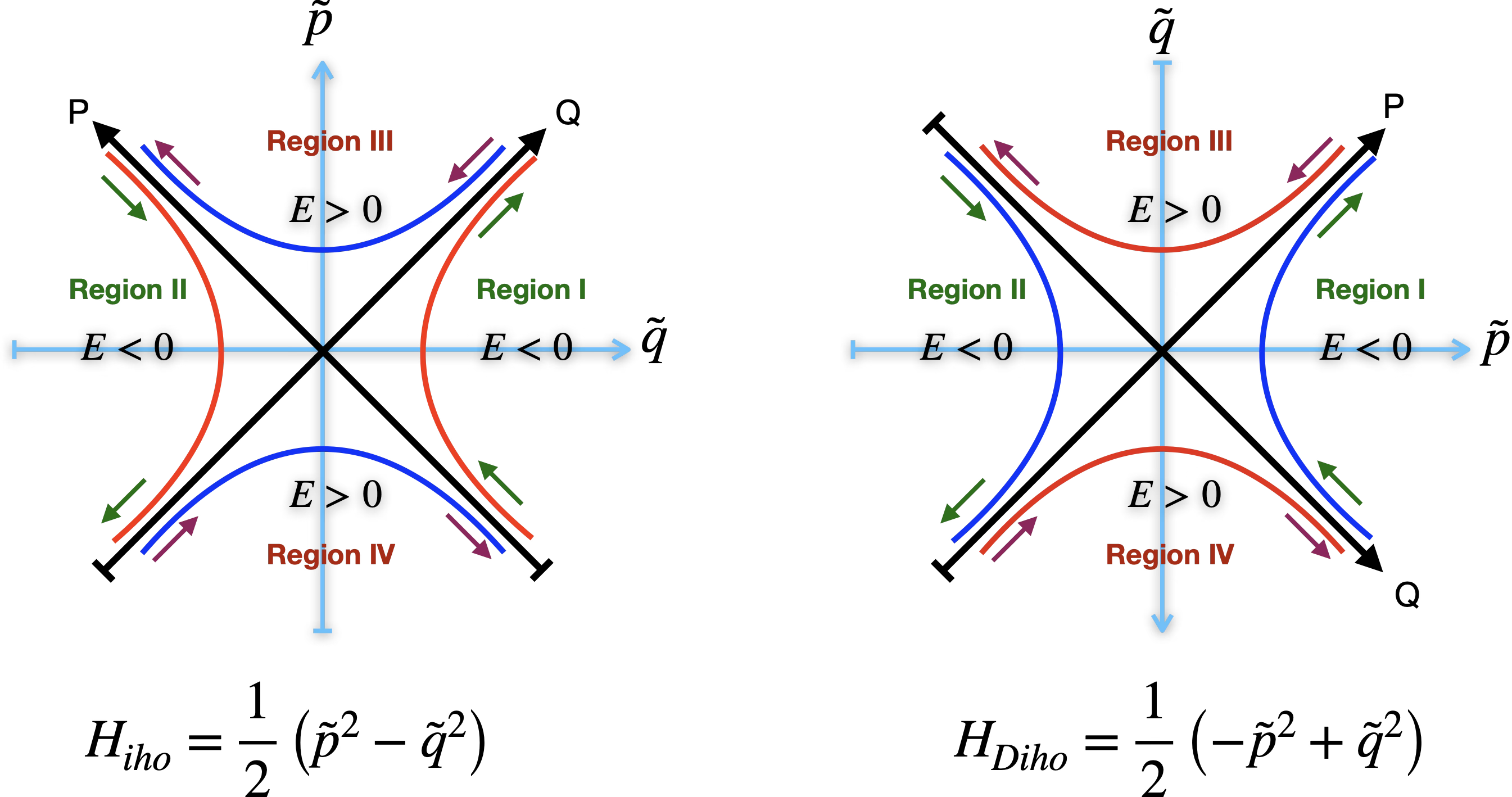}
    \caption{Phase space of the IHO (left panel) and the dual-IHO (right panel) representing doubly degenerate positive and negative energy solutions in \eqref{sol2BK}. The negative energy trajectories are given by $Q>0,\, P<0$ and $Q<0,\, P>0$ whereas the positive energy trajectories are $Q>0,\,P>0$ and $Q<0,\,P<0$. These double degenerate trajectories are related by \eqref{discreteTPQ} whereas the positive and negative energy regions are related by \eqref{BKPE}.}
    \label{fig:IHO-DualIHO}
\end{figure}

It is worth noticing that the IHO phase space can also be mimicked by another equivalent physical system which we call the dual-IHO, whose Hamiltonian is given by 
\begin{equation}
	H_{Diho} =
\frac{ \omega}{2} \LF -\Tilde{p}^2+\Tilde{q}^2 \RF= -\frac{ \omega}{2}\LF Q\cdot P+P\cdot Q  \RF
	\label{DIHOhamil}
\end{equation}
The dual-IHO Hamiltonian \eqref{DIHOhamil} is indefinite similar to the IHO Hamiltonian \eqref{IHOhamil}. The phase space of the dual-IHO can be understood as a $90^\circ$ (clockwise) rotated version of IHO (See Fig.~\ref{fig:IHO-DualIHO}), where in each region the role of position and the momentum gets swapped (i.e., $E\to -E$) since the classical solutions of dual-IHO are (See the right panel of Fig.~\ref{fig:IHO-DualIHO})
\begin{equation}
    \begin{aligned}
        \tilde q & = \sqrt{\vert E\vert }\sinh\LF \omega t \RF,\,\tilde p = \sqrt{\vert E\vert }\cosh\LF \omega t \RF,\,t: -\infty\to \infty\,(\rm Region\,I,\,E<0)  \\
          \tilde q & = \sqrt{\vert E\vert }\sinh\LF \omega t \RF,\,\tilde p = -\sqrt{\vert E\vert }\cosh\LF \omega t \RF,\,t: \infty\to -\infty\,(\rm Region\,II,\,E<0) \\ 
        \tilde q & = -\sqrt{\vert E\vert }\cosh\LF \omega t \RF,\,\tilde p = -\sqrt{\vert E\vert }\sinh\LF \omega t \RF,\,t: -\infty\to \infty\,(\rm Region\,III,\,E>0)\\
          \tilde q & = \sqrt{\vert E\vert }\cosh\LF \omega t \RF,\,\tilde p = -\sqrt{\vert E\vert }\sinh\LF \omega t \RF,\,t: \infty\to -\infty\,(\rm Region\,IV,\,E>0) \,,
    \end{aligned}
    \label{dualphihoeq}
\end{equation}
We witness an important symmetry here in both \eqref{phihoeq} and \eqref{dualphihoeq}, the $\Pc\Tc$ (Parity $\Pc$ and Time reversal $\Tc$), i.e., $t\to -t,\,\tilde{q}\to -\tilde{q}$.
The phase space of IHO and dual-IHO are depicted in the left and right panels of Fig.~\ref{fig:IHO-DualIHO} respectively. The regions ${\rm I (III)}$ and ${\rm II (IV)}$ are $\Pc\Tc$ conjugates of each other. 
The lines of $\tilde p = \pm \tilde q$ are called separatrices, which divide the classical trajectories into four distinct domains. We can notice that these regions come as degenerate pairs with opposite arrows of time with $\Pc\Tc$ symmetry
\begin{equation}
    Q\to -Q,\quad P\to -P,\quad t\to -t.  
    \label{discreteTPQ}
\end{equation}
The positive and negative energy regions of IHO and dual-IHO in Fig.~\ref{fig:IHO-DualIHO} are related by the following transformations 
\begin{equation}
    Q\to \mp P,\quad P\to \pm Q\implies \tilde{p}\to \mp \tilde{q},\quad \tilde{q}\to \pm \tilde p\,.
    \label{BKPE}
\end{equation}
The solutions of the \eqref{IHOhamil} and \eqref{DIHOhamil} can also be written in canonically rotated coordinates as
\begin{equation}
	Q= Q_0 e^{\omega t},\quad P= P_0 e^{-\omega t},\quad H_{iho} = \frac{\omega}{2}E = Q_0P_0 = - H_{Diho},
 \label{sol2BK}
\end{equation}
where we can notice that exponentially growing or decaying behavior of $Q,\, P$ such that they complement each other. 

The position- and momentum-space wavefunctions of the IHO Hamiltonian (which are equivalent for dual-IHO, too)
(for $Q>0$ and $E<0$, in units $\omega=1$) are given by
\cite{Sierra:2016rgn,Berry1999,Ullinger:2022xmv}
\begin{equation}
    \Psi(Q)=\frac{C}{\sqrt{2\pi\hbar}}|Q|^{-\frac12+\frac{i|E|}{\hbar}},
    \qquad
    \Psi(P)=\frac{1}{\sqrt{2\pi\hbar}}|P|^{-\frac12-\frac{i|E|}{\hbar}}
    (2\hbar)^{\frac{i|E|}{\hbar}}
    \frac{\Gamma\!\left(\frac14+\frac{i|E|}{2\hbar}\right)}
         {\Gamma\!\left(\frac14-\frac{i|E|}{2\hbar}\right)} .
\end{equation}
These states satisfy the appropriate orthogonality and completeness
relations \cite{Ullinger:2022xmv,Sundaram:2024ici}. Explicit expressions
in $(\tilde q,\tilde p)$ and a discussion of nonsingular probability
densities at phase-space horizons can be found in
\cite{BARTON1986322,Sundaram:2024ici,Ullinger:2022xmv}. Under time
evolution the IHO wavefunction becomes delocalized, precluding a
standard particle interpretation, in close analogy with quantum field
theory in curved spacetime \cite{Gaztanaga:2025awe}. 

\subsection{Normalizability of Eigenstates and existence of ground state in HO, Ghost, and (dual) IHO} 

Here we point out an important distinctions between HO, ghost and (dual) IHO physical systems with regard to the normalizability of the wave functions and the existence of ground state.

The Hamiltonian $H_{\rm HO} = \frac{\omega}{2}(\tilde p^2 + \tilde q^2)$ is bounded below,
with discrete spectrum $E_n = \omega\!\left(n + \tfrac{1}{2}\right)$.
The eigenstates are Hermite functions,
\begin{equation}
    \psi_n(\tilde q) = \mathcal{N}_n\, H_n(\tilde q)\, e^{-\tilde q^2/2},
\end{equation}
which belong to $L^2(\mathbb{R})$: they decay as $e^{-\tilde q^2/2}$ and are
square-integrable.  Every eigenstate, including the ground state
$\psi_0 = \pi^{-1/4}e^{-\tilde q^2/2}$, satisfies $\int|\psi_n|^2\,d\tilde q = 1$.
The spectrum is bounded below and the ground state is the unique
normalizable state of lowest energy.

The ghost Hamiltonian $H_{\rm ghost} = -\tfrac{\omega}{2}(\tilde p^2+\tilde q^2) = -H_{\rm HO}$
simply reverses the sign of every energy level:
$E_n^{\rm ghost} = -\omega\!\left(n+\tfrac{1}{2}\right)$.
Crucially, the eigenstates are identical to those of the HO, i.e. the same Hermite functions $\psi_n(\tilde q)$.  Reversing the sign of the Hamiltonian does not change the eigenfunctions, only the eigenvalues.  Therefore,
the ghost has a normalizable highest energy state $\psi_0$ with $E_0 = -\omega/2$, which serves as the reference vacuum. So All ghost eigenstates are in $L^2(\mathbb{R})$. The spectrum is bounded above (by $-\omega/2$) and unbounded below. It is important to note that the pathology of the ghost is not in the wave functions, they are
perfectly normalizable Hermite functions, but in the inner-product
structure. Since the Hamiltonian is negative definite, canonical quantization condition is $[a,a^\dagger]=-1$,
and the $n$-particle state has norm $\langle n|n\rangle = (-1)^n$.
The vacuum has positive norm $+1$, but the one-particle state has negative
norm $-1$, the two-particle state positive, and so on.  The metric is
indefinite.  

The IHO Hamiltonian
$H_{\rm IHO} = \tfrac{\omega}{2}(\tilde q^2 - \tilde p^2) = \tfrac{\omega}{2}(\tilde q^2 + \partial_{\tilde q}^2)$
is fundamentally different from both the HO and the ghost.
The eigenvalue equation reads
\begin{equation}
    \frac{\omega}{2}(\tilde q^2+\partial_{\tilde q}^2)\,\psi_E = E\,\psi_E
    \qquad\Longrightarrow\qquad
    \partial_{\tilde q}^2\,\psi_E = \!\left(\tilde q^2 - \frac{2E}{\omega}\right)\psi_E.
\end{equation}
The solutions are parabolic cylinder functions $D_\nu(\tilde q)$ with
$\nu = \frac{iE}{\omega} - \tfrac{1}{2}$.  Their large argument asymptotics are \cite{Sundaram:2024ici}
\begin{align}
    D_\nu(\tilde q) &\sim \tilde q^\nu\, e^{-\tilde q^2/4}
    \quad\text{as}\quad \tilde q\to+\infty, \nonumber \\[4pt]
    D_\nu(\tilde q) &\sim \tilde q^\nu\, e^{-\tilde q^2/4}
             + \frac{\Gamma(-\nu)}{\sqrt{2\pi}}\,\tilde q^{-\nu-1}\,e^{+\tilde q^2/4}
    \quad\text{as}\quad \tilde q\to-\infty.
    \label{eq:IHO-eigen}
\end{align}
The second term grows as $e^{+\tilde q^2/4}$ from $-\infty$ and is not
square-integrable.  Normalizability requires $\Gamma(-\nu)=0$, i.e.\
$\nu = 0,1,2,\ldots$, which gives
\begin{equation}
    E = -i\omega\!\left(n+\tfrac{1}{2}\right), \qquad n = 0,1,2,\ldots
\end{equation}
These energies are purely imaginary which means the IHO Hamiltonian must be non-Hermitian. If we demand the IHO Hamiltonian to be Hermitian (which we do in this work), then only real values of energies are allowed, for any real energies the IHO eigenmodes are non-normalizable.  However, every eigenstate is only Dirac-delta normalizable:
$\langle E|E'\rangle = \delta(E-E')$.
This is categorically different from both the HO and the ghost.  As we argued before and also in Fig.~\ref{fig:IHO-DualIHO}, since the dual-IHO is an identical physical system as IHO, the conclusions about ground state non-normalizability apply straightforwardly.
We summarize all these concepts in Table.~\ref{CompareHs}. 

\begin{table}[h]
\centering
\renewcommand{\arraystretch}{1.6}
\begin{tabular}{llll}
\hline\hline
\textbf{System} & \textbf{Hamiltonian} & \textbf{Nature} & \textbf{Ground state in $L^2(\mathbb{R})$} \\
\hline
Harmonic oscillator
    & $\dfrac{1}{2}\!\left(\tilde{p}^2+\tilde{q}^2\right)$
    & Positive definite
    & Yes \\[6pt]
IHO
    & $\dfrac{1}{2}\!\left(\tilde{p}^2-\tilde{q}^2\right)$
    & Indefinite
    & No  \\[6pt]
Ghost
    & $-\dfrac{1}{2}\!\left(\tilde{p}^2+\tilde{q}^2\right)$
    & Negative definite
    & Yes \\[6pt]
dual-IHO
    & $\dfrac{1}{2}\!\left(-\tilde{p}^2+\tilde{q}^2\right)$
    & Indefinite
    & No \\[6pt]
\hline\hline
\end{tabular}
\caption{Comparison of the HO, IHO, ghost, and dual-IHO Hamiltonians
(in units $\omega=1$).  The IHO and dual-IHO are identical physical
systems sharing the same phase space (see Fig.~\ref{fig:IHO-DualIHO}). }
\label{CompareHs}
\end{table}

\subsection{Arrow of time in standard quantum theory and the challenge from IHO}

In QM, Schr\"odinger equation is a first-order differential equation in time
\begin{equation}
	i\frac{\partial \vert \Psi\rangle }{\partial t_p} = \hat{H}\vert \Psi \rangle = \Ec\vert\Psi\rangle,\quad t_{p}: -\infty \to \infty,
	\label{eq:Sch1}
\end{equation}
where one defines an arrow of time with respect to which a positive energy state is defined 
\begin{equation}
	\vert \Psi\rangle_{t_p} = e^{-i\Ec t_p}\vert \Psi\rangle_0,\quad \Ec>0,\quad t_p: -\infty \to \infty
\end{equation}
The construction of QFT entirely stands on this assumption on the arrow of time \cite{Coleman:2018mew,Srednicki:2007qs,Donoghue:2019ecz}. Quantum theory is known to be time symmetric \cite{Schrodinger1956,tHooft:2018jeq,Donoghue:2019ecz}. If we replace everywhere in quantum theory "$i\to -i$", we would change the arrow of time $t: -\infty \to \infty$ to $t: \infty \to -\infty$. This means Sc
hr\"odinger equation with opposite arrow of time would become  
\begin{equation}
	-i\frac{\partial \vert \Psi\rangle }{\partial t_p} = \hat{H}\vert \Psi \rangle = \Ec\vert\Psi\rangle,\quad t_{p}: \infty \to -\infty,
	\label{eq:Sch1}
\end{equation}
where a positive energy state would then be defined as 
\begin{equation}
	\vert \Psi\rangle_{t_p} = e^{i\Ec t_p}\vert \Psi\rangle_0,\quad \Ec>0,\quad t_p: \infty \to -\infty
\end{equation}
Changing the arrow of time does not change the results in quantum theory. Therefore, it often appears that fixing an arrow of time is just an observer's choice that can be freely made. However, it is important to notice that the (quantum) time symmetry inherent in many physical systems, although classically entropy always grows (i.e., unidirectional thermodynamic arrow of time). Minkowski spacetime $ds^2 = -dt_p^2+d\textbf{x}^2$ is $\Pc\Tc$ symmetric, therefore quantum fields behave the same whatever the arrow of time we choose. A question we can pose here is, can we formulate a quantum theory without choosing an arrow of time? An answer to this is the next section, the direct-sum quantum theory. 

The IHO phase spaces in Fig.~\ref{fig:IHO-DualIHO} pose us a new challenge where parity conjugate worlds split into the domains with opposite arrow of time. It has been an open problem since Berry and Keating \cite{Berry1999} first looked at quantization of this physical system which poses a challenge to the standard Shr\"{o}dinger equation's practice of fixing an arrow of time in the first place. Berry and  Keating suggested an antipodal identification $\LF Q,\,P \RF \leftrightarrow \LF -Q,\,-P \RF$ so that one can achieve periodicity with the $\Pc\Tc$ symmetry hyperbolic trajectories. This proposal, in an ad-hoc way, justified the matching of the IHO spectrum with the non-trivial zeros of the Riemann zeta function; the physical understanding of this proposal was an open question until our recent study \cite{Gaztanaga:2025awe}. harmonic oscillator (IHO) physics underlies the Higgs sector through the tachyonic instability, practical Standard Model (SM) calculations only involve fluctuations of the Higgs field around the true vacuum, where ordinary harmonic-oscillator (HO) physics and a particle interpretation apply. For this reason, the physical Higgs boson is understood as a fluctuation about the minimum of the potential.
To explicitly see this, let us write down the Higgs potential 
\begin{equation}
V_H = -\frac{\mu_H^2}{2} H_{\rm sm}^\dagger H_{\rm sm}+ \frac{\lambda_H}{4}\bigl(H_{\rm sm}^\dagger H_{\rm sm}\bigr)^2 ,
\label{Higgs-pot}
\end{equation}
where $H_{\rm sm}=\frac{1}{\sqrt{2}}he^{i\theta_h}$ is the complex ${\rm SU}(2)$ Higgs doublet, $\mu_H^2>0$ is not the physical Higgs mass, and $\lambda_H>0$ denotes the self-interaction. The potential possesses two degenerate minima $v=\pm\sqrt{\mu_H^2/\lambda_H}$ due to the $\mathbb Z_2$ symmetry $H{\rm sm}\to -H_{\rm sm}$. Writing $\mu_H^2=\lambda_H v^2$ and adding an irrelevant constant yields the Mexican-hat form
\begin{equation}
V_H=\frac{\lambda_H}{4}\bigl(H_{\rm sm}^\dagger H_{\rm sm}-v^2\bigr)^2 ,
\end{equation}
which still contains a negative mass-squared (tachyonic) term.

After spontaneous symmetry breaking, expanding around the minimum in the unitary gauge,
\begin{equation}
H_{\rm sm}=\frac{1}{\sqrt{2}}
\begin{pmatrix}
0\\ v+\phi_h
\end{pmatrix},
\qquad
m_h^2=2\lambda_H v^2,
\end{equation}
one obtains the physical Higgs field $\phi_h$ with positive mass squared. Thus, prior to symmetry breaking, the Higgs sector may be viewed as an infinite collection of self-interacting IHOs, while after symmetry breaking, it reduces to ordinary harmonic oscillators describing the massive Higgs mode. The associated tachyonic instability is typically treated as a non-perturbative transition from a false to a true vacuum via Euclidean instantons \cite{Coleman:1977py}.

In practice, detailed IHO quantum mechanics is unnecessary for SM phenomenology, since calculations are performed perturbatively about the true vacuum. Nevertheless, the quantum theory of the IHO is highly non-trivial, originating from the seminal work of Berry and Keating \cite{Berry1999}, and remains an active subject across a wide range of problems in theoretical physics \cite{Subramanyan:2020fmx}.

Quantum mechanically, the inverted harmonic oscillator (IHO) admits two main formulations \cite{Sierra:2007du}. 
(i) The Berry and Keating  quantization identifies antipodal phase-space points $(Q,P)\!\sim\!(-Q,-P)$ and imposes dilatation based boundary conditions, yielding a discrete spectrum whose counting function matches the average distribution of the non trivial Riemann zeros. 
(ii) The scattering quantization treats the IHO as a tunneling problem, but phase-space horizons obstruct tunneling unless one allows unphysical evolution between negative  and positive energy sectors \cite{BALAZS1990123}, and the link to Riemann zeros remains unclear.

For $Q>0$ and $E<0$, the position and momentum wavefunctions are (in the $\omega=1$) \cite{Sierra:2016rgn,Berry1999,Ullinger:2022xmv}
\begin{equation}
\Psi(Q)=\frac{C}{\sqrt{2\pi\hbar}}|Q|^{-1/2+i|E|/\hbar},\qquad
\Psi(P)=\frac{1}{\sqrt{2\pi\hbar}}|P|^{-1/2-i|E|/\hbar}(2\hbar)^{i|E|/\hbar}
\frac{\Gamma\!\left(\frac14+\frac{i|E|}{2\hbar}\right)}
{\Gamma\!\left(\frac14-\frac{i|E|}{2\hbar}\right)},
\end{equation}
which form a complete orthogonal set. These wavefunctions delocalize under time evolution, precluding a particle interpretation and mirroring the situation in curved-spacetime QFT.
Introducing quantum cutoffs $|Q|\!\ge\!\ell_Q$ and $|P|\!\ge\!\ell_P$ with $\ell_Q\ell_P=2\pi\hbar$ discretizes the spectrum, giving the state count
\begin{equation}
N(E)=\frac{|E|}{2\pi\hbar}\!\left(\ln\frac{|E|}{2\pi\hbar}-1\right)+\frac{7}{8},
\end{equation}
which matches the asymptotic density of Riemann zeros under the identification $\bar T\!\leftrightarrow\!|E|/\hbar$.
The deep origin of this correspondence lies in the discrete phase-space symmetries of the IHO, which form the dihedral group $D_4$ and enforce the BK boundary condition
\begin{equation}
Q^{1/2}\zeta\!\left(\tfrac12-\tfrac{iE}{\hbar}\right)\Psi(Q)
+P^{1/2}\zeta\!\left(\tfrac12+\tfrac{iE}{\hbar}\right)\Psi(P)=0,
\label{boundarycondiBK}
\end{equation}
implying that $\Psi(P)$ is the time-reversal of $\Psi(Q)$ \cite{Berry1999}, though its global geometric meaning remains unresolved.
The IHO wavefunction is also an eigenfunction of the Weyl reflected Laplace--Beltrami operator
\begin{equation}
L_R=-Q^2\partial_Q^2-2Q\partial_Q
=\Big(\tfrac12-\tfrac{i\hat H_{\rm IHO}}{\hbar\omega}\Big)
\Big(\tfrac12+\tfrac{i\hat H_{\rm IHO}}{\hbar\omega}\Big),
\label{LBOp}
\end{equation}
with eigenvalue $\frac14+\frac{E^2}{\hbar^2}$.
The Berry and Keating identification $(Q,P)\!\sim\!(-Q,-P)$ parallels antipodal maps in de Sitter and black-hole spacetimes \cite{Schrodinger1956,tHooft:2016qoo}, effectively forming a mathematical bridge between phase-space regions with opposite arrows of time.
Finally, the IHO is invariant under hyperbolic dilatations
\begin{equation}
Q\to e^\lambda Q,\qquad P\to e^{-\lambda}P,
\end{equation}
generated by the Hamiltonian vector field
\begin{equation}
X_H=Q\partial_Q-P\partial_P.
\end{equation}
The associated eigenfunctions take the universal form
\begin{equation}
\Psi_D(Q,P)=\Big(\frac{Q}{P}\Big)^\lambda g(QP),
\end{equation}
and the Wigner function correspondingly reads
\begin{equation}
W_E(Q,P)=\mathcal N\Big(\frac{Q}{P}\Big)^{iE/\hbar}g_W(QP).
\end{equation}

These results show that IHO physics is fundamentally tied to dilatation symmetry, antipodal phase-space geometry, and time-reversal pairing features that reappear in QFT of black-hole \cite{Kumar:2023hbj}, de Sitter \cite{Kumar:2024oxf}, and Rindler spacetimes \cite{Kumar:2023ctp}.

\subsection{Occurrence of IHO physics in QFT in curved spacetime} 

Conundrum between GR and QM first emerged when one first tried to apply them together at the gravitational horizons. The seminal work of Einstein and Rosen (ER) in 1935 on particle problem in GR elucidates the conundrum and hints on the breakdown of particle interpretation at the black hole (BH) horizon. On the top of that, ER paper does suggest a much needed necessity of understanding a particle in GR using mathematical bridges between two sheets of spacetime with opposite arrows of time. Schr\"{o}dinger's 1956 monograph \cite{Schrodinger1956} on expanding universes (in de Sitter  space) suggested again a description of particle to be by "rods" (analogous to ER's mathematical bridges) connecting antipodal points in spacetime. Schr\"{o}dinger insisted this antipodal identification to preserve unitarity, the framework also known as Elliptic interpretation of de Sitter (dS) space \cite{Parikh:2002py}. Schr\"{o}dinger's and ER's ideas also independently reflected in recent works by 't Hooft in the context of BH physics \cite{tHooft:2016qoo}  (See \cite{Gaztanaga:2025awe} for the detailed discussion on these historical developments). 

The inverted harmonic oscillator (IHO) provides a canonical example of a quantum system with an unbounded quadratic potential, whose classical instability does not preclude a consistent quantum description. Rather than admitting bound states, the IHO is naturally formulated in terms of scattering states with continuous spectra and hyperbolic time evolution. Such systems arise generically in the presence of horizons, saddle points of effective potentials, and tachyonic mass terms, and have appeared in diverse areas of theoretical physics. In the following, we summarize the key quantum properties of the inverted harmonic oscillator and clarify why it offers a natural language for describing unstable but non-pathological degrees of freedom in gravitational and field theoretic settings.

We can witness the arrow of time symmetry in dS by considering foliation of it in the flat Friedmann-Lemaître-Robertson-Walker (FLRW) metric of the following form 
\begin{equation}
    ds^2 = -dt^2+a^2(t)d\textbf{x}^2=\frac{1}{H^2\tau^2}\LF -d\tau^2+d\textbf{x}^2 \RF,\quad a(t)=e^{Ht}=-\frac{1}{H\tau},\quad R_{dS}=12H^2\,. 
    \label{FLRWdS}
\end{equation}
Here $R_{dS}$ denotes the curvature scalar of de Sitter space, and $\tau= \int \frac{dt}{a}$ is the conformal time, and $H=\frac{\dot a}{a}$ is the Hubble parameter.
The Schr\"{o}dinger's realization of expanding universes is
\begin{equation}
	{\rm Expansion\,\, of \,\,Universe}  \implies \begin{cases}
	 H>0,\quad t: -\infty \to \infty \implies \tau<0 \\ 
	 H<0,\quad t: \infty \to -\infty \implies \tau>0
	\end{cases}
 \label{tdSsym}
\end{equation}
Similarly, we can realize the contracting Universes as well. A massless Klein-Gordon (KG) field action in \eqref{FLRWdS} after a canonical rescaling $\phi = -\frac{1}{H\tau}\Phi $ would reads
\begin{equation}
    S_{KG}^{\rm fdS} = \frac{1}{2} \int d\tau \, d^3x \left[ (\partial_\tau \Phi)^2 - (\nabla \Phi)^2 -(- \tilde{\mu}_{\text{eff}}^2(\tau))\, \Phi^2 \right],\quad   \tilde{\mu}_{\text{eff}}^2(\tau)  
= \frac{2}{\tau^2}  
    \label{flatdSKG}
\end{equation}
Comparing \eqref{flatdSKG} with \eqref{IHOfield}, we can immediately notice that the field $\Phi$ is an "IHO field" in the dS expanding Universe with a time-dependent effective (negative) mass square term that is symmetric under $\tau\to -\tau$. 

Even in the context of the Schwarzschild BH, there exist two arrows of time to characterize the exterior and interior of the BH. We could see this by the Schwarzschild BH metric
\begin{equation}
    ds^2 = -\LF 1-\frac{2GM}{r} \RF dt^2 + {\LF 1-\frac{2GM}{r} \RF}^{-1} dr^2+ r^2 d\Omega^2
    \label{eq:SBH}
\end{equation}
in Kruskal coordinates\footnote{\begin{equation}
       U = \pm \,4GM \sqrt{\Big\vert 1- \frac{r}{2GM}\Big\vert } \exp\LF -\frac{t-r}{4GM} \RF,\, V = \pm \,4GM \sqrt{\Big\vert 1-\frac{r}{2GM}\Big\vert } \exp\LF \frac{t+r}{4GM} \RF
       \label{Kruskalcoord}
   \end{equation}
   which obey 
   \begin{equation}
   	\begin{aligned}
   		UV& = 16G^2M^2\LF 1-\frac{r}{2GM}\RF \exp\LF {\frac{r}{2GM}}\RF,\quad \frac{U}{V} =  \pm\, \exp\LF{-\frac{t}{2GM}}\RF 
   	\end{aligned}
   	\label{eq:KS}
   \end{equation}} $(U,\,V)$  
\begin{equation}
    ds^2 = \frac{2GM}{r}e^{1-\frac{r}{2GM}}\LF -dT^2+dX^2\RF + r^2d\Omega^2 
    \label{KSmetr}
\end{equation}
where
\begin{equation}
T = \frac{U+V}{2\sqrt{e}},\,X=\frac{V-U}{2\sqrt{e}}
\label{TXdef}
\end{equation}
 From \eqref{eq:KS} we can notice that
\begin{equation}
    r>2GM \implies \begin{cases}
        U<0,\, V>0 \\ 
        U>0,\,V<0
    \end{cases},\quad r<2GM \implies \begin{cases}
        U>0,\, V>0 \\ 
        U<0,\,V<0
    \end{cases}
    \label{eq:extint}
\end{equation}
where we can notice the following discrete symmetry in both regions $r<2GM$ and $r>2GM$ 
\begin{equation}
   T\to -T \implies U\to -U,\, V\to -V \implies X\to -X
   \label{TtomT}
\end{equation}
As demonstrated in \cite{Gaztanaga:2025awe} quantum fields in BH spacetime has close correspondence with IHO physics. The exponentially growing and decaying character of $\LF U,
 V \RF$ \eqref{Kruskalcoord} generate the dilatation symmetries akin to the phase space of IHO we discussed in the last section (e.g., compare \eqref{sol2BK} and \eqref{Kruskalcoord}). It is explicitly proved \cite{Betzios:2020xuj,Betzios:2020wcv,Ullinger:2022xmv} that the wave function of a single state at the BH Horizon is characterized by the Berry and Keating type of IHO Hamiltonian \eqref{BKHamilt}. 

\subsection{Inflationary (quasi-de Sitter) quantum fluctuations}

\label{sec:inflaIHO}

Inflationary quantum fluctuations also contain IHO physics. This is an interesting example where gravity is dynamical, and we can see how gravitational fluctuations propagate through the exponentially expanding spacetime and leave their imprints as temperature fluctuations in the CMB and gravitational waves. Inflation is approximately quasi-de Sitter \cite{Starobinsky:1980te}, with mild departures encoded in the slow-roll parameters (defined in the Einstein frame)
\begin{equation}
    \epsilon \equiv \frac{d}{dt}\!\left(\frac{1}{H}\right) = -\frac{\dot H}{H^2}\,,\qquad 
    \eta \equiv \frac{\dot\epsilon}{H\epsilon}\,,
    \label{epsiloneta}
\end{equation}
where $H=\dot a/a$. Realistic inflation requires an additional scalar degree of freedom (e.g. an inflaton field or modified gravity such as $R+R^2$). Scalar and tensor fluctuations about a flat FLRW background can be parameterized as
\begin{equation}\label{ADMmetric}
ds^2=a^2(\tau)\Big[-(1+2\Phi)d\tau^2+\big((1-2\Psi)\delta_{ij}+h_{ij}\big)dx^idx^j\Big],
\end{equation}
with $\Phi,\Psi$ the Bardeen potentials (Newtonian gauge) and $h_{ij}$ transverse-traceless. The linearized constraint equations imply
\begin{equation}
\Phi=\Psi,\qquad \Psi' + \mathcal{H}\Phi = \frac{1}{2}\,\bar\phi'\,\delta\phi,
\end{equation}
leading to the standard second-order action for the curvature perturbation $\zeta \equiv \Psi+\frac{\dot{\bar\phi}}{H}\delta\phi$ \cite{Mukhanov:2005sc,Baumann:2018muz}:
\begin{equation}\label{scalar}
\delta^{(2)}S_{s}=\frac{1}{2}\int d\tau d^3x\, a^2\frac{\dot{\bar\phi}^{\,2}}{H^2}\left[\zeta^{\prime 2}-(\partial\zeta)^2\right].
\end{equation}
In canonical form, with $V_{MS}\equiv a\frac{\dot{\bar\phi}}{H}\zeta$,
\begin{equation}
\delta^{(2)}S_{s}=\frac{1}{2}\int d\tau d^3x\, V_{MS}\Big[-\partial_\tau^2+\partial_i^2-(-\mu_{\rm eff}^2)\Big]V_{MS},
\qquad 
\mu_{\rm eff}^2=\frac{1}{\tau^2}\big(2+2\epsilon+\eta\big),
\label{ihoinf}
\end{equation}
so that in quasi-dS ($\epsilon,\eta\ll1$) the Mukhanov--Sasaki mode experiences an effective negative mass squared, i.e. IHO-type dynamics \cite{Albrecht:1992kf}. The tensor sector is analogous:
\begin{equation}
\delta^{(2)}S_{h}=\frac{1}{2}\int d\tau d^3x\, u_{h}\Big[-\partial_\tau^2+\partial_i^2-(-\mu_{\rm eff}^2)\Big]u_{h},
\qquad 
\mu_{\rm eff}^2=\frac{1}{\tau^2}\big(2+2\epsilon\big),
\label{ihoinfhij}
\end{equation}
with
\begin{equation}
h_{ij}=\sum_{s=\times,+}\polarizationtensor^{s}_{ij}\,u_{s}.
\label{hijexp}
\end{equation}

\section{The Direct-Sum Quantum (Field) Theory}
\label{sec:DQFT}

\begin{figure}
    \centering
    \includegraphics[width=0.5\linewidth]{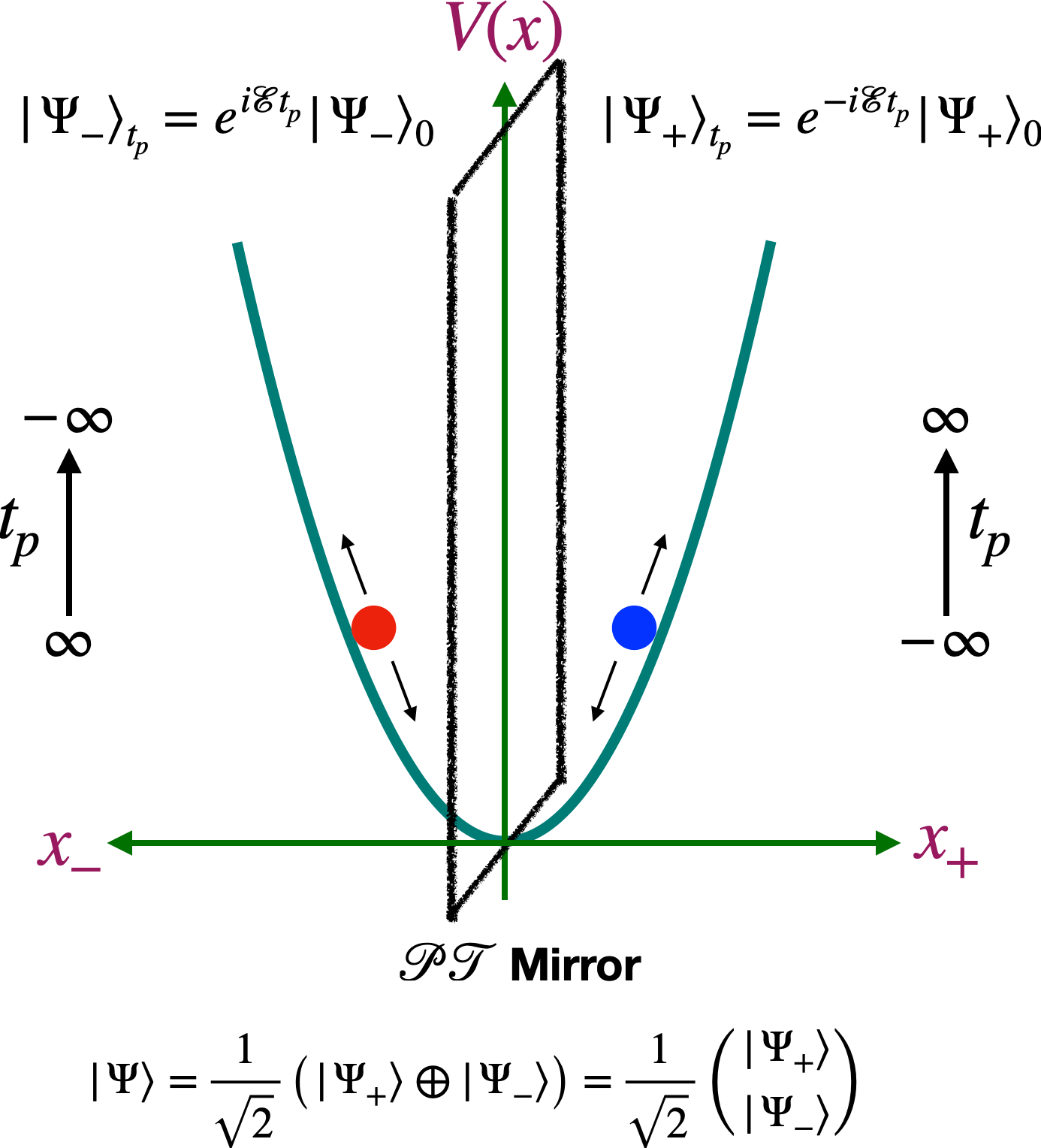}
    \caption{The figure illustrates a new formulation of the quantum harmonic oscillator in a direct-sum Hilbert space. In quantum theory, time appears as a parameter, whereas spatial position is represented by an operator. Within this framework, a quantum state is described as a direct-sum of two components localized at parity-conjugate points in physical space.}
    \label{fig:HO}
\end{figure}

In the previous section, we discussed the role of IHO physics involved in the Higgs mechanism description and in the context of QFTCS. In particular, we have identified that the physics of IHO requires understanding the physical world with two arrows of time. The purpose of this section is to summarize the details of direct-sum quantum theory that has been developed in a series of works \cite{Kumar:2023ctp,Kumar:2023hbj,Kumar:2024ahu,Kumar:2024oxf,Gaztanaga:2024vtr,Gaztanaga:2024nwn,Gaztanaga:2025awe}, which establishes quantum theory with two arrows of time using the concept of geometric superselection sectors. In a nutshell, the whole program is to reposition ourself the understanding of quantum theory without choosing an arrow of time. We represent a quantum state as a {direct-sum} (not a superposition) of two orthogonal components\footnote{The direct-sum operation differs from superposition.}
\begin{equation}
\ket{\Psi}=\frac{1}{\sqrt{2}}\left(\ket{\Psi_+}\oplus\ket{\Psi_-}\right)
=\frac{1}{\sqrt{2}}
\begin{pmatrix}
\ket{\Psi_+}\\[2pt]
\ket{\Psi_-}
\end{pmatrix},
\end{equation}
where $\ket{\Psi_\pm}$ are positive-energy states at parity-conjugate points, evolving with opposite arrows of time. Their dynamics is governed by the direct-sum Schr\"odinger equation \cite{Kumar:2023hbj}
\begin{equation}
i\frac{\partial}{\partial t_p}
\begin{pmatrix}
\ket{\Psi_+}\\
\ket{\Psi_-}
\end{pmatrix}
=
\begin{pmatrix}
\hat H_+ & 0\\
0 & -\hat H_-
\end{pmatrix}
\begin{pmatrix}
\ket{\Psi_+}\\
\ket{\Psi_-}
\end{pmatrix},
\label{schdisum}
\end{equation}
defined on $\mathcal H=\mathcal H_+\oplus\mathcal H_-$, with $\mathcal H_\pm$ being geometric superselection sectors. The Hamiltonian splits as
\begin{equation}
\hat H=\hat H_+(\hat x_+,\hat p_+)\oplus \hat H_-(\hat x_-,\hat p_-),
\end{equation}
and the operators become
\begin{equation}
\hat x=\tfrac{1}{\sqrt{2}}(\hat x_+\oplus \hat x_-),\qquad
\hat p=\tfrac{1}{\sqrt{2}}(\hat p_+\oplus \hat p_-),
\end{equation}
with $x_+=x\gtrsim 0,\; x_-=x\lesssim 0$ and $\hat p_\pm=\mp i\,\partial/\partial x_\pm$.
Equivalently,
\begin{equation}
\begin{pmatrix}
 i\partial_{t_p}\ket{\Psi_+}\\
 -i\partial_{t_p}\ket{\Psi_-}
\end{pmatrix}
=
\begin{pmatrix}
 \hat H_+ & 0\\
 0 & \hat H_-
\end{pmatrix}
\begin{pmatrix}
 \ket{\Psi_+}\\
 \ket{\Psi_-}
\end{pmatrix},
\label{schdisum1}
\end{equation}
where the minus sign shows that $\ket{\Psi_-}$ evolves backward in time.
The wavefunction and normalization are
\begin{equation}
\Psi(x)=\frac{1}{\sqrt{2}}
\begin{cases}
\Psi_+(x_+)e^{-i\mathcal E t}, & x_+=x\gtrsim 0,\\
\Psi_-(x_-)e^{i\mathcal E t}, & x_-=x\lesssim 0,
\end{cases}
\end{equation}
\begin{equation}
\frac{1}{2}\int_{-\infty}^0 dx_-\,|\Psi_-|^2
+\frac{1}{2}\int_0^\infty dx_+\,|\Psi_+|^2=1.
\end{equation}
The canonical commutators are
\begin{equation}
[\hat x_\pm,\hat p_\pm]=\pm i,\qquad
[\hat x_+,\hat x_-]=[\hat p_+,\hat p_-]=[\hat x_+,\hat p_-]=[\hat p_+,\hat x_-]=0.
\end{equation}
Parity $\mathcal P$ and time-reversal $\mathcal T$ act independently within each sector: $\mathcal T$ flips energy and momentum, while $\mathcal P$ flips momentum only. Shifting the spatial origin leaves the direct-sum structure invariant. Figure~\ref{fig:HO} illustrates this new direct-sum interpretation of the quantum harmonic oscillator.

Since, Minkowski spacetime $ds^2=-dt_p^2+d\mathbf{x}^2$ is $\mathcal{P}\mathcal{T}$ symmetric under $t_p\!\to\!-t_p$ and $\mathbf{x}\!\to\!-\mathbf{x}$. This allows a direct extension of the first-quantized direct-sum Schr\"odinger framework to quantum fields, which we call {direct-sum quantum field theory} (DQFT) \cite{Kumar:2023ctp,Kumar:2024oxf}. In this picture, every quantum field is expressed as direct-sum of components in the $\Pc\Tc$ conjugate geometric superselection sector Fock spaces. 

For a real Klein--Gordon field, the operator becomes a direct sum of two $\mathcal{P}\mathcal{T}$-conjugate components,
\begin{equation}
\hat\phi(x)=\frac{1}{\sqrt{2}}
\begin{pmatrix}
\hat\phi_+ & 0\\
0 & \hat\phi_-
\end{pmatrix},
\label{minQuant}
\end{equation}
with mode expansions
\begin{equation}
\hat\phi_+(x)=\int\frac{d^3k}{(2\pi)^{3/2}\sqrt{2|k_0|}}
\Big(a_{(+)\mathbf{k}}e^{ik\cdot x}+a^\dagger_{(+)\mathbf{k}}e^{-ik\cdot x}\Big),
\end{equation}
\begin{equation}
\hat\phi_-(-x)=\int\frac{d^3k}{(2\pi)^{3/2}\sqrt{2|k_0|}}
\Big(a_{(-)\mathbf{k}}e^{-ik\cdot x}+a^\dagger_{(-)\mathbf{k}}e^{ik\cdot x}\Big),
\end{equation}
where $k\!\cdot\!x=-k_0 t_p+\mathbf{k}\!\cdot\!\mathbf{x}$. The operators satisfy
\begin{equation}
[a_{(\pm)\mathbf{k}},a^\dagger_{(\pm)\mathbf{k}}]=1,\qquad
[a_{(\pm)\mathbf{k}},a^\dagger_{(\mp)\mathbf{k}}]=0,
\end{equation}
implying a new causality condition
\begin{equation}
[\hat\phi_+(x),\hat\phi_-(-y)]=0,
\end{equation}
in addition to the standard spacelike commutator
\begin{equation}
[\hat\phi_\pm(x),\hat\phi_\pm(y)]=0,\qquad (x-y)^2>0.
\end{equation}
Minkowski vacuum splits as
\begin{equation}
\ket{0}_M=\begin{pmatrix}\ket{0_+}_M\\ \ket{0_-}_M\end{pmatrix},\qquad
a_{(+)\mathbf{k}}\ket{0_+}_M=0,\quad a_{(-)\mathbf{k}}\ket{0_-}_M=0,
\end{equation}
with positive-energy modes evolving as $e^{-i\mathcal{E}t}$ in $\ket{0_+}_M$ and $e^{+i\mathcal{E}t}$ in $\ket{0_-}_M$. The DQFT Fock space is therefore a direct sum
\begin{equation}
\mathcal{F}=\mathcal{F}_+\oplus\mathcal{F}_-
\end{equation}
of geometric superselection sectors.
The two-point function becomes
\begin{equation}
\langle 0|\hat\phi(x)\hat\phi(x')|0\rangle
=\tfrac12\Big(
\langle 0_+|\hat\phi_+(x)\hat\phi_+(x')|0_+\rangle
+\langle 0_-|\hat\phi_-(-x)\hat\phi_-(-x')|0_-\rangle
\Big),
\end{equation}
and the propagator, which is a time-ordered product of the fields $\phi_\pm$ in their respective vacua 
\begin{equation}
\begin{aligned}
\Delta^{\pm}_F(x-y) & \equiv    \langle 0_\pm \vert T \{\hat \phi_\pm\hat \phi_\pm\} \vert 0_\pm \rangle \\  \\ & = \theta(x^0 - y^0)\,\hat{\phi}_{\pm}(\pm x)\hat{\phi}_{\pm}(\pm y)
+
\theta(y^0 - x^0)\,\hat{\phi}_{\pm}(\pm y)\hat{\phi}_{\pm}(\pm x)\\
& =  \int \frac{d^4x}{\LF 2\pi \RF^4} \frac{\mp i}{k^2+m^2\mp i\epsilon} e^{ik\cdot\LF x-y \RF}
\end{aligned}
\end{equation}
In the DQFT, the fields $\hat{\phi}_{\pm}$ act on
superselection sectors (corresponding to parity conjugate regions) with opposite temporal orientation. 
This ensures that correlation functions in the two sectors correspond to
opposite arrows of time, while maintaining $\mathcal{PT}$ symmetry of the full theory.
The Feynman propagators in DQFT are defined sectorially, and there is no meaningful way to define a global Feynman propagator because parity conjugate regions share an opposite arrow of time. We can split the Feynman propagator in the following way 
\begin{equation}
    \Delta^{\pm}_F(k) = \frac{\mp i}{k^2+m^2\mp i\epsilon} = \mp i {\rm PV}\LF \frac{1}{k^2+m^2} \RF + \pi \delta\LF k^2+m^2 \RF 
\end{equation}
where $\rm PV$ here stands for the Cauchy principal value. 
Interactions in the DQFT picture split block-diagonally, e.g.
\begin{equation}
\frac{\lambda}{3}\hat\phi^3
=\frac{\lambda}{3}
\begin{pmatrix}
\hat\phi_+^3 & 0\\
0 & \hat\phi_-^3
\end{pmatrix},
\end{equation}
so there is no mixing between sectors. As a result, all standard QFT amplitudes are reproduced in Minkowski space as an average of identical $\mathcal{P}\mathcal{T}$ symmetric quantities.
In the case of a scattering process, observables are the squares of amplitudes integrated over the phase space, giving cross-sections, decay rates, and transition probabilities \cite{Coleman:2018mew,Srednicki:2007qs}. The total amplitude square in the DQFT is the average of quantities evaluated in the parity conjugate superselection sectors 
\begin{equation}
\vert A_{N\to M} \vert^2
=\frac{\vert A^{N\to M}_+(p_a,-p_b)\vert^2+\vert A^{N\to M}_-(-p_a,p_b)\vert^2}{2},
\qquad
\vert A^{N\to M}_+\vert^2=\vert A^{N\to M}_-\vert^2,
\end{equation}
where $p_a$ and $p_b$, with $a=1,\cdots N$ and $b=1,\cdots M$, denote the four-momenta of all states participating in the scattering process. The quantities $A_{\pm}$ represent the scattering amplitudes, expressed as functions of the four-momenta of the initial and final states, computed in the two vacua $\vert 0_{SM\pm}\rangle$. Note that the in- and out-states defined with respect to $\vert 0_{SM\pm}\rangle$ appear with opposite signs, reflecting the fact that the arrows of time in the two vacua are reversed relative to each other. Owing to the $\Pc\Tc$ symmetry of Minkowski spacetime, the amplitudes $A_{\pm}$ are identical at every order in perturbation theory. $\mathbb{CPT}$ invariance holds separately in each sector, so standard SM results, including $\mathbb{CP}$ violation, remain unchanged.
All Standard-Model fields admit the same direct-sum structure. For example, complex scalar field ($\phi_c$), fermion ($\psi$) and the gauge field ($A_\mu$) operators are split as $\Pc\Tc$ mirror components 
\begin{equation}
\hat\phi_c=\tfrac{1}{\sqrt{2}}(\hat\phi_{c+}\oplus\hat\phi_{c-}),\qquad
\hat\psi=\tfrac{1}{\sqrt{2}}(\hat\psi_+\oplus\hat\psi_-),\qquad
\hat A_\mu=\tfrac{1}{\sqrt{2}}(\hat A_{+\mu}\oplus\hat A_{-\mu}),
\end{equation}
with canonical (anti)commutation relations enforced independently in each geometric sector. All interaction operators similarly split as
\begin{equation}
\mathcal{O}_{\rm SM}
=\mathcal{O}_{{\rm SM}+}\oplus\mathcal{O}_{{\rm SM}-}.
\end{equation}
The key distinction is that $\mathcal{P},\mathcal{T}$ in DQFT are {geometric} spacetime transformations defining superselection sectors, whereas $\mathbb{C},\mathbb{P},\mathbb{T}$ act on scattering states in momentum space. With this separation, the SM vacuum remains $\mathbb{CPT}$ invariant,
\begin{equation}
(\mathbb{CPT})\,\ket{0_{\rm SM}}\,(\mathbb{CPT})^{-1}
=\ket{0_{\rm SM}}. 
\end{equation}
Thus, DQFT reposition us with a new understanding of quantum theory with geometric superselection sectors which play a key role in preserving unitarity and information paradox in curved spacetime. 

\subsection{Quantization of IHO or dual-IHO in direct-sum quantum theory}

\label{sec:IHOQM}

Let us now define the new quantum construction of the IHO and dual-IHO using a direct-sum framework \cite{Gaztanaga:2025awe}, consistent with the key insights of Berry and Keating and Aneva \cite{Berry1999,Aneva:1999fy}. 
The IHO and dual-IHO phase spaces (See Fig.~\ref{fig:IHO-DualIHO}) can be organized into regions related by a discrete set of transformations that form the dihedral group $D_4$ of order eight \cite{Aneva:1999fy},
\begin{equation}
\begin{aligned}
T_1^\pm:\quad & Q'=-\frac{h}{Q},\qquad P'=\pm\frac{P Q^2}{h},\\
T_2^\pm:\quad & Q'=-\frac{h}{P},\qquad P'=\mp\frac{Q P^2}{h},
\end{aligned}
\end{equation}
together with $-T_1^\pm$ and $-T_2^\pm$, which include the transformations in \eqref{discreteTPQ} and \eqref{BKPE}. These maps preserve dilatations and the quantization cutoffs $Q\ge \ell_Q,\;P\ge \ell_P$.

The existence of phase-space regions related by discrete symmetries motivates a formulation with {geometric superselection sectors}. Guided by the relation between the IHO Hamiltonian \eqref{IHOhamil} and the Weyl reflected Laplace--Beltrami operator \eqref{LBOp}, we write positive- and negative-energy states as a direct sum on a split Hilbert space,
\begin{equation}
\ket{\Psi_{\rm iho}}
=\ket{\Psi_{\rm iho}^{(+E)}}\oplus\ket{\Psi_{\rm iho}^{(-E)}}
=\begin{pmatrix}
\ket{\Psi_{\rm iho}^{(+E)}}\\[2pt]
\ket{\Psi_{\rm iho}^{(-E)}}
\end{pmatrix},
\qquad
\mathcal H_{\rm iho}=\mathcal H_{\rm iho}^{(+E)}\oplus\mathcal H_{\rm iho}^{(-E)} .
\label{PNEIHO}
\end{equation}
Because classical trajectories are doubly degenerate. In the context of IHO each energy sector further splits as
\begin{equation}
\begin{aligned}
\ket{\Psi_{\rm iho}^{(-E)}}&=\frac{1}{\sqrt{2}}\big(\ket{\Psi_I^{(-E)}}\oplus\ket{\Psi_{II}^{(-E)}}\big),
\qquad
\mathcal H^{(-E)}=\mathcal H_I^{(-E)}\oplus\mathcal H_{II}^{(-E)},\\
\ket{\Psi_{\rm iho}^{(+E)}}&=\frac{1}{\sqrt{2}}\big(\ket{\Psi_{III}^{(+E)}}\oplus\ket{\Psi_{IV}^{(+E)}}\big),
\qquad
\mathcal H^{(+E)}=\mathcal H_{III}^{(+E)}\oplus\mathcal H_{IV}^{(+E)} ,
\end{aligned}
\label{ERBIHO}
\end{equation}
where regions III and IV individually contain parity-conjugate points $\pm\tilde q$, while regions I and II jointly cover $\pm\tilde q$ (Fig.~\ref{fig:IHO-DualIHO}).  

Accordingly, the position and momentum operators decompose as
\begin{equation}
\begin{aligned}
\hat Q&=\frac{1}{\sqrt{2}}(\hat Q_{(+E)}\oplus\hat Q_{(-E)}),\qquad
\hat Q_{(+E)}=\frac{1}{\sqrt{2}}(\hat Q_{III}\oplus\hat Q_{IV}),\qquad
\hat Q_{(-E)}=\frac{1}{\sqrt{2}}(\hat Q_I\oplus\hat Q_{II}),\\
\hat P&=\frac{1}{\sqrt{2}}(\hat P_{(+E)}\oplus\hat P_{(-E)}),\qquad
\hat P_{(+E)}=\frac{1}{\sqrt{2}}(\hat P_{III}\oplus\hat P_{IV}),\qquad
\hat P_{(-E)}=\frac{1}{\sqrt{2}}(\hat P_I\oplus\hat P_{II}),
\end{aligned}
\label{posmomiho}
\end{equation}
with non-vanishing commutators
\begin{equation}
[\hat Q_{III},\hat P_{III}]=i\hbar,\quad
[\hat Q_{IV},\hat P_{IV}]=-i\hbar,\quad
[\hat Q_I,\hat P_I]=i\hbar,\quad
[\hat Q_{II},\hat P_{II}]=-i\hbar .
\end{equation}
The IHO Hamiltonian \eqref{IHOhamil} then splits into four block-diagonal components describing the four phase-space regions.
Time evolution is governed by the direct-sum Schr\"odinger equation
\begin{equation}
i\hbar\frac{\partial}{\partial t}
\begin{pmatrix}
\ket{\Psi_{\rm iho}^{(+E)}}\\
\ket{\Psi_{\rm iho}^{(-E)}}
\end{pmatrix}
=
\begin{pmatrix}
\hat H_{\rm iho}^{(+E)} & 0\\
0 & \hat H_{\rm iho}^{(-E)}
\end{pmatrix}
\begin{pmatrix}
\ket{\Psi_{\rm iho}^{(+E)}}\\
\ket{\Psi_{\rm iho}^{(-E)}}
\end{pmatrix},
\label{di-sumiho}
\end{equation}
which further reduces, using the degeneracies $I\!\leftrightarrow\!III$ and $II\!\leftrightarrow\!IV$, to
\begin{equation}
\begin{aligned}
i\hbar\frac{\partial}{\partial t}
\begin{pmatrix}
\ket{\Psi_{III}^{(+E)}}\\
\ket{\Psi_{IV}^{(+E)}}
\end{pmatrix}
&=
\begin{pmatrix}
\hat H_{III}^{(+E)} & 0\\
0 & -\hat H_{IV}^{(+E)}
\end{pmatrix}
\begin{pmatrix}
\ket{\Psi_{III}^{(+E)}}\\
\ket{\Psi_{IV}^{(+E)}}
\end{pmatrix},\\[4pt]
i\hbar\frac{\partial}{\partial t}
\begin{pmatrix}
\ket{\Psi_I^{(-E)}}\\
\ket{\Psi_{II}^{(-E)}}
\end{pmatrix}
&=
\begin{pmatrix}
\hat H_I^{(-E)} & 0\\
0 & -\hat H_{II}^{(-E)}
\end{pmatrix}
\begin{pmatrix}
\ket{\Psi_I^{(-E)}}\\
\ket{\Psi_{II}^{(-E)}}
\end{pmatrix},
\end{aligned}
\end{equation}
with each Hamiltonian a function of the corresponding operators in \eqref{posmomiho}.
Since the positive- and negative-energy sectors are related by \eqref{BKPE}, Fourier transforms in regions $I,II$ become inverse Fourier transforms in $III,IV$ and vice versa. Solving \eqref{di-sumiho} (setting $\omega=1$) yields the geometric boundary conditions
\begin{equation}
\begin{aligned}
Q_I^{1/2}\zeta\!\left(\tfrac12-\tfrac{i|E|}{\hbar}\right)\Psi_{(-E)}(Q_I)
+Q_{III}^{1/2}\zeta\!\left(\tfrac12+\tfrac{i|E|}{\hbar}\right)\Psi_{(+E)}(Q_{III})&=0,\\
Q_{II}^{1/2}\zeta\!\left(\tfrac12-\tfrac{i|E|}{\hbar}\right)\Psi_{(-E)}(Q_{II})
+Q_{IV}^{1/2}\zeta\!\left(\tfrac12+\tfrac{i|E|}{\hbar}\right)\Psi_{(+E)}(Q_{IV})&=0,
\end{aligned}
\label{boundarycondi}
\end{equation}
which generate the Riemann zeros $\zeta(\tfrac12\pm i\bar T)$. Although formally similar to the Berry–Keating condition \eqref{boundarycondiBK}, \eqref{boundarycondi} now admits a {geometric interpretation} via superselection sectors, a key element missing in the original proposal \cite{Berry1999,Aneva:1999fy}. In the context of dual-IHO, we must apply $E\to -E$ for the components of all regions in \eqref{ERBIHO}-\eqref{boundarycondi}.

In summary, direct-sum quantization resolves the quantum chaos pathology of the IHO (or dual-IHO) \cite{Berrychaos,Berry1999} by geometrically splitting phase space into superselection sectors and connecting them through mathematical bridges (direct-sums) between sheets with opposite arrows of time. This yields a physically transparent realization of the Berry and Keating program and embeds the Riemann zero structure into a consistent quantum framework. The concept of mathematical bridges here echoes very well with the Einstein-Rosen bridges, which are discussed in detail in \cite{Gaztanaga:2025awe}.

\section{Stelle or quadratic gravity and Renormalizability } 
\label{sec:QG-Renorm}

Quadratic curvature modification of Einstein's gravity is well known to improve the ultraviolet behavior of the theory and to render the gravitational action perturbatively renormalizable. However, when linearized around maximally symmetric backgrounds, such theories generically exhibit additional propagating degrees of freedom beyond the massless graviton. Among these, the appearance of a massive spin--2 mode with a tachyonic or negative mass-squared term has traditionally been interpreted as signaling a fundamental instability of the theory \cite{Shapiro2015}. This interpretation has motivated extensive efforts to eliminate or evade such modes through parameter tuning, nonlocal modifications\footnote{The non-local modifications of quadratic gravity are aimed to resolve the ghost problem by the choice of infinite derivative form factors which can be fine-tuned further to achieve superrenormalizability \cite{Krasnikov:1987yj,Tomboulis:1997gg,Modesto:2011kw,Biswas:2011ar}. However, the non-local theories equations of motion are overly complicated which is a huge price to pay  \cite{Koshelev:2017tvv,Koshelev:2018rau}. Furthermore, recent investigations have indicated the need for selection of formfactors with background dependence to avoid ghosts \cite{Koshelev:2023elc,Koshelev:2022olc}.  }, or alternative quantization prescriptions. In this section, we revisit the physical meaning of the spin--2 tachyon in quadratic gravity and argue that its presence does not necessarily indicate a pathological instability. Instead, we propose that this mode admits a consistent reinterpretation as an inverted harmonic oscillator degree of freedom, naturally associated with spacelike momentum support and horizon-related quantum dynamics. Thus, the IHO-like spin--2 instability is rather healthier in contrast with pathological spin--2 ghost excitation case.

Based on the principles of QFT renormalizability, UV completion can be achieved once we expand a low-energy theory up to the terms involving dimensionless couplings. Applying the same principles to GR, the only theory of quantum gravity in 4D is quadratic gravity or Stelle gravity 
\begin{equation}
S = \int d^4x\,\sqrt{-g}
\left[
\frac{M_p^2}{2}R
+ \frac{\alpha}{2}R^2
+ \frac{\beta}{2}W_{\mu\nu\rho\sigma}W^{\mu\nu\rho\sigma}
\right]\,.
\label{StelleQG}
\end{equation}
In 1977, Stelle \cite{Stelle:1976gc} showed that the quadratic gravity action \eqref{StelleQG} becomes perturbatively
renormalizable once quadratic curvature terms are included in the action.  In fact, quadratic gravity remains the only known four-dimensional theory of gravity that is perturbatively renormalizable within the standard framework of quantum field theory \cite{Buoninfante:2025dgy}.
The key observation is that adding $R^2$ and $W_{\mu\nu\rho\sigma}W^{\mu\nu\rho\sigma}$ terms raises
The derivative order of the graviton kinetic operator is from two to four.
As a result, the graviton propagator scales as $1/k^4$ at large momentum,
rather than $1/k^2$, leading to a substantial improvement of ultraviolet
behavior.
Using standard power-counting arguments together with covariant gauge fixing
and (Faddeev-Poppov) ghost fields, Stelle demonstrated that the superficial degree of divergence of any Feynman diagram is bounded independently of loop order and
that all ultraviolet divergences correspond to local operators containing at
most four derivatives of the metric. Consequently, divergences can be absorbed
into a finite set of counterterms of the same form as those already present in
the action, namely a cosmological constant term\footnote{We do not write here the cosmological constant $\Lambda_{\rm cc}$ term in \eqref{StelleQG} since it can be set even to zero or any other value to explain 'dark energy' effects. Our focus here is solely the UV scales, therefore for simplicity we drop the $\Lambda_{\rm cc}$ which anyway does not run with the energy scales \cite{Donoghue:2024uay}. }, the Einstein--Hilbert term,
and curvature-squared terms. This establishes perturbative renormalizability
by power counting. Stelle further noted that although the theory is renormalizable, the higher-derivative structure introduces an additional massive spin--2 pole with negative residue in the propagator when quantized in a conventional Hilbert space, leading to an apparent conflict with unitarity. The proof of renormalizability itself, however, is independent of the interpretation of
this extra degree of freedom.

Conformally transforming the action \eqref{StelleQG} into the Einstein frame by the field redefinition $\tilde g_{\mu\nu} = \frac{M_p^2+2\alpha R}{M_p^2} g_{\mu\nu}$ renders an Einstein frame action 
\begin{equation}
S_{E-QG}
= \int d^4x\,\sqrt{-\tilde g}\,
\left[
\frac{M_p^2}{2}\,\tilde R
- \frac{1}{2}\,\tilde g^{\mu\nu}\partial_\mu\phi\,\partial_\nu\phi
- V(\phi)
+ \frac{\beta}{2}\,
\tilde W_{\mu\nu\rho\sigma}\tilde W^{\mu\nu\rho\sigma}
\right] .
\label{EinsteinFrameAction}
\end{equation}
with the well-known Starobinsky potential 
\begin{equation}
V(\phi)
=
\frac{M_p^4}{8\alpha}
\left(
1 - e^{-\sqrt{\frac{2}{3}}\frac{\phi}{M_p}}
\right)^2 .
\label{StarobinskyPotential}
\end{equation}
for the field 
\begin{equation}
\phi(R)
= \sqrt{\frac{3}{2}}\, M_{\rm Pl}\,
\ln\!\left(1 + \frac{2\alpha R}{M_p^2}\right)
\label{phiR}
\end{equation}
Clearly the quadratic gravity naturally gives a scalar field (called 'scalaron') as an integral part of gravity that leads the well known Starobinsky inflation \cite{Starobinsky:1980te,Starobinsky:1981vz,Mukhanov:1990me}. The presence of Weyl square term does not change the background evolution in Starobinsky inflation since the Weyl tensor vanishes for any FLRW background but it could change the behavior of perturbations which we would discuss later. It is important to note that the inflationary cosmology requires $\alpha>0$ but the sign of the Weyl square coefficient $\beta$ would determine the nature of additional spin--2 degree of freedom. In this paper, we stick to the case of $\alpha>0$ in favor of the inflationary cosmology. 

The full graviton propagator of \eqref{StelleQG} around Minkowski spacetime, 
in de~Donder gauge using the spin projectors $\LF P^{(2)}_{\mu\nu,\rho\sigma}\RF$ (spin--2)
and $\LF P^{(0\text{s})}_{\mu\nu,\rho\sigma}\RF $ (spin-0), can be deduced as \cite{Stelle:1976gc,Buoninfante:2025dgy}
\begin{equation}
D_{\mu\nu,\rho\sigma}(k)
=
-\frac{i}{M_p^2}
\left[
\frac{P^{(2)}}{k^2}
-\frac{P^{(2)}}{k^2 + m_2^2}
-\frac{1}{2}\frac{P^{(0\text{s})}}{k^2}
+\frac{1}{2}\frac{P^{(0\text{s})}}{k^2 + m_0^2}
\right].
\label{eq:fullprop}
\end{equation}
where
\begin{equation}
m_2^2 = -\frac{M_p^2}{\beta}, \qquad
m_0^2 = \frac{M_p^2}{6\alpha}.
\end{equation}
where $m_0^2>0$ is mass square of the scalaron, and $m_2^2\lessgtr 0$ is the "mass-scale" associated with the additional spin--2 mode of the theory. 

For $\beta>0$, one has $m_2^2<0$. At first sight, this may appear
tachyonic. However, the resulting spin--2 sector does not describe a
tachyonic particle excitation. Instead, we later show that its Hamiltonian takes the
form of a dual inverted harmonic oscillator with hyperbolic phase
space. The corresponding modes possess spacelike momentum support
and therefore cannot appear as on-shell intermediate states in
scattering amplitudes. As we will show, this structure eliminates
the ghost pathology while preserving the renormalizability of the
theory.

Here $P^{(i)}_{\mu\nu,\alpha\beta}(p)$ denote the Barnes--Rivers spin projectors acting
on the space of symmetric rank-2 tensors in momentum space. The explicit Barnes--Rivers operators are
\begin{equation}
\begin{aligned}
P^{(2)}_{\mu\nu,\rho\sigma} &=
\frac12\!\left(\theta_{\mu\rho}\theta_{\nu\sigma}
+\theta_{\mu\sigma}\theta_{\nu\rho}\right)
-\frac13\,\theta_{\mu\nu}\theta_{\rho\sigma},\\
P^{(0s)}_{\mu\nu,\rho\sigma} &= \frac13\,\theta_{\mu\nu}\theta_{\rho\sigma}
\end{aligned}
\end{equation}
where 
\begin{equation}
\theta_{\mu\nu}(k)\equiv \eta_{\mu\nu}-\frac{k_\mu k_\nu}{k^2}
\label{eq:theta}
\end{equation}
Irrespective of the sign of $\beta$, quadratic gravity \eqref{StelleQG} is renormalizable by power counting. 
It is worth emphasizing that there is no unique action of quantum gravity in 4D that is known to be renormalizable except for quadratic gravity. Thus, before we worry about the unitarity problem of quadratic gravity, we first deal with understanding of power-counting renormalizability, which doesn't depend on the nature of spin--2 degree of freedom in the UV. 

Below, we illustrate the power-counting renormalizability of quadratic gravity. To see this, first, we expand the metric around a background (taken to be flat for power counting).

\begin{equation}
g_{\mu\nu} = \eta_{\mu\nu} + \kappa\, h_{\mu\nu},
\qquad \kappa \equiv M_p^{-1}.
\end{equation}
The curvature scales schematically as
\begin{equation}
R \sim \partial^2 h + \kappa (\partial h)^2 + \cdots ,
\end{equation}
so the curvature-squared terms $R^2$ and
$W_{\mu\nu\rho\sigma}W^{\mu\nu\rho\sigma}$
generate interaction vertices containing four derivatives, while the
Einstein--Hilbert term generates vertices with two derivatives. At large
momentum, the four-derivative sector dominates the kinetic operator,
implying that the graviton propagator scales as
\begin{equation}
D^{(2)}(k)
\simeq
\frac{-i\,m_2^2}{M_p^2\,k^4}
P^{(2)}_{\mu\nu,\rho\sigma}
+\mathcal{O}\!\left(\frac{1}{k^6}\right),\quad D^{(0)}(k)
\simeq
\frac{i\,m_0^2}{2M_p^2\,k^4}
P^{(0\text{s})}_{\mu\nu,\rho\sigma}
+\mathcal{O}\!\left(\frac{1}{k^6}\right),
\qquad k^2 \to \infty .
\label{eq:spin2asym}
\end{equation}
Consider a connected Feynman graph with $L$ loops, $I$ internal graviton
propagators, $V_4$ four-derivative vertices, and $V_2$ two-derivative
vertices. The superficial degree of divergence obtained by momentum
counting is
\begin{equation}
D = 4L - 4I + 4V_4 + 2V_2 .
\end{equation}
Using the topological identity for connected graphs,
\begin{equation}
L = I - (V_4 + V_2) + 1,
\end{equation}
one finds
\begin{equation}
D = 4 - 2V_2 \le 4 ,
\end{equation}
since $V_2 \ge 0$.
Therefore, ultraviolet divergences are bounded by operators of mass
dimension at most four. Radiative corrections can generate only local
counterterms of the form
\begin{equation}
\int d^4x \sqrt{-g}
\left\{
1,\; R,\; R^2,\; R_{\mu\nu}R^{\mu\nu},\;
R_{\mu\nu\rho\sigma}R^{\mu\nu\rho\sigma}
\right\},
\end{equation}
or equivalently, in four dimensions,
$\sqrt{-g}$, $\sqrt{-g}R$, $\sqrt{-g}R^2$, and
$\sqrt{-g}W_{\mu\nu\rho\sigma}W^{\mu\nu\rho\sigma}$,
up to the Euler (Gauss--Bonnet) density. This demonstrates that the theory
is power-counting renormalizable in four spacetime dimensions.

\subsection{Quadratic gravity's spin--2 TT sector: the ghost problems}

We proceed by explicitly deriving the splitting of the transverse--traceless (TT) spin--2 mode into massless and massive sectors from the quadratic variation of \eqref{StelleQG}. This illustrates the clear meaning of the spin--2 part of the propagator \eqref{eq:fullprop}. 

\subsubsection{Spin--2 sector and diagonalization}

We consider the spin--2 sector of quadratic gravity \eqref{StelleQG} expanded around
Minkowski spacetime $\bar g_{\mu\nu}=\eta_{\mu\nu}$, keeping only the TT part of the tensor perturbations, the perturbed Minkowski metric can be written as
\begin{equation}
ds^2=-dt^2+\big(\delta_{ij}+2h_{ij}\big)dx^i dx^j,
\qquad
\partial_i h_{ij}=0,\qquad h_{ii}=0 .
\end{equation}
The quadratic action for the TT sector reads
\begin{equation}
S_{\rm TT}^{(2)}
=\frac12\int d^4x\,
\Big[
F\,\partial_\lambda h_{ij}\partial^\lambda h_{ij}
-\beta\,(\Box h_{ij})(\Box h_{ij})
\Big],
\qquad
\Box\equiv -\partial_t^2+\nabla^2 ,
\label{eq:STT_start}
\end{equation}
where $F=M_p^2+2\alpha\bar R=M_p^2$ (since the background here is Minkowski).
Decomposing into helicities and Fourier modes,
\begin{equation}
h_{ij}(t,\mathbf x)=\sum_{s=\pm}\int\frac{d^3k}{(2\pi)^3}\,
e^{i\mathbf k\cdot\mathbf x}\,
e^{(s)}_{ij}(\hat{\mathbf k})\,h_s(t,k),
\end{equation}
the action becomes a sum over $(s,\mathbf k)$, with Lagrangian
\begin{equation}
L_k[h]
=\frac12\Big[
F(\dot h^{\,2}-k^2 h^2)
+\beta(\ddot h+k^2 h)^2
\Big].
\label{eq:Lk_4th}
\end{equation}
Introducing an auxiliary variable
\begin{equation}
q=\ddot h+k^2 h ,
\end{equation}
the Lagrangian can be written as
\begin{equation}
L_k[h,q]
=\frac12\Big[
F(\dot h^{\,2}-k^2 h^2)
-\beta q^2
+2\beta q\,\ddot h
+2\beta k^2 q h
\Big].
\label{eq:Lk_with_q}
\end{equation}
Integrating by parts to remove $\ddot h$ and dropping a total derivative yields
\begin{equation}
L_k[h,q]
=\frac12\Big[
F\dot h^{\,2}
-2\beta \dot q\,\dot h
-Fk^2 h^2
+2\beta k^2 qh
-\beta q^2
\Big].
\label{eq:Lk_2nd}
\end{equation}
The kinetic terms can be completed into a perfect square,
\begin{equation}
F\dot h^{\,2}-2\beta\dot q\,\dot h
=
F\left(\dot h-\frac{\beta}{F}\dot q\right)^2
-\frac{\beta^2}{F}\dot q^{\,2}.
\end{equation}
Defining the shifted field $
\tilde h \equiv h-\frac{\beta}{F}q $
the Lagrangian becomes
\begin{equation}
L_{TT}
=\frac12\Big[
F\dot{\tilde h}^{\,2}
-\frac{\beta^2}{F}\dot q^{\,2}
-
\big(
Fk^2\tilde h^2
-\frac{\beta^2}{F}k^2 q^2
+\beta q^2
\big)
\Big].
\label{eq:Lk_diag}
\end{equation}
Introducing canonically normalized variables
\begin{equation}
u\equiv \sqrt{F}\,\tilde h,
\qquad
v\equiv \frac{|\beta|}{\sqrt{F}}\,q ,
\end{equation}
the Lagrangian diagonalizes to
\begin{equation}
L_k
=
\frac12\big(\dot u^{\,2}-k^2 u^2\big)
-\frac12\big(\dot v^{\,2}-(k^2+m_2^2)v^2\big),
\qquad
m_2^2=-\frac{F}{\beta}.
\label{eq:Lk_uv_final}
\end{equation}
Restoring the helicity sum and momentum integral, the final diagonal action is
\begin{equation}
S_{\rm TT}^{(2)}
=
\frac12\sum_{s=\pm}\int dt\int\frac{d^3k}{(2\pi)^3}
\big(\dot u_s^{\,2}-k^2 u_s^{\,2}\big)
-
\frac12\sum_{s=\pm}\int dt\int\frac{d^3k}{(2\pi)^3}
\big(\dot v_s^{\,2}-(k^2+m_2^2)v_s^{\,2}\big),
\label{eq:STT_uv_final}
\end{equation}
\eqref{eq:STT_uv_final} makes explicit that the transverse--traceless sector of the four-derivative action is dynamically equivalent, mode-by-mode, to a pair of decoupled second-order systems. For each helicity $s=\pm$ and Fourier momentum $\mathbf{k}$, the canonically normalized variable $u_s(t,k)$ describes the standard massless tensor mode with dispersion $\omega_u^2=k^2$, while $ v_s(t,k)$ captures the additional propagating tensor degree of freedom generated by the $(\Box h_{ij})^2$ term, with dispersion $\omega_v^2=k^2+m_2^2$ and mass parameter $m_2^2=-F/\beta$. The sum over helicities and integral over $\mathbf{k}$ therefore exhibits the TT action as a direct sum of independent quadratic oscillators, and in the GR limit $\beta\to 0$ the extra mode is pushed to infinite mass and decouples, leaving only the $u_s$ sector. Note that for $2\to2$ scattering of physical gravitons, the amplitude depends only on the spin--2 part of the propagator, and it is an element for our next discussion on the status of unitarity in quadratic gravity.

\subsubsection{Optical theorem and unitarity}

A central consistency requirement of any quantum field theory is unitarity of the $S$-matrix, which ensures conservation of probability and underlies the optical theorem relating the imaginary part of forward scattering amplitudes to physical intermediate states \cite{Weinberg:1995mt,Cutkosky1960}. In quadratic gravity, it is traditionally regarded as violated due to the presence of spin--2 ghost modes.

\begin{figure}[t]
\centering
\begin{tikzpicture}[baseline=(current bounding box.center)]

  \node at (-6.2,0) {\Large $2\,\mathrm{Im}$};

  \begin{scope}[xshift=-3.2cm]
    \begin{feynman}
     \vertex (i1) at (-2.0,  1.0) ;
     \vertex (i2) at (-2.0, -1.0) ;
     \vertex (vL) at (-1.1,  0.0) ;
     \vertex (c1) at (-0.1,  0.0) ;
     \vertex (c2) at ( 0.9,  0.0) ;
     \vertex (vR) at ( 1.9,  0.0) ;
     \vertex (o1) at ( 2.8,  1.0) ;
     \vertex (o2) at ( 2.8, -1.0) ;

      \diagram*{
        (i1) -- [fermion] (vL) -- [fermion] (i2),
        (o1) -- [fermion] (vR) -- [fermion] (o2),

        (vL) -- [photon] (c1),
        (c2) -- [photon] (vR),

        (c1) -- [fermion, dashed, half left,  looseness=1.25] (c2),
        (c2) -- [fermion, dashed, half left,  looseness=1.25] (c1),
      };

      \draw[dashed, line width=0.6pt] (0.40,-1.35) -- (0.40,1.35);
    \end{feynman}
  \end{scope}

   \node at (0.75,0) {\Large $=\;\int d\Pi$};

  \begin{scope}[xshift=4cm]

   \def\BARX{2.15}   
   \def\AMPW{1.85}   

    \draw[line width=0.9pt] (-\BARX,-1.55) -- (-\BARX,1.55);
    \draw[line width=0.9pt] ( \BARX,-1.55) -- ( \BARX,1.55);
    \node at (\BARX+0.18,1.28) {\Large $^{2}$};

    \begin{feynman}
      \vertex (a1) at (-\AMPW,  1.00) ;
      \vertex (a2) at (-\AMPW, -1.00) ;
      \vertex (v)  at (-1.05,  0.00) ;
      \vertex (x)  at ( 0.25,  0.00) ;
      \vertex (b1) at ( \AMPW,  1.00) ;
      \vertex (b2) at ( \AMPW, -1.00) ;

      \diagram*{
        (a1) -- [fermion] (v) -- [fermion] (a2),
        (v)  -- [photon] (x),
        (x)  -- [fermion, dashed] (b1),
        (x)  -- [fermion, dashed] (b2),
      };
    \end{feynman}
  \end{scope}
 \end{tikzpicture}
 \caption{Diagrammatic optical theorem / cutting rule.}
\label{fig:Opth}
\end{figure}
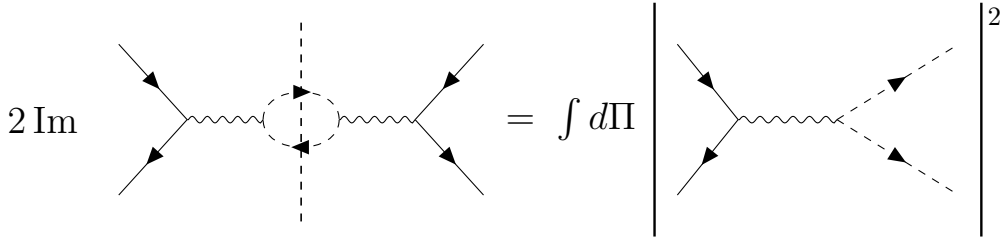

In the context of the optical theorem, unitarity can be checked if the $S$--matrix is written as
\begin{equation}
S = \mathbf{1}+ iT,
\qquad
S^\dagger S=\mathbf{1}.
\end{equation}

\begin{equation}
(\mathbf{1}-iT^\dagger)(\mathbf{1}+iT)=\mathbf{1}
\quad\Rightarrow\quad
i(T^\dagger-T)=T^\dagger T.
\end{equation}
Taking the matrix element between an initial asymptotic state $|i\rangle$ and inserting a resolution of the identity over {physical asymptotic states} $\{|f\rangle\}$,
\begin{equation}
\mathbf{1}=\sum_f |f\rangle\langle f|,
\end{equation}
gives the operator-form optical theorem
\begin{equation}
2\,\text{Im}\langle i|T|i\rangle
=
\sum_f \langle i|T^\dagger|f\rangle\langle f|T|i\rangle
=
\sum_f \big|\langle f|T|i\rangle\big|^2>0.
\label{eq:OT-operator}
\end{equation}
where 
\begin{equation}
    \langle f | T | i \rangle = (2\pi)^4 \delta^4\!\left(\sum_{j \in \mathrm{final}} p_j - \sum_{i \in \mathrm{initial}} p_i\right) \cdot \mathcal{M}_{fi}
\end{equation}
Expressing the optical theorem \eqref{eq:OT-operator} in terms of the invariant amplitudes $\mathcal{M}_{fi}$ and the Lorentz-invariant phase-space measure $d\Pi_f$, we get 
\begin{equation}
2\,\text{Im}\,\mathcal{M}(i\to i)
=
\sum_f \int d\Pi_f\,\big|\mathcal{M}(i\to f)\big|^2>0,
\qquad
d\Pi_f = (2\pi)^4\delta^{(4)}(P_f-P_i)\prod_{j\in f}\frac{d^3p_j}{(2\pi)^3\,2E_{p_j}}.
\label{eq:OT-phase-space}
\end{equation}
Importantly, the sum/integral is over {on-shell physical states} only. The optical theorem \eqref{eq:OT-phase-space} is diagrammatically depicted in Fig.~\ref{fig:Opth}. 

For the optical theorem, an important point is that the imaginary part of the invariant amplitude gets a contribution from the delta function part of the Feynman propagator (here we assume a case of a massive scalar field) via
\begin{equation}
2\,\text{Im}\!\left(\frac{1}{q^2+m^2-i\epsilon}\right)=2\pi\,\delta(q^2+m^2).
\label{eq:im-scalar}
\end{equation}
When terms of the form \eqref{eq:im-scalar} appear under loop integrals, the $\delta$--functions localize the integration to the on-shell hypersurfaces $q^2=-m^2$; with the positive-energy restriction $\theta(q^0)$, we obtain the Lorentz-invariant one-particle phase-space measure:
\begin{equation}
\int d^4q\,\delta(q^2+m^2)\,\theta(q^0)\,F(q)
=
\int \frac{d^3\mathbf{q}}{2E_{\mathbf{q}}}\,F(E_{\mathbf{q}},\mathbf{q}),
\qquad
E_{\mathbf{q}}=\sqrt{\mathbf{q}^2+m^2}.
\label{eq:phase-space-from-delta}
\end{equation}
where  $F(q)$ is a generic function of the momenta that depends on the nature of the loops and vertices in a particular physical process. 

For any Feynman diagram $\mathcal{M}$ built from a Hermitian interaction Lagrangian, we find that the discontinuity $2\,\text{Im}\left(\mathcal{M}\right)$ is given by a sum over {cuts} of the diagram, where each cut propagator is replaced by an on-shell factor proportional to $\delta(q^2+m^2)\theta(q^0)$. This is the Cutkosky cutting rule and is the diagrammatic implementation of inserting $\sum_f|f\rangle\langle f|$ in \eqref{eq:OT-operator}. Thus, for a given internal mode, {whether it contributes to the optical theorem is determined by whether its on-shell delta-function has support on the physical kinematics.}

\subsubsection{The spin--2 ghost and unitarity violation}

The Hamiltonian corresponding to the Lagrangian \eqref{eq:Lk_uv_final} is 
\begin{equation}
    H_{TT}
=\frac12\sum_{s=\pm}\int\frac{d^3k}{(2\pi)^3}\Big(\dot u_s^2+k^2u_s^2\Big)
-\frac12\sum_{s=\pm}\int\frac{d^3k}{(2\pi)^3}\Big(\dot v_s^2+(k^2+m_2^2)v_s^2\Big).
\label{HamilQG}
\end{equation}
where we can witness that the field $v_s$ is a ghost if $m_2^2>0\,\LF \beta<0 \RF$. 
This means the theory contains a massive spin--2 degree of freedom whose Hamiltonian is negative definite. This is the case that has been widely considered in the literature since Stelle's seminal work in 1977 \cite{Stelle:1976gc,Salam:1978fd,Salvio:2018crh}. Given the ghost degree of freedom spoils unitarity, the prospects of quadratic gravity \eqref{StelleQG} as a consistent quantum gravity have been given up since then; at the same time, there was a surge in the developments of alternative frameworks such as string theory and loop quantum gravity, which diverted the focus away from this theory for decades. 
From \eqref{HamilQG} we can clearly witness the splitting of spin--2 field into the massless and the "massive" modes, which is consistent with the propagator structure of quadratic gravity. In the case of $m_2^2>0$, the massive mode $v_s$ Hamiltonian is negative definite, and therefore, one cannot conceive it as a healthy particle on-shell. If we try to quantize this mode, unitarity is violated because the Hilbert space of the ghost is defined with negative norm. Thus, 
we end up with "negative probabilities" or "negative norms" which break the Born rule of quantum mechanics. If, instead, we demand positive norms, then the ghost mode with a negative definite energy spectrum would cause pathological instabilities when they are coupled to modes with a positive energy spectrum (which coincides with understanding so-called Ostrogradsky instabilities) \cite{Woodard:2006nt,Woodard:2015zca,Woodard:2023tgb}.

It is important to review how the optical theorem is violated with a ghost.

\subsection*{Tree-level unitarity violation:}

Let us consider the case of 2-2 (massless) graviton scattering with massive (ghost) spin--2. The invariant amplitude in this case becomes 
\begin{equation}
    i\mathcal{M} = \left(-iV^{\mu\nu\rho\sigma}(k_1,k_2)\right) \times \frac{i\,P^{(2)}_{\mu\nu\alpha\beta}}{M_p^2(k^2+m_2^2-i\epsilon)} \times \left(-iV^{\alpha\beta}_{\rho\sigma}(k_3,k_4)\right)
    \label{Inamplitude}
\end{equation}
where $V^{\mu\nu\rho\sigma}$ are vertex factors comes from the 3rd order perturbation of the quadratic gravity action \eqref{StelleQG} around Minkowski $g_{\mu\nu}= \eta_{\mu\nu}+\kappa h_{\mu\nu}$.
Simplifying further \eqref{Inamplitude} gives 
\begin{equation}
    \mathcal{M} = -\frac{\,\mathcal{N}}{M_p^2(k^2+m_2^2-i\epsilon)} = \frac{\mathcal{N}}{M_p^2}\left[\mathrm{PV}\left(\frac{1}{k^2+m_2^2}\right) + i\pi\delta(k^2+m_2^2)\right]
    \label{Iampdelta}
\end{equation}
where $\mathcal{N} = V^{\mu\nu\rho\sigma} P^{(2)}_{\mu\nu\alpha\beta} V^{\alpha\beta}_{\rho\sigma} > 0$ collects all real positive kinematic factors from the vertices and spin-2 projector contracted on physical external graviton polarizations. Let us now verify the optical theorem \eqref{eq:OT-phase-space} below 
\begin{equation}
    \underbrace{2\,\mathrm{Im}(\mathcal{M}^{\mathrm{tree,ghost}}_{gg\to gg})}_{\mathrm{LHS}} 
= -\frac{2\pi\,\mathcal{N}}{M_p^2}\delta(k^2+m_2^2) 
\neq 
+\frac{2\pi\,\mathcal{N}}{M_p^2}\delta(k^2+m_2^2) 
= \underbrace{\int d\Pi_1\,|\mathcal{M}^{\mathrm{tree}}_{gg\to v}|^2}_{\mathrm{RHS}}
\end{equation}
from which we can conclude that the optical theorem is clearly violated. 

\subsection*{Loop-level unitarity violation:}
Consider a forward elastic amplitude mediated (at one loop) by a pair of massive spin--2 modes,
schematically as in Fig.~3. In a standard unitary theory, the discontinuity across the branch cut
is obtained by putting the intermediate states on shell (Cutkosky rules), yielding
\begin{equation}
2\,\mathrm{Im}\,\mathcal{M}^{(\text{loop})}_{i\to i}
=
\int d\Pi_2 \, |\mathcal{M}^{(\text{tree})}_{i\to (v\,v)}|^2
\;\ge\;0.
\label{eq:cut_healthy}
\end{equation}
However, for a ghost the sign flips. The reason is elementary: each cut propagator contributes a
spectral density proportional to the residue of the corresponding pole. For a normal particle,
\begin{equation}
\frac{1}{k^2+m_2^2-i\epsilon}
=
\mathrm{PV}\!\left(\frac{1}{k^2+m_2^2}\right)
+i\pi\,\delta(k^2+m_2^2),
\label{eq:disc_normal}
\end{equation}
where $\rm PV$ here stands for the Principal Value. 
So, the discontinuity (or the cut) is {positive} and produces \eqref{eq:cut_healthy}.
For a ghost, the propagator comes with an overall minus sign (negative residue),
\begin{equation}
-\frac{1}{k^2+m_2^2-i\epsilon}
=
-\mathrm{PV}\!\left(\frac{1}{k^2+m_2^2}\right)
-\;i\pi\,\delta(k^2+m_2^2),
\label{eq:disc_ghost}
\end{equation}
which implies that the cut contribution carries the {opposite} sign relative to a unitary
sum over physical states.
More explicitly, cutting the two ghost lines gives
\begin{equation}
2\,\mathrm{Im}\,\mathcal{M}^{(\text{loop})}_{i\to i}\Big|_{\text{ghost cut}}
=
-\int d\Pi_2 \, |\mathcal{M}^{(\text{tree})}_{i\to (v\,v)}|^2,
\label{eq:cut_ghost_negative}
\end{equation}
where the minus sign is the unitarity-violating signal. The right-hand side of
\eqref{eq:cut_ghost_negative} is a sum of non-negative contributions in any
positive-norm Hilbert space; hence \eqref{eq:cut_ghost_negative} cannot be reconciled with
\eqref{eq:OT-phase-space}. In other words, the imaginary part of the forward amplitude fails to equal a positive sum over physical intermediate states. In addition, the same obstruction appears if one interprets the cut as an inclusive rate.
In a standard theory, the optical theorem implies positivity of total probability,
\begin{equation}
\sum_n \int d\Pi_n \, |\mathcal{M}_{i\to n}|^2 \ge 0.
\end{equation}
For the ghost channel, \eqref{eq:cut_ghost_negative} corresponds to an effective negative
contribution to an inclusive rate, i.e.\ a negative ``probability'', which is incompatible with
the Born rule in a positive-definite Hilbert space.
Therefore, when the massive spin--2 mode is a genuine ghost ($\beta<0$ so that $m_2^2>0$ and the
massive pole has negative residue), the theory violates unitarity in the standard sense of
admitting a unitary $S$-matrix acting on a positive-norm asymptotic Hilbert space.
This is the precise meaning in which the ghost spoils unitarity.

To see unitarity violation, and how the minus sign emerges from the $\delta$-part (explicitly), in the most transparent way, it is enough to isolate the
distributional (on-shell) part of a propagator. 
In other words, the following sign flip is the microscopic origin of the negative contribution to the optical theorem.
\begin{equation}
\Big(\pi\,\delta(k^2+m_2^2)\Big)_{\text{ghost}} \;=\; - \Big(\pi\,\delta(k^2+m_2^2)\Big)_{\text{healthy}}.
\label{eq:delta-signflip}
\end{equation}

Let us consider a one-loop example, with two ghost lines and the cut (Fig.~\ref{fig:Opth}).
Take a standard bubble integral with two internal lines (here chosen scalar for clarity; in the case of gravity the
spin--2 projector factors just multply these terms):
\begin{equation}
i\mathcal{M}^{\text{(bubble)}}(P)
=
\int\!\frac{d^4k}{(2\pi)^4}\;
\big[-iV_L\big]\;\Delta_1(k)\;\Delta_2(P-k)\;\big[-iV_R\big],
\label{eq:bubble-def}
\end{equation}
where $V_{L,R}$ are real tree-level vertices and $P$ is the external (forward) momentum.
For the ghost (spin--2) particles,
\begin{equation}
\Delta_1(k)=\frac{i}{k^2+m_2^2-i\epsilon},\qquad
\Delta_2(P-k)=\frac{i}{(P-k)^2+m_2^2-i\epsilon}.
\end{equation}
The absorptive contributions come from product of propagators contains a term
\begin{equation}
\Delta_1(k)\Delta_2(P-k)
\supset
\Big(\pi\,\delta(k^2+m_2^2)\Big)\Big(\pi\,\delta((P-k)^2+m_2^2)\Big),
\label{eq:double-delta-healthy}
\end{equation}
which is real and positive as a distribution.
This is precisely the term that yields the imaginary part of the loop amplitude after the
$k^0$-integration: the two $\delta$-functions enforce on-shell conditions and produce the
Lorentz-invariant phase space.

Cutting two ghost lines gives an overall factor $(-1)^2=+1$ from the propagators,
but the crucial sign relative to the optical theorem comes from how the ghost couples to the
Hilbert-space metric: the intermediate state sum effectively inserts an extra minus sign,
equivalently traced to the negative residue/spectral weight.\footnote{In operator language,
the completeness relation carries the metric $\eta_n=\pm1$ for an indefinite-norm state space,
$\sum_n \eta_n |n\rangle\langle n|=1$. This is exactly the same sign that appears in the cut.}

Concretely, the absorptive part contributed by ghost intermediate states takes the form
\begin{equation}
2\,\mathrm{Im}\,\mathcal{M}^{\text{(bubble)}}(P)\Big|_{\text{ghost}}
=
-\int d\Pi_2(P)\;
\Big|\mathcal{M}^{\text{(tree)}}_{i\to (g\,g)}(P)\Big|^2,
\label{eq:ghost-cut-negative}
\end{equation}
where the minus sign signals that the discontinuity is not a positive sum over
physical intermediate states.

The optical theorem expressed as diagrammatic rule in Fig.~\ref{fig:Opth} can now be read literally as:
the dashed cut replaces propagators by their $\delta$-parts. If the internal mode has negative
residue (ghost), its $\delta$-part comes with the wrong sign, and the cut contributes with the
wrong sign to $2\,\mathrm{Im}\,\mathcal{M}_{i\to i}$, contradicting the positivity of
$\sum_n \int d\Pi_n |\mathcal{M}_{i\to n}|^2$ in a positive-norm theory.

\subsection{The saga of quantum prescriptions to save unitarity}

\label{sec:prescriptions}

In the previous subsection, we learned that in a naïve Feynman quantization the massive spin--2 pole appears with an opposite sign residue, and if one treats it as an ordinary particle degree of freedom the usual probabilistic interpretation is endangered: the would be “ghost” can show up as a negative norm (or negative probability) excitation and can also enter the Cutkosky rules as an on-shell intermediate state. In other words, the obstruction is not merely aesthetic; it is localized in the absorptive parts of amplitudes, i.e., in the pieces that the optical theorem reads as sums over physical intermediate states.

All proposed “solutions” so-far in the literature share a single strategic move retaining the ghost degree of freedom (with negative definite Hamiltonian $\beta<0,\, m_2^2>0$):
\begin{center}
    " The extra spin--2 mode must not be counted as an ordinary asymptotic particle" 
\end{center}
What differs is how that move is realized, by dynamics (instability), by prescription (modified cutting rules), or by a different Hilbert-space structure. All the resolutions to the ghost problem are reviewed in detail in \cite{Buoninfante:2025klm}, we summarize here the key technical details of this review and draw common aspects across various proposals. Since our investigation in the next section is to achieve unitarity by the alternative choice $\beta>0,\, m_2^2<0$, we restrict ourselves away from the ghost resolutions (with $\beta<0,\, m_2^2>0$). 

One way to resolve ghost issue is to treat the massive spin--2 ghost in higher-derivative theories as a Lee–Wick–type resonance that becomes unstable (the idea of unstable ghost resonance) once loop effects are included \cite{Donoghue:2021cza,Donoghue:2019fcb}. In that context, radiative corrections generate a width, the would be ghost is removed from the asymptotic spectrum, and perturbative unitarity can be organized by summing only over cuts of stable states; when one nevertheless works in a narrow-width approximation that treats the unstable mode as if it were cut, one must adopt the original Lee–Wick contour prescription to recover the correct result. However, a key technical obstacle to interpreting this as ordinary “decay” lies in the location of the dressed poles. For an ordinary unstable particle, the complex poles migrate to the second Riemann sheet (so the one-particle pole disappears from the physical sheet), whereas for a ghost resonance the dressed propagator generically develops a complex-conjugate pair of poles in the first (physical) sheet, which signals that the ghost cannot be eliminated as an asymptotic degree of freedom in the same way. This is precisely the point emphasized in recent analysis \cite{Buoninfante:2025klm} on ghost resonances where it was shown that above threshold, the ghost’s complex poles remain on the physical sheet (unlike standard resonances), leading to the conclusion that the naive “ghost decays away” picture is not technically justified in the usual analytic structure, and motivating the “no-decay/anti-instability” interpretation \cite{Kubo:2023lpz,Kubo:2024ysu}. A complementary viewpoint \cite{Holdom:2016xfn,Holdom:2002hp} is the nonperturbative/IR-emergence perspective: rather than “fixing” the spin--2 ghost by a Lee–Wick contour or by treating it as a decaying resonance, it is argued that the perturbative pole structure is not the right guide to the physical spectrum once quadratic gravity runs strong. In this picture, motivated by a QCD analogy, the would-be ghost can be removed or reshaped by nonperturbative dynamics, while GR emerges as the effective IR description. A related classical/semiclassical theme is that the dangerous higher-derivative instability can be parametrically slow and can be damped in expanding backgrounds, so that the problematic mode need not appear as a physical asymptotic excitation even before invoking any explicit pole prescription.

A second class of resolutions is to modify the quantization prescription so that the extra pole is
projected out of the physical spectrum by construction. This is the ``fakeon'' framework \cite{Anselmi:2018tmf,Anselmi:2018bra,Anselmi:2018ibi}, in which the problematic pole can
propagate internally but is not allowed to appear as a physical on-shell intermediate state.
In the unitarity language, the net effect is precisely that the would-be $\delta(k^2+m_2^2)$ support (with internal lines of time-like momenta) associated with that pole is not interpreted as a physical cut contribution. The price of this
construction is that microcausality is not the same as in standard local QFT above the fakeon
scale, the UV completion of the theory can only be in Euclidean space with a contrived contour prescription that projects away ghosts as on-shell states.

A third, conceptually different route is to keep the higher-derivative dynamics but change the
Hilbert-space structure so that ``wrong-sign residue'' does not imply negative norm. This is the
$\mathcal{PT}$-symmetric quantization \cite{Bender:2008vh,Bender:2007wu,Kuntz:2024rzu} (famously for the Pais--Uhlenbeck model), where a nontrivial inner product
renders time evolution unitary with positive norm. In this view, one does not remove the extra
mode from the spectrum by prescription; rather, one changes what ``Hermitian'' and ``positive
probability'' mean at the fundamental level.

With these distinctions in mind, the point of the our construction (with IHO-like spin--2 sector) that we present in the next section is that the optical theorem
obstruction is avoidedalready at the level relevant for the forward unitarity cut: the kinematics
and/or the adopted physical prescription implies the absence of physical support
for the spin--2 intermediate state in the forward channel, so that no additional Lee--Wick contour,
fakeon projection, or $\mathcal{PT}$-Hilbert-space (with non-Hermitian Hamiltonians) reconstruction is required for the optical theorem statement employed here. 

\section{Unitary quadratic quantum gravity with IHO-like spin--2 sector}
\label{sec:UQQG-IHOrenorm}

Having shown that the standard quantization of quadratic gravity violates the optical theorem due to the massive spin--2 ghost, we now present an alternative unitary formulation based on the dual-IHO spin--2 sector and its quantization scheme (See Sec .~\ref {sec:IHOQM}). Worth emphasizing that IHO physics is universal as it is prevalent in SM, also in other areas of physics, chemistry, and mathematics. Here, we see that it is highly relevant to quantum gravity. 
In this approach, the IHO-like instability arises from an extra spin--2 mode with spacelike-momentum support that does not appear in asymptotic states or unitarity cuts, yet remains consistent with virtual (purely quantum) processes. In this section, we demonstrate how this framework leads to a consistent realization of QQG, in which renormalizability and unitarity coexist and the additional spin--2 sector is understood as a dynamical instability rather than a fundamental inconsistency.

With $\beta>0$, the mass scale associated with the additional spin--2 field would become
\begin{equation}
\mu_2^2 \equiv \frac{M_p^2}{\beta} = -\,m_2^2 > 0.
\end{equation} 
Then, the spin--2 sector of quadratic gravity would be described by the following Hamiltonian 
\begin{equation}
   H_{TT}
=\frac12\sum_{s=\pm}\int\frac{d^3k}{(2\pi)^3}\Big(\dot u_s^2+\vec k^2u_s^2\Big)
-\frac12\sum_{s=\pm}\int\frac{d^3k}{(2\pi)^3}\Big(\dot v_s^2+(\vec k^2-\mu_2^2)v_s^2\Big).
\label{HTTsplitting}
\end{equation}
The field $u_s$ is massless spin--2 field whereas the field $v_s$ is now a dual-IHO spin--2 field, which is nothing but an infinite collection of dual-IHO (See Table.~\ref{CompareHs}). 
The graviton propagator \eqref{eq:fullprop} in this context becomes
\begin{equation}
D_{\mu\nu,\rho\sigma}(k)
=
-\frac{i}{M_p^2}
\left[
\frac{P^{(2)}}{k^2}
-\frac{P^{(2)}}{k^2 - \mu_2^2}
-\frac{1}{2}\frac{P^{(0\text{s})}}{k^2}
+\frac{1}{2}\frac{P^{(0\text{s})}}{k^2 + m_0^2}
\right].
\label{eq:tachprop}
\end{equation}
A key issue in assessing the physical viability of dual-IHO spin--2 modes is whether they contaminate observable scattering processes and violate the optical theorem. We argue that this need not be the case since such modes are supported exclusively in the spacelike region of momentum space. Spacelike excitations do not define asymptotic particle states and therefore cannot appear as external legs in the $S$-matrix. Their contributions are restricted to internal lines and are kinematically separated from the timelike sector describing ordinary matter and radiation. This separation allows the theory to maintain unitarity and causal propagation for physical (on-shell) degrees of freedom, while accommodating spacelike modes (dual-IHO spin--2) as effective degrees of freedom associated with instabilities, horizons, or IHO dynamics.

\subsection{Quantization of IHO spin--2 field with geometric superselection sectors}

For $\beta>0$, each transverse--traceless spin--2 polarization $v_s(t,\vec x)$
is governed by the quadratic Hamiltonian density
\begin{equation}
\mathcal{H}_s
= -\frac12 \pi_s^2
- \frac12 (\nabla v_s)^2
+ \frac12 \mu_2^2 v_s^2 ,
\qquad
\mu_2^2=\frac{M_P^2}{\beta}>0,
\label{dualIHO}
\end{equation}
where $\pi_s=-\dot v_s$.
This structure differs from a ghost: the Hamiltonian is not negative definite, but rather indefinite.
Rather, it possesses a hyperbolic quadratic form in phase space.
The defining feature of the dual-IHO is the negative sign in the kinetic term,
while the spatial gradient and mass terms remain positive. The field equation reads
\begin{equation}
\ddot v_s - \nabla^2 v_s - \mu_2^2 v_s = 0 ,
\end{equation}
which coincides with the classical equation obtained from the quadratic action.
Expanding in eigenfunctions of $-\nabla^2$,
\begin{equation}
-\nabla^2 f_\lambda(\vec x)
=
\lambda\, f_\lambda(\vec x),
\qquad
\lambda \ge 0,
\end{equation}
we write
\begin{equation}
v_s(t,\vec x)
=
\sum_\lambda q_{s,\lambda}(t) f_\lambda(\vec x),
\qquad
\pi_s(t,\vec x)
=
\sum_\lambda p_{s,\lambda}(t) f_\lambda(\vec x).
\end{equation}
The Hamiltonian becomes
\begin{equation}
H
=
\frac12 \sum_\lambda
\left(
- p_{s,\lambda}^2
- \lambda\, q_{s,\lambda}^2
+ \mu_2^2 q_{s,\lambda}^2
\right).
\end{equation}
For each mode, the quadratic form is hyperbolic in phase space.
Defining
\begin{equation}
\Omega_{s,\lambda}^2
=
-\lambda+\mu_2^2,
\end{equation}
the reduced Hamiltonian reads
\begin{equation}
H_{s,\lambda}
=
-\frac12 p_{s,\lambda}^2
+
\frac12 \Omega_{s,\lambda}^2 q_{s,\lambda}^2 .
\end{equation}
This is the dual-IHO:
the instability can be seen easily in momentum space rather than configuration space.
The energy shells satisfy
\begin{equation}
- p_{s,\lambda}^2
+ \Omega_{s,\lambda}^2 q_{s,\lambda}^2
= 2E ,
\end{equation}
with separatrices
\begin{equation}
p_{s,\lambda}
=
\pm
\Omega_{s,\lambda}
q_{s,\lambda}.
\end{equation}
Upon canonical quantization, the Hilbert space therefore decomposes into four dynamically disconnected geometric superselection sectors (See the right panel of Fig.~\ref{fig:IHO-DualIHO})
\begin{equation}
\mathcal{H}_{\text{IHO}}
=
\mathcal{H}_{I}
\oplus
\mathcal{H}_{II}
\oplus
\mathcal{H}_{III}
\oplus
\mathcal{H}_{IV}.
\end{equation}
These are geometric superselection sectors: time evolution acts within each sector but does not mix them.
Parity-time symmetry exchanges opposite quadrants,
\begin{equation}
(I) \leftrightarrow (II),
\qquad
(III) \leftrightarrow (IV).
\end{equation}
The minimal $\Pc\Tc$-symmetric physical construction is therefore a direct-sum of conjugate sectors.
Therefore, we quantize each transverse--traceless polarization of the massive spin--2 sector
as a dual-IHO. There is no global Fock vacuum that exists, no positive-frequency split can be defined, and no additional asymptotic particle states arise as we discussed earlier. Furthermore, we cannot interpret the dual-IHO spin--2 mode as a particle, just as quantum fields in curved spacetime cannot have a particle interpretation. This exploits the elegance of QFTCS in shedding light on the nature of quantum gravity. In both physical contexts, the healthy instability, like IHO, plays a universal role. In the context of QFTCS, DQFT splits a quantum state as a direct-sum of 4-components with pair of parity conjugate regions inside and outside the horizon. Note that the concept of time is different from the interior to exterior, therefore, they form geometric superselection sectors and a single quantum state is projected as "mirror" images on either side of the gravitational horizons. It is exactly this construction shown to save unitarity and resolve information paradoxes in curved spacetime \cite{Kumar:2023ctp,Kumar:2023hbj,Kumar:2024oxf}.  

\subsection{Källén--Lehmann representation and the principal value propagator of the dual-IHO spin-2 field}
\label{sec:KL-PV}

Here we focus on understanding the propagator of the dual-IHO spin--2 sector which is fundamentally different from the Feynman propagator in the standard QFT. 

According to the Källén--Lehmann (KL) representation \cite{Kallen:1952zz,Lehmann:1954xi,Srednicki:2007qs}, the full non-perturbative Feynman propagator of a scalar field (with $-i\epsilon$ prescription) is
\begin{equation}
\tilde{\Delta}(k^2) = \frac{-i}{k^2+m^2-i\epsilon} + \int_{0}^{\infty} ds\, \frac{-i\,\rho(s)}{k^2+s-i\epsilon}
\label{eq:KL}
\end{equation}
{Strictly speaking, the isolated first term in \eqref{eq:KL} is the stable one-particle contribution to the spectral measure. Equivalently, the spectral density contains a term
\begin{equation}
\rho_{\rm full}(s)=Z\,\delta(s-m^2)+\rho(s),
\end{equation}
where $Z$ is a normalization factor associated with one particle state. The propagator can now be written as
\begin{equation}
\tilde{\Delta}(k^2)=\int_0^\infty ds\,\frac{-i \, \rho_{\rm full}(s)}{k^2+s-i\epsilon}.
\end{equation}
where} the spectral density that corresponds to the multiparticle states is defined as:
\begin{equation}
\rho(s) \equiv \sum_n |\langle k,n|\varphi(0)|0\rangle|^2\, \delta(s-M_n^2)
\label{eq:rho}
\end{equation}
This packages all the multiparticle matrix elements into a single function of $s\geq 4m^2$. The $M_n\geq 2m$ here is the invariant mass of the multiparticle states. It satisfies $\rho(s) \geq 0$ and $\rho(s) = 0$ for $s < 4m^2$.

Here $\vert 0\rangle$ is the vacuum state and $\vert n\rangle$ runs over
all physical one-particle states.
Three properties of \eqref{eq:KL}--\eqref{eq:rho} are essential for
what follows.

\begin{enumerate}

\item The spectral domain is timelike.
The integration variable $m^2 \geq 0$ corresponds to
$k^2 = -m^2 \leq 0$, i.e.\ to timelike or lightlike momenta.
A propagator pole at spacelike momentum $k^2 = \mu^2 > 0$
corresponds to $m^2 = -\mu^2 < 0$, which lies
outside the domain of integration in \eqref{eq:KL}.
No standard Feynman propagator with a spacelike pole can be
represented in the form \eqref{eq:KL}.
{Equivalently, the Wightman spectrum condition requires \cite{Streater1964,Haag1996}
\begin{equation}
{\rm spec}\,P^\mu\subset \overline V_+
=\{k^\mu:k^2\leq0,\;k^0\geq0\},
\label{eq:spectrum}
\end{equation}
so that the KL variable is always
\begin{equation}
s=-k^2\geq0.
\end{equation}
The dual-IHO pole at $k^2=\mu_2^2>0$ would require $s=-\mu_2^2<0$, which is outside the physical spectral support.}

\item $\rho > 0$ requires a normalizable vacuum.
The matrix element $\langle 0|\phi(0)|n\rangle$ is only defined
if the bra $\langle 0|$ exists as a normalizable state,
$\langle 0| 0\rangle = 1$.
If no such state exists, the spectral density \eqref{eq:rho} is
undefined; one must set $\rho = 0$. {Note that setting the spectral weight of the dual-IHO sector to zero does not mean that the internal Green function vanishes. It means that the physical on-shell spectral contribution is absent. The off-shell Green function must still be obtained by inverting the quadratic kinetic operator, as we further show below. Furthermore,} the distinction from plane waves is worth emphasizing.
Free field plane wave states $|{\bf k}\rangle = a_{\bf k}^\dag|0\rangle$ are not in $L^2(\mathbb{R})$, they satisfy
$\langle{\bf k}|{\bf k}'\rangle = \delta^{(3)}({\bf k}-{\bf k}')$, yet the free scalar field has a perfectly well-defined Feynman propagator.  The reason is that the vacuum $|0\rangle$ is normalizable: $\langle 0|0\rangle = 1$, so the KL matrix element $\langle 0|\phi(0)|{\bf k}\rangle$ is defined. For the dual-IHO, there is no normalizable state to place in the bra position, regardless of the excited states.  The decisive condition is the non-normalizability of the ground state, not of the excited states. 

\item The $\delta$-function piece requires a normalizable
one-particle state.
Each term in \eqref{eq:rho} contributes a delta function
$\delta(s - M_n^2)$ to $\rho$ only if the state $|n\rangle$ is a
normalizable eigenstate with a definite mass $M_n$.
Non-normalizable (Dirac-delta normalised) scattering states
contribute a continuous measure to $\rho$, not isolated
delta functions; but even this continuous contribution requires a normalizable vacuum $|0\rangle$ for the matrix element to be defined.

\end{enumerate}
Note that we have argued in Sec.~\ref{sec:IHO-SM} that the dual-IHO ground state is not in $L^2(\mathbb{R})$. This result holds regardless of whether additional quantization
conditions are imposed.  Those conditions select a discrete set of allowed energies, but they do not alter the asymptotic behaviour of the eigenfunctions within each sector.  The wave functions at the discrete energies are still parabolic cylinder functions, which diverge and are still not in $L^2(\mathbb{R})$.

{ The conclusion here $\rho=0$ for the dual IHO spin--2 field does not by itself imply that the entire internal propagator is zero, nor does it directly produce the principal value. Rather, it says that the on-shell spectral contribution at $k^2=\mu_2^2$ is absent. The internal propagator is then fixed by solving the Green-function equation for the quadratic operator and imposing this absence of spectral weight.}

{The dual IHO Greens function can be defined through (suppressing spin--2 projectors and overall normalization factors) 
\begin{equation}
\big(\square+\mu_2^2\big)\,\Delta_{\rm dIHO}(x-y)=-i\,\delta^{(4)}(x-y),
\label{eq:green-position}
\end{equation}
Fourier transforming,
\begin{equation}
\Delta_{\rm dIHO}(x-y)=\int\!\frac{d^4k}{(2\pi)^4}\,
e^{ik\cdot(x-y)}\,\tilde\Delta_{\rm dIHO}(k),
\end{equation}
and using $\Box\to-(k^2)$ in the signature $k^2=-k_0^2+\mathbf k^2$, Eq.~\eqref{eq:green-position}
becomes the algebraic distributional relation
\begin{equation}
(k^2-\mu_2^2)\,\tilde\Delta_{\rm dIHO}(k)=i,
\qquad X\equiv k^2-\mu_2^2 .
\label{eq:green-momentum}
\end{equation}
}
{This is simply the momentum-space statement that $\tilde\Delta_{\rm dIHO}$ is a right inverse of the kinetic operator. It requires no reference to a vacuum, a Fock space, or a KL representation.
A particular solution of \eqref{eq:green-momentum} is $i\,\rm PV(1/X)$, since the Cauchy principal value satisfies, as distributions,
\begin{equation}
X\,\rm PV\frac{1}{X}=1 .
\end{equation}
The general solution differs by a homogeneous distribution $T$ obeying $X\,T(X)=0$. The
only distributions supported at the single point $X=0$ are $\delta(X)$ and its
derivatives, and since
\begin{equation}
X\,\delta(X)=0,
\qquad
X\,\delta^{(n)}(X)=-n\,\delta^{(n-1)}(X)\quad(n\ge1),
\end{equation}
the kernel of multiplication by $X$ is spanned by $\delta(X)$ alone. Hence, the most general Lorentz-invariant Green distribution is
\begin{equation}
\tilde\Delta_{\rm dIHO}(k)
=i\,{\rm PV}\frac{1}{k^2-\mu_2^2}
+C\,\delta(k^2-\mu_2^2),
\qquad C\in\mathbb R .
\label{eq:family}
\end{equation}
The coefficient $C$ is the entire content of any propagator prescription. By the Sokhotski--Plemelj relations \cite{GelfandShilov1964,Hormander1983},
\begin{equation}
\frac{i}{X-i\epsilon}=i\,\rm PV\frac1X-\pi\,\delta(X),
\qquad
\frac{i}{X+i\epsilon}=i\,\rm PV\frac1X+\pi\,\delta(X),
\label{eq:SP}
\end{equation}
the Feynman, anti-Feynman, and principal value propagators are precisely the choices
$C=-\pi$, $C=+\pi$, and $C=0$ within the family \eqref{eq:family}. The $\delta(X)$ term
is the on-shell, cut-producing, absorptive part; the $\rm PV$ term is purely dispersive. The question is therefore not whether a pole exists,it does, as a direct consequence of the field equation \eqref{eq:green-position} but what value $C$ takes.}

\subsubsection{Caution on splitting modes for tachyonic and dual-IHO fields}

{A potential concern is that the mode Hamiltonians appear to admit a momentum space splitting into regions with qualitatively different behavior. One may therefore wonder whether a subset of modes can be quantized as ordinary harmonic oscillators with normalizable ground states, thereby restoring a conventional particle interpretation. We now show that such a splitting is not Lorentz invariant and therefore cannot define a physical decomposition of the theory. Let us emphasize one structural similarity between tachyonic (IHO) and dual-IHO field. The (indefinite) Hamiltonian densities of IHO and dual-IHO fields read as 
\begin{equation}
\begin{aligned}
\mathcal{H}_T
& = \frac12 \pi_t^2
+ \frac12 (\vec k^2 
-  \mu^2) v_t^2 , \\
\mathcal{H}_{\rm dIHO}
& = -\frac12 \pi_s^2
- \frac12 (\vec k^2 
-  \mu_2^2) v_s^2 ,
\label{tachyonH}
\end{aligned}
\end{equation}
Both cases interestingly have similar structure as we discussed in Sec.~\ref{sec:IHO-SM}. 
We caution that one cannot split the modes based on the the frame dependent $\textbf{k}^2$ terms i.e., one may be tempted to divide momentum space into regions
\begin{equation}
    \textbf{k}^2 \gtrless \mu^2,\quad   \textbf{k}^2 \gtrless \mu_2^2
    \label{splitmodes}
\end{equation}
and interpret the Hamiltonian densities as positive definite (in the case of tachyon when $\textbf{k}^2>\mu^2$) or negative definite (in the case of dual-IHO when $\textbf{k}^2<\mu_2^2$). In this way, one may think to define a normalizable ground state for the dual IHO field and adopt the KL representation. Such a decomposition cannot define distinct physical sectors because the inequalities \eqref{splitmodes} are not Lorentz invariant. Physical sectors in a relativistic theory must be characterized by Lorentz invariant quantities, such as the orbit determined by the invariant four momentum $k^2$, rather than by the frame dependent three momentum magnitude $|\textbf{k}|$. To be more precise,  let us apply Lorentz boost transformations in the \textbf{x}-direction 
\begin{equation}
\begin{split}
k'^0 &=\gamma\,(k^0-v\,k^1)\\
k'^1 &=\gamma\,(-vk^0+\,k^1), \qquad \gamma=(1-v^2)^{-1/2}.  
\label{eq:boost}
\end{split}
\end{equation}
by taking $k^\mu = (k^0, k^1,0,0)$. Since only $k_{\nu}k^{\nu}$ is Lorentz invariant, a mode satisfying $(k^1){^2}>\mu^2$ (or $(k^1){^2}<\mu_2^2$ in one frame can satisfy $(k^{\prime 1}){^2}<\mu^2$ (or $(k^{\prime 1}){^2}>\mu_2^2$) in another, for a suitable choice of $v<1$. The Lorentz group, therefore mixes the two regions. Neither region is separately invariant and neither can be assigned an independent physical interpretation. Therefore the dual-IHO sector cannot be viewed as a mixture of ordinary oscillators and inverted oscillators classified  by $\textbf{k}^2 \lessgtr \mu_2^2$ (or $\textbf{k}^2 \gtrless  \mu^2$). Such a classification is frame dependent and is not preserved by Lorentz transformations. The physically relevant object is the full Lorentz invariant orbit associated with the spacelike pole, not a frame dependent partition of its Fourier modes. A normalizable vacuum constructed from the subset $\textbf{k}^2 < \mu_2^2$ (or $\textbf{k}^2 > \mu^2$)  would itself fail to be Lorentz invariant because Lorentz boosts map modes into and out of that subset. The construction of \cite{Jacobson:1987ap} relies on selecting by hand the subset $\textbf{k}^2>\mu^2$ in order to define a normalizable vacuum. Since this selection is not Lorentz invariant, that framework is conceptually different from the present investigation, where Lorentz covariance is maintained throughout.}

\subsubsection{Pole at the spacelike momenta and KL representation}

{Let us discuss why there is complete failure of the spectrum condition on a spacelike shell. A field whose propagator pole lies at spacelike momentum $k^2=\mu_2^2>0$ violates the Wightman spectrum condition \eqref{eq:spectrum} not marginally but completely,
and no Lorentz frame places the supporting momentum inside $\overline V_+$. The condition \eqref{eq:spectrum} requires every physical state to have $k^2\le0$ and $k^0\ge0$, both Lorentz invariant. A spacelike pole has $k^2=\mu_2^2>0$, so the first condition $k^2\le0$ already fails, and it fails in every frame because $k^2$ is Lorentz invariant. There is no boost transformation into $\overline V_+$: the orbit of a spacelike momentum
under the Lorentz group is the entire spacelike shell $k^2=\mu_2^2>0$, which is
disjoint from $\overline V_+$. Considering the arguments presented before, even the weaker hope, an invariant $k^0>0$ half is unavailable.
This is because, in \eqref{eq:boost} we can choose any boost velocity satisfying 
\begin{equation}
    \frac{k^0}{k^1}<v<1 
\end{equation}
one can turn $k^0>0$ into $k^{\prime 0}<0$. It is worth noting that for the timelike (or lightlike) shell $k^0>0$ is Lorentz invariant. 
Therefore, for the dual-IHO case, the spectral condition violation is total, not just for a state at the edge of the cone, but a shell lying entirely outside it. In the KL spectral integral, the variable is $s=-k^2$, with support $s\ge0$ enforced by
\eqref{eq:spectrum}. A spacelike pole at $k^2=\mu_2^2>0$ would demand $s=-\mu_2^2<0$,
outside the spectral domain for all frames. The dual-IHO spin--2 field is thus excluded from the KL representation a priori by the spectrum condition alone, independently of the non-normalizability of its ground state. The two exclusions reinforce one another, the spacelike pole removes the field from the spectral domain, and the absence of a normalizable vacuum removes the matrix elements that would define the spectral density in the first place. Either suffices to force $\rho=0$.}

\subsubsection{Unitarity of the physical $S$-matrix forces $C=0$}

{ 
For any such amplitude, the optical theorem \eqref{eq:OT-phase-space}
is the sum running over physical asymptotic states $f$. The dual IHO spin--2 cannot be physical asymptotic state since its pole is space-like and its spectral density is zero as we argued earlier. 
Suppose $C\neq0$ in \eqref{eq:family}, then cutting the dual-IHO line then contributes to the left-hand side of \eqref{eq:OT-phase-space} a term localized on the shell $k^2=\mu_2^2$. To be matched on the right-hand side, this would require a physical state $|f\rangle$ whose total momentum $k$ satisfies $k^2=\mu_2^2>0$, i.e.\ a spacelike total momentum with imaginary invariant mass $M_f^2=-\mu_2^2<0$. No such asymptotic state exists for the dual IHO spin--2, since every physical state must lie in $\overline V_+$ for that to happen, which is impossible. This means, a nonzero $C$ would deposit absorptive weight on the left-hand side of \eqref{eq:OT-phase-space} with no physical state available to saturate it on the right. Thus, unitarity of the graviton scalaron sector in quadratic gravity (with $\beta>0$) manifestly enforces
\begin{equation}
C=0\,.
\end{equation}
We stress that this conclusion uses only the spectrum condition on the genuine
asymptotic states and unitarity of the embedding amplitude. It does not assume that the dual-IHO field admits a KL representation or a vacuum of its own. 
The KL representation enters only as an independent and fully consistent corroboration. An on-shell term $C\,\delta(k^2-\mu_2^2)$ would correspond to KL spectral weight
\begin{equation}
\rho(s)\;\propto\;C\,\delta(s+\mu_2^2),
\end{equation}
supported at $s=-\mu_2^2<0$, outside the spectral domain $s\ge0$ fixed by
\eqref{eq:spectrum}. Demanding KL-representability gives $C=0$, thus we end up with the unique propagator of dual-IHO as 
\[
\tilde\Delta_{\rm dIHO}(k)
=
\operatorname{PV}\,\frac{i}{k^{2}-\mu_{2}^{2}} .
\label{eq:PV-propagator}
\]
Therefore, we can conclude the dual-IHO line never appears as an external leg because its propagator becomes $\rm PV$ not by a prescription but by a theoretical consistency (See \cite{Kumar:2026qnw} for more details). 
Thus, the dual-IHO spin--2 only appears on internal lines of
amplitudes whose asymptotic states are the massless graviton and the scalaron. These genuine asymptotic states possess a normalizable vacuum and satisfy the spectrum condition \eqref{eq:spectrum}. 
}

\subsection{The dual-IHO spin--2 field, unitarity cuts and renormalization}

After gauge fixing, the graviton propagator admits a Barnes--Rivers decomposition into spin projectors. For physical external gravitons, one uses transverse-traceless (TT) polarizations $\varepsilon_{\mu\nu}(k)$ satisfying
\begin{equation}
p^\mu\varepsilon_{\mu\nu}(k)=0,\qquad
\varepsilon^\mu{}_\mu(k)=0,\qquad
\varepsilon_{\mu\nu}=\varepsilon_{\nu\mu}.
\label{eq:TT}
\end{equation}
Thus, the spin--2 sector has the schematic form (in the DQFT)
\begin{equation}\label{eq:spin2-prop-PV}
D^{\pm(2)}_{\mu\nu\rho\sigma}(k)
=\frac{\mp i}{M_p^2}\,P^{(2)}_{\mu\nu\rho\sigma}
\left[
\frac{1}{k^2\mp i\epsilon}
-\mathrm{PV}\!\left(\frac{1}{k^2-\mu_2^2}\right)
\right]
\,,
\end{equation}
where $\pm$ correspond to the parity conjugate geometric superselection sectors with opposite arrows of time (See Sec.~\ref{sec:DQFT}). 
Notice that the dual-IHO spin--2 no $\delta(k^2-\mu_2^2)$ support from the extra spin--2 pole. 
This implies, in quadratic gravity with a spin--2 dual-IHO pole at $k^2=\mu_2^2>0$, it does not contribute to the optical theorem at tree-level for physical graviton scattering since it cannot appear as an on-shell intermediate state in any unitarity cut (See panel (a) in Fig.~\ref{fig:cut-vs-renorm}). To be more precise, For $2\to2$ a (forward) elastic scattering of physical massless external gravitons with momenta $p_1,p_2\to p_3,p_4$ obeying
$p_i^2=0$ and $p_i^0>0$, one has the kinematic inequalities
\begin{equation}
s \le 0,\qquad t \ge 0,\qquad u \ge 0,
\label{eq:kin-inequalities}
\end{equation}
with strict $s<0$ in the center-of-mass frame (since $k_s=(E_1+E_2,\mathbf{0})$ implies
$s=k_s^2=-(E_1+E_2)^2$). Although the momentum transfers $k_t,k_u$ are spacelike
($t,u\ge0$), the optical theorem for forward $2\to2$ scattering is controlled by the
cut the momentum flowing through the intermediate state, which is the total incoming
momentum $k=k_s=p_1+p_2$. Therefore the on-shell condition for the dual-IHO pole,
$k^2=\mu_2^2>0$, cannot be satisfied in the physical forward region.
Therefore, the dual-IHO spin--2 propagator yields no contribution to
$2\,\mathrm{Im}(M)$ in the forward unitarity cut. 
Since the optical theorem is precisely
the statement that $2\,\mathrm{Im}(M)$ equals the sum over cut (on-shell) intermediate states,it follows that the IHO mode does not appear in that sum. 
Nevertheless, this spin--2 mode can still contribute off shell to tree amplitudes and to loop renormalization (real parts and UV divergences) through the principal value pieces of propagators, but it cannot contribute to the unitarity (optical theorem) sum because it never goes on shell in physical kinematics and it also its propagator must be defined by the principal value as we learned in the previous section.

Let us consider the one-loop bubble (self-energy) diagram with two identical IHO lines and external momentum $P^\mu$ flowing through the two-point function (the lower panel of Fig.~\ref{fig:cut-vs-renorm}). Up to tensor numerators (spin projectors), the scalar skeleton integral reads
\begin{equation}
i\Pi(P)\;=\;\int\!\frac{d^4k}{(2\pi)^4}\,\mathcal{N}(k,P)
{\rm PV}\Big[\frac{i}{\big(k^2-\mu_2^2\big)}\Big]{\rm PV}\Big[\frac{i}{\big((P-k)^2-\mu_2^2\big)}\Big]
\label{eq:bubble}
\end{equation}
where $\Nc(k,\,P)$ here contains the vertex factors.
The location of branch points and cuts is controlled entirely by the Landau (pinch) conditions, i.e. by the possibility of putting the internal lines simultaneously on shell. For the present bubble these are
\begin{equation}
k^2=\mu_2^2,\qquad (P-k)^2=\mu_2^2.
\label{eq:onshell}
\end{equation}
Subtracting the two equations in \eqref{eq:onshell} gives
\begin{equation}
(P-k)^2-k^2 = 0
\quad\Longrightarrow\quad
P^2-2P\!\cdot\! k=0
\quad\Longrightarrow\quad
P\!\cdot\! k=\frac{P^2}{2}.
\label{eq:Pdotk}
\end{equation}
Now decompose $k^\mu$ into components parallel and orthogonal to $P^\mu$:
\begin{equation}
k^\mu = a\,P^\mu + k_\perp^\mu,
\qquad P\!\cdot\! k_\perp = 0.
\end{equation}
Then \eqref{eq:Pdotk} fixes $a=\tfrac12$, i.e.
\begin{equation}
k^\mu = \frac{1}{2}P^\mu + k_\perp^\mu,
\qquad P\!\cdot\! k_\perp = 0.
\label{eq:kdecomp}
\end{equation}
Imposing the on-shell condition $k^2=\mu_2^2$ and using \eqref{eq:kdecomp} yields
\begin{equation}
\mu_2^2 = k^2 = \left(\frac{P}{2}\right)^2 + k_\perp^2
= \frac{P^2}{4} + k_\perp^2
\quad\Longrightarrow\quad
k_\perp^2 = \mu_2^2 - \frac{P^2}{4}.
\label{eq:kperp2}
\end{equation}
The pinch singularity (and hence the branch point) occurs when the orthogonal momentum is forced to vanish, $k_\perp^\mu\to 0$, because then the two poles in the $k^0$-integration pinch the contour. Setting $k_\perp^2=0$ in \eqref{eq:kperp2} gives the branch point
\begin{equation}
{\,P^2_\ast = 4\mu_2^2\,}.
\label{eq:branchpoint}
\end{equation}
that lies on the spacelike side of the invariant axis. 
\eqref{eq:branchpoint} denotes the threshold, i.e., the minimum invariant energy required for a given internal configuration to become kinematically possible on-shell.
Therefore, for physical timelike external momentum in the $s$-channel (where $P^2<0$), the simultaneous on-shell conditions \eqref{eq:onshell} cannot be met on the physical sheet, and the IHO spin--2 bubble does not generate an absorptive part in that timelike kinematic region. This is the most crucial difference in comparison with the fakeon prescription \cite{Anselmi:2018bra,Anselmi:2018tmf} where the pole of the fakeon appears for timelike momenta $P^2 = -m_2^2<0$ (the on-shell condition) where as the threshold would become $P^2_\ast = -4m_2^2$. To project out the massive ghost poles theory, it must first be defined in Euclidean space, and specific contours must be defined. As a consequence, causality violation is inevitable at microscopic scales.  
In the IHO-type spin--2 case, unitarity cuts do not hit poles in the time-like momenta. 

Therefore, for ghosts, the pole and the two-particle threshold sit in the physical timelike region, so the mode enters unitarity cuts with the wrong sign; for the dual-IHO spin--2, both the pole and the branch point are spacelike, so the theory itself forbids absorptive contributions without any prescription. {In the companion paper \cite{Kumar:2026qnw}, we have explicitly shown that the mixed bubble 1-loop case where one of the internal lines is a dual IHO while the other is a massless graviton or massive scalaron does not generate any non-local counter terms in contrast to the cases of spin--2 ghost ($\beta<0$) and fakeonic spin--2 and Feynman-Wheeler like spin--2 \cite{Anselmi:2020tqo}. The prime reason is all the other possibilities, other than dual IH0 spin--2, the poles, and the physical Landau singularities occur at timelike momenta.}

It is important to see how the IHO (principal value) line enters loop integrals. A generic one-loop contribution to an amplitude (self-energy, vertex correction, or $2\to 2$ scattering) has the form 
\begin{equation}\label{eq:generic-loop}
\mathcal{M}^{\pm}_{1\text{-loop}}(p_i)
\;\sim\;
\int\!\frac{d^4q}{(2\pi)^4}\;
\frac{N(q,p_i)}{\prod_{a}\big(D_a(q,p_i)\mp i\epsilon\big)}.
\end{equation}
Here $p_i$ are external momenta, and:
\begin{enumerate}
\item $N(q,p_i)$ is the numerator produced by the tensor structure of vertices and propagators
(e.g., contractions of polarization tensors, projectors, and factors of momenta arising from derivatives).
In gravity, it is typically a polynomial in $q$ and the external momenta $p_i$.
\item The product over $a$ runs over the internal denominators (propagators) in the diagram; for instance
$D_a(q,p_i)$ could be $q^2$, $(q+p)^2$, $q^2-\mu_2^2$, etc., depending on the topology and routing.
\end{enumerate}
To isolate the effect of a single dual-IHO  spin--2 line, it is convenient to rewrite \eqref{eq:generic-loop} as
\begin{equation}
\mathcal{M}^{\pm ({\rm dual-IHO)}}_{1\text{-loop}}
\;\sim\;
\int\!\frac{d^4q}{(2\pi)^4}\;
F(q;p_i)\;
{\rm PV}\Big[\frac{1}{q^2-\mu_2^2}\Big],
\label{eq:loop-dualIHO}
\end{equation}
Here, $F(q;p_i)$ is simply all the remaining in the integrand after factoring out the IHO-like spin--2 in the denominator.
For fixed external kinematics, $F(q;p_i)$ is a (typically smooth) function of $q$ away from the poles of the
other denominators.
Now, we proceed to identify the exact principal value and $\delta$ decomposition at the level of the loop.

Let us now verify that, in the UV limit, the principal value prescription preserves the renormalizability of quadratic gravity.
For $|k^2|\gg \mu_2^2$ one finds
\begin{equation}\label{eq:PV-UV}
\mathrm{PV}\!\left(\frac{1}{k^2-\mu_2^2}\right)
=
\frac{1}{k^2}\left(1+\frac{\mu_2^2}{k^2}+O(k^{-4})\right),
\qquad |k^2|\to\infty.
\end{equation}
Therefore, at large loop momentum, the dual-IHO line falls off as $1/k^2$ (up to subleading powers), just like an
ordinary propagator. Consequently, the UV degree of divergence of \eqref{eq:loop-dualIHO} is controlled by the same power counting as in the standard Feynman prescription; the IHO mode still contributes to UV divergences, counterterms, and running couplings.

The difference between the Feynman and principal value prescriptions is
\begin{equation}
\frac{1}{k^2-\mu_2^2-i\epsilon}
-\mathrm{PV}\!\left(\frac{1}{k^2-\mu_2^2}\right)
=
i\pi\,\delta(k^2-\mu_2^2),
\end{equation}
which is supported only on the (finite) hypersurface $k^2=\mu_2^2$ in momentum space.
This on-shell term generates the discontinuities (absorptive contributions) associated with cuts through that line,
but it does not control the $|k|\to\infty$ region that produces UV divergences. Hence the dual-IHO prescription
removes the on-shell (cut-producing) part of the dual-IHO line while retaining its off-shell contribution to
renormalization and real (dispersive) loop corrections. Consequently, the dual-IHO mode can and does contribute {virtually} through the
principal value part of its propagator. This yields:
\begin{enumerate}
\item  real part corrections to amplitudes (dispersive effects), e.g.\ modifying the
finite parts of $2\to2$ graviton scattering at tree level and in loops;
\item UV improvement and counterterms, since the spin--2 sector effectively behaves as a
Pauli--Villars like subtraction,
\begin{equation}
D^{(2)}(k)\;\propto\;\left(\frac{1}{k^2}-{\rm PV}\LT\frac{1}{k^2-\mu_2^2}\RT\right)\,,
\end{equation}
so that for $k^2\to\infty$ one has $D^{(2)}(k)\sim \mu_2^2/k^4$, which is precisely the mechanism
behind the $1/k^4$ UV falloff and power-counting renormalizability.
\end{enumerate}
{Note that loop
diagrams may, of course, acquire imaginary parts from ordinary physical graviton, scalaron,
or matter lines, but not from cutting the dual-IHO spin--2 line. Thus, the dual-IHO sector
contributes virtually to real dispersive amplitudes and UV counterterms, while opening no
new physical intermediate-state channel in the optical theorem.}
Importantly, these virtual contributions affect the {running} of the local couplings
($M_P^2$, $\alpha$, $\beta$ and matter couplings (and $\Lambda_{\rm cc}$) if included) through standard
UV divergences.

In summary, the dual-IHO prescription enforces a clean split between what is allowed and what is forbidden. The IHO-like spin--2 part is off-shell as there is no $\delta$-function support, thus its propagator is understood as a principal value, so loop integrals generate only real contributions, exactly the pieces needed for renormalization (counterterms and the real parts of self-energies/vertices).  Since the $\delta$-supported part is what normally produces an absorptive contribution and hence a unitarity cut (via the optical theorem), the corresponding physical channel involving the spin--2 dual-IHO never opens. The dual-IHO spin--2 mode is therefore present as a virtual, renormalizing sector but is excluded from the spectrum of on-shell intermediate states. We also provided a further justification for this using the spectral representation of the dual-IHO propagator. {In the Appendix.~\ref{app:BRST}, we discuss the BRST quantization and physical Hilbert space structure and argue that the virtual dual IHO spin-2 sector is consistent with unitarity.}

\begin{figure}[t]
\centering

\begin{subfigure}{0.95\textwidth}
\centering
\resizebox{0.70\linewidth}{!}{%
\begin{tikzpicture}[font=\small]
\tikzset{lbl/.style={font=\scriptsize, fill=white, inner sep=1.2pt, rounded corners=1pt}}

\begin{feynman}
  \vertex (i1) at (-2, 1) {$p_1$};
  \vertex (i2) at (-2,-1) {$p_2$};
  \vertex (v1) at (-0.7,0) ;
  \vertex (v2) at (0.7,0) ;
  \vertex (f1) at (2, 1) {$p_3$};
  \vertex (f2) at (2,-1) {$p_4$};

  \diagram*{
    (i1) -- [graviton] (v1) -- [graviton, very thick, edge label={\scriptsize $k$}] (v2) -- [graviton] (f1),
    (i2) -- [graviton] (v1),
    (v2) -- [graviton] (f2),
  };

  \draw[dashed, thick] (0,1.35) -- (0,-1.35);
  \node[lbl] at (-0.40,1.50) {cut};

  \node[lbl] at ($(v1)!0.5!(v2) + (0,-0.38)$) {dual-IHO};
  \node[red] at (0.02,0.05) {\Large $\times$};

  \node[lbl, align=center, text width=6.2cm] at (0,-1.72)
  {On cut: $k^0>0\Rightarrow k^2\le 0$\\
   \textcolor{red}{\Large $\times$}\; On-shell $k^2=\mu_2^2>0$ (no support)};
\end{feynman}

\end{tikzpicture}%
}
\caption{Unitarity cuts have causal support and cannot hit the spacelike dual-IHO pole.}
\end{subfigure}

\vspace{1.0em}

\begin{subfigure}{0.95\textwidth}
\centering
\resizebox{0.62\linewidth}{!}{%
\begin{tikzpicture}[font=\small]
\begin{feynman}
  \vertex (a) at (-2.0,0) ;
  \vertex (b) at (-0.7,0) ;
  \vertex (c) at (0.7,0) ;
  \vertex (d) at (2.0,0) ;

  \diagram*{
    (a) -- [graviton] (b),
    (b) -- [half left, very thick, graviton, edge label={\scriptsize $D^{(2)}(k)$}] (c),
    (c) -- [half left, very thick, graviton] (b),
    (c) -- [graviton] (d),
  };
\end{feynman}
\end{tikzpicture}%
}

\vspace{0.2em}
{\normalsize
\[
D^{(2)}(k)\ \propto\ \left(\frac{1}{k^{2}}-\frac{1}{k^{2}-\mu_2^{2}}\right)\ \sim\ \frac{\mu_2^{2}}{k^{4}},
\qquad
\Rightarrow\ \text{power counting:}\ D=4-2V_{2}\le 4.
\]
}

\caption{UV improvement via $1/k^4$ falloff.}
\end{subfigure}

\caption{Dual-IHO spin--2: absent from physical unitarity cuts (top), but improves UV behavior via $1/k^4$ falloff (bottom).}
\label{fig:cut-vs-renorm}
\end{figure}

\section{A safe beginning of the Universe with the unitary quadratic quantum gravity}
\label{sec:safe-begin}

Any consistent quantum gravity theory has to consistently resolve the singularities that are inevitable in GR \cite{Misner:1969qxx}. In this section, we discuss how the unitary QQG can lead to "a safe beginning of the Universe'' which is singularity-free and can lead to the Universe that we perceive today.

The action functional plays a foundational role in both the classical and quantum formulations of gravity. Classically, it determines the field equations through a variational principle; quantum mechanically, it governs the weighting of configurations in the path integral. In this sense, the action is not merely a register device, but the central object controlling the dynamical and probabilistic structure of the theory. The finite-action principle, originally introduced by Barrow \cite{Barrow1987,Barrow2019} in the context of higher-derivative gravity and anisotropic cosmology, proposes that physically admissible cosmological histories must possess a finite gravitational action when integrated over spacetime. In the early Universe regime, this requirement can be expressed as the on-shell action of a given solution of the equations of motion
\begin{equation}
   \int_{t\to 0} dt \, d^3x \, \sqrt{-g}\, \mathcal{L}_\text{grav} \longrightarrow \text{finite}\; ,
\end{equation}
as the initial time is approached. This condition acts as a global constraint on the space of solutions and is independent of the detailed local behavior of the equations of motion. It is particularly relevant in the study of singular geometries, where curvature invariants may diverge, and higher-derivative terms can dominate the dynamics. From the path-integral perspective,
\begin{equation}
    \mathcal{Z} = \int Dg_{\mu\nu}\, e^{iS[g_{\mu\nu}]},
\end{equation}
configurations contribute according to their action phase. When the on-shell action diverges, the phase factor fails to admit a controlled semiclassical interpretation, and such configurations do not define acceptable saddle points of the functional integral. Consequently, geometries with divergent on-shell action are excluded from the class of configurations that can consistently contribute to the quantum theory. In this way, the finite-action principle provides a well-defined selection criterion rooted in the structure of the gravitational path integral.  Recently, this idea has been sharpened and systematically applied to quadratic gravity by Lehners and Stelle \cite{Lehners:2023fud}.
 
In GR, the generic approach to a spacelike singularity
is described by the Belinskii–Khalatnikov–Lifshitz (BKL) dynamics
\cite{Belinskii1970}. The essential feature of BKL evolution is that,
as $t \to 0$, time derivatives dominate over spatial derivatives and
the dynamics becomes effectively ultralocal. The geometry evolves
through a sequence of anisotropic Kasner-like epochs in which shear
dominates over matter and curvature terms.

A representative local form of the metric during such an epoch is the
Kasner solution
\begin{equation}
ds^2 = -dt^2 + \sum_{i=1}^3 t^{2p_i} dx_i^2,
\qquad
\sum_i p_i = 1,
\qquad
\sum_i p_i^2 = 1,
\end{equation}
where the exponents $p_i$ encode anisotropic expansion or contraction
along the spatial directions.
In these geometries the Weyl tensor is dominated by the shear and
scales as
\begin{equation}
W^2 \equiv { W}_{\mu\nu\rho\sigma} { W}^{\mu\nu\rho\sigma}
\sim \frac{\sigma^2}{t^4},
\end{equation}
with $\sigma^2 = \mathcal{O}(1)$ during BKL evolution, since the
anisotropy parameters remain finite while the overall curvature blows
up.
The effective spatial volume behaves as
\begin{equation}
a^3(t) \sim t,
\end{equation}
so that in quadratic gravity the Weyl-squared contribution to the
action contains
\begin{equation}
S_{\text{quad}}
\supset
\int dt\, a^3 W^2
\sim
\int dt\, t \cdot \frac{1}{t^4}
=
\int dt\, \frac{1}{t^3}.
\end{equation}
This integral diverges as $t \to 0$, demonstrating that BKL-type
anisotropic singularities generically produce infinite action in
quadratic gravity.
Thus, generic BKL and also the Mixmaster singularities \cite{Belinskii1982,Misner:1969qxx,Misner1969,Barrow1982} that are pathological and produce finite on-shell action in GR, lead to infinite on-shell action in quadratic
gravity. Under a finite-action selection principle in the gravitational
path integral, such configurations are excluded. This is the key point
sharpened by Lehners and Stelle: once higher-derivative terms are
present, the GR notion of a ``generic'' anisotropic singularity is no
longer admissible. 

What remains is a restricted class of early-time behaviors in which
anisotropy is dynamically regulated so that $a^3 W^2$ remains integrable.
In this regime, the additional spin--2 sector of quadratic gravity plays
a crucial structural role. For $\beta>0$, the massive spin--2 mode is
dynamically a dual-IHO and does have any asymptotic particle
excitations, it does not enlarge the physical spectrum. Instead, it
acts as a non-particle instability channel in the UV regime,
redistributing anisotropic energy and suppressing the shear growth that
would otherwise drive $W^2$ and the quadratic action to divergence.

Taken together, three elements form a coherent framework:
\begin{enumerate}
\item Stelle's renormalizable UV behavior of quadratic gravity
\item Lehners--Stelle finite-action criterion excluding BKL-like divergences in quadratic gravity
\item The dual-IHO realization of the spin--2 sector (the unitary quadratic quantum gravity)
\end{enumerate}

These components mutually reinforce one another. Ultraviolet control
and cosmological regularity arise from the same higher-derivative
structure. Within this perspective, quadratic gravity offers a credible route to
ultraviolet completion in which cosmological regularity is not imposed
by hand, but emerges from the same higher-derivative structure that
controls the UV behavior. In particular, the GR conclusion that anisotropic BKL-type singularities
are generically admissible no longer holds in quadratic gravity.
This follows directly from the finite-action criterion and the
higher-derivative structure of the theory, which modifies the UV
scaling of curvature invariants. 

Furthermore, it is interesting to note that finite action principle might also resolve black hole singularities. It is well-known that the interior of the Schwarzschild metric is analogous to Kantowski-Sachs cosmological spacetime (which is spatially homogeneous but anisotropic) \cite{Doran:2006dq}. Given finite action principle in quadratic gravity resolves anisotropic singularities. Therefore, in this respect the black holes in quadratic gravity could be non-singular (a similar argument is made in \cite{Holdom:2002xy} using perturbative numerical approaches). However, this subject requires a separate dedicated investigation which we defer for the future. 

\subsection{Starobinsky inflation as an emergent scenario}

Given that the finite action principle effectively excludes singular solutions in quadratic gravity, the UV instability triggered by the IHO spin-2 field would likely drive the Universe towards minimizing the Weyl-square part of the action. This generically lands us with the FLRW Universe in which the Weyl tensor is zero due to its conformal flatness. The $R^2$ part of the action \eqref{StelleQG} governs the dynamics of the Universe, in the form of what is known as the Starobinsky theory of cosmic inflation \cite{Starobinsky:1980te}. According to this, the Universe emerges from a highly symmetric configuration, the de Sitter phase \cite{Starobinsky:1981vz}.  As discussed earlier, the scalar fields in the de Sitter phase would again acquire IHO instability, the scalaron degree of freedom from $R^2$ modification of gravity drives the Universe gradually away from the de Sitter phase. 

Starobinsky inflation \cite{Starobinsky:1980te} is a quasi-de Sitter evolution of the Universe within the theory of $R+R^2$ gravity (which is quadratic gravity without the Weyl square term). This is the first proposed concrete framework of the early Universe motivated by the foundational aspects of quantum gravity \cite{Starobinsky:1981vz}. 
The Ricci scalar during Starobinsky inflation is an eigenmode of the d'Alembertian operator 
\begin{equation}
    \square R = m_0^2 R
    \label{staro}
\end{equation}
which is the trace equation of quadratic gravity in the absence of any matter energy-momentum tensor $T_{
\mu\nu}=0$. The Starobinsky framework's quasi-de Sitter evolution naturally exits to scalaron (matter) domination and subsequently to gravitational particle production (also known as reheating) \cite{Starobinsky:1981vz,Vilenkin:1985md,Gorbunov:2010bn}. 
Inflationary quasi-de Sitter evolution (in a flat FLRW Universe) is
\begin{equation}
	\begin{aligned}
		a(t) & \approx a_0 m_0^{-1/6}\LF t_f-t \RF^{-1/6}e^{-\frac{m_0^2\LF t_f-t \RF^2}{12}}  \, , \\
		H(t) =\frac{\dot{a}}{a} & \approx \frac{m_0^2}{6}\LF t_f-t \RF+ \frac{1}{6\LF t_f-t \RF}+\dots \, , \\
		R(t)\approx 12H^2+6\dot{H} &\approx \frac{m_0^2\LF t_f-t \RF^2}{3}-\frac{m_0^2}{3}+\frac{4}{3\LF t_f-t \RF^2}+\dots \, .
	\end{aligned}
	\label{solAS}
\end{equation}
This evolution is quasi-de Sitter since the slowly varying  Hubble parameter and the Ricci scalar during inflation $t\ll t_f$, which is expressed by the following slowly varying parameters 
\begin{equation}
    \epsilon_1 = -\frac{\dot H}{H^2},\quad \epsilon_2 = \frac{\dot F}{2HF},\quad F= M_p^2+2\alpha \bar R
    \label{slrdef}
\end{equation}
that are related as
\begin{equation}
\epsilon_2
=
\frac{6\alpha H}{F}
\left[
-\dot{\epsilon}_1
+ H\left(-4\epsilon_1 + 2\epsilon_1^2\right)
\right] 
\end{equation}
For the Starobinsky inflationary solution
\begin{equation}
    \epsilon_1 \approx \frac{1}{2N},\quad \epsilon_2 \approx -\epsilon_1+ \frac{1}{2}\epsilon_1^2 +\Oc\LF \epsilon_1^3 \RF
\end{equation}
After the quasi-de Sitter (inflationary) phase, the Universe enters the scalaron (matter) dominated era $t\gg t_f$
\begin{equation}
	\begin{aligned}
		a(t) & \approx a_r \LT m_0\LF t-t_f \RF \RT^{2/3}\LT 1+\frac{2}{3m_0\LF t-t_f \RF}\sin m_0\LF t-t_1 \RF \RT \, , \\ 
		H(t)&\approx  \frac{2}{3\LF t-t_f \RF}\Bigg( 1+\cos m_0\LF t-t_1 \RF \Bigg) \,,
	\end{aligned}
	\label{bmetric}
\end{equation}
where the oscillatory factors later result in the particle production by the decay of the scalaron, which renders the Universe into the radiation phase \cite{Starobinsky:1981vz,Jeong:2023zrv}. 
It is important to note that the quadratic gravity is renormalizable, even with couplings to matter fields \cite{Buoninfante:2025dgy,Buchbinder:2021wzv,Buchbinder:1992gdx}. However, for the UV renormalizability, the Higgs field must be coupled non-minimally to quadratic gravity \cite{Buchbinder:2021wzv,Buchbinder:1992gdx}. Coupling of matter fields to quadratic gravity can be written as 
\begin{equation}
    S_{\rm QG+SM} = \int d^4x \sqrt{-g}  \Bigg[ \frac{M_p^2}{2}R + \frac{\alpha}{2}R^2+ \frac{\beta}{2}W_{\mu\nu\rho\sigma}W^{\mu\nu\rho\sigma}+ \Lc_{\rm SM} \Bigg] 
    \label{QGSM}
\end{equation}
where $\Lc_{\rm SM}$ is the standard model Lagrangian. In \eqref{QGSM} the scalaron field is coupled to all SM degrees of freedom. To explicitly see this, 
we can convert the action into the Einstein frame via the conformal transformation as 
\begin{equation}
\begin{aligned}
    S_{\rm E-QG+SM}
= & \int d^4x\,\sqrt{-\tilde g}\,
\Bigg[
\frac{M_p^2}{2}\tilde R
- \frac{1}{2}\,\tilde g^{\mu\nu}\partial_\mu\phi\,\partial_\nu\phi
- V(\phi)
\\& + \frac{\beta}{2}\,
\tilde W_{\mu\nu\rho\sigma}\tilde W^{\mu\nu\rho\sigma}+ e^{-2\sqrt{\frac{2}{3}}\frac{\phi}{M_p}}\tilde\Lc_{SM}
\Bigg]
\end{aligned}
\label{SMQG}
\end{equation}
The equations of motion of quadratic gravity are given by 
\begin{equation}
\begin{aligned}
\frac{M_{\rm P}^2}{2} G_{\mu\nu}
+\frac{\alpha}{2}\,E^{(R^2)}_{\mu\nu}
+\frac{\beta}{2}\,B_{\mu\nu}
=\frac12\,T_{\mu\nu}\,,
\qquad
T_{\mu\nu}\equiv -\frac{2}{\sqrt{-g}}\frac{\delta S_{\rm m}}{\delta g^{\mu\nu}}\,.
\label{QQGeq}
\end{aligned}
\end{equation}
where 
\begin{align}
E^{(R^2)}_{\mu\nu}
\;=\;
2R\,R_{\mu\nu}
-\frac12\,g_{\mu\nu}R^2
+2\Big(g_{\mu\nu}\Box-\nabla_\mu\nabla_\nu\Big)R\,,
\qquad
\Box\equiv g^{\rho\sigma}\nabla_\rho\nabla_\sigma\,.
\label{QGeq}
\end{align}

\begin{equation}
\begin{aligned}
B_{\mu\nu}
\;\equiv\;
\nabla^\rho\nabla^\sigma C_{\mu\rho\nu\sigma}
+\frac12\,R^{\rho\sigma}C_{\mu\rho\nu\sigma},
\quad g^{\mu\nu}B_{\mu\nu}=0\,.
\end{aligned}
\label{bach}
\end{equation}
In the absence of matter, the trace equation of quadratic gravity \eqref{QGeq} is exactly the eigen-value equation \eqref{staro} for the Ricci scalar since the Bach tensor \eqref{bach} is traceless. 

It is vital to notice that the part of the Weyl square is invariant under conformal transformation.
 During the high curvature regime $i.e., \phi \gg M_p$, a part of the SM degrees of freedom are exponentially suppressed, and what remains is purely the role of quadratic gravity. 
 This answers the question of why, at very high curvature scales, the dynamics of scalarons with massless gravitons, UV-controlled by the IHO-like off-shell spin--2 mode, dominate the Universe.  We could perceive \eqref{SMQG} as the unification of SM and gravity towards the Planck scales. 
In this paper, we refrain from detailed discussions related to the two unknown elements of nature, known as dark energy and dark matter. These topics are beyond the scope of this paper. However, we stress that the compelling framework for dark energy is the cosmological constant $\Lambda_{\rm cc}$, which may not be an issue of the UV but rather of infrared (IR) scales \cite{Donoghue:2024uay}. One possibility to perceive cosmological constant as inverse of the Schwarzschild radius ($\Lambda_{\rm cc} = \frac{3}{r_S^2}$ associated with the entire (finite) Universe with respect to a superior observer \cite{Gaztanaga:2022fhp,Gaztanaga:2023hkm,Gaztanaga:2025cun}. 
The dark matter, on the other hand, could be a consequence of a much more intricate understanding of GR that has been recently investigated in the context of matter horizons, which are a new class of causal boundaries that separate local structures from cosmic expansion \cite{Umeh:2026ajv,Umeh:2026dm}. There are many other ideas such as entropic gravity \cite{Verlinde:2016toy} for dark energy and dark matter that do not invoke any exotic physics beyond the SM of particle physics. Thus, in this respect, quadratic gravity could be the last missing element in our understanding of the Universe and its origins. It is worth emphasizing that the Starobinsky scalaron could also drive a bounce in closed FLRW and smoothly lead to standard inflation in the flat FLRW as it was found in \cite{Muller:2025tfs}.  This bouncing scenario is also expected from the aspects of quantum theory \cite{Gaztanaga:2025cun} within a finite Universe. Thus the unitary QQG could potentially led to explanation of our Universe emerging from a quantum bounce. 

\subsection{1-loop beta function, asymptotic freedom and the running of couplings}

An often explored question in the literature about quadratic gravity is whether it is asymptotically free or safe \cite{Fradkin:1981iu,Avramidi:1985ki,Holdom:2015kbf}. This question actually arose from our understanding of QCD, which is asymptotically free. Quadratic gravity is often compared with QCD and expected to be asymptotically free. The 1-loop beta functions of quadratic gravity are given by \cite{Fradkin:1981iu,Avramidi:1985ki}
\begin{equation}
\begin{aligned}
\beta_{f_2}
& \equiv
\frac{d f_2}{d\ln\mu}
=
\frac{1}{(4\pi)^2}\frac{133}{10}\,f_2^{\,2} \\ 
\beta_{f_0}
&\equiv
\frac{d f_0}{d\ln\mu}
=
\frac{5}{18(4\pi)^2}
\left(
18 f_2^2
+18 f_2 f_0
+ f_0^2
\right).
\end{aligned}
\label{betaf}
\end{equation}
that indicate to us the running of the $R^2$ and $W^2$ coefficients with respect to the renormalization scale $\mu$ in the action \eqref{StelleQG}. The scale $\mu_0$ is some reference scale at which we specify the value of $f_2(\mu_0)$. 
Here $f_2 = \frac{M_p^2}{\beta}$ and $f_0 = \frac{M_p^2}{\alpha} $ indicate the physical coupling constants. Solving the first equation gives 
\begin{equation}
f_2(\mu)
=
\frac{f_2(\mu_0)}
{1 - a f_2(\mu_0)\ln\!\left(\frac{\mu}{\mu_0}\right)},\quad a = \frac{1}{\LF4\pi\RF^2}\frac{133}{10}\,.
\end{equation}
The coupling diverges when the denominator vanishes at a scale $\mu_\ast$, which is called the Landau pole 
\begin{equation}
1 - a f_2(\mu_0)\ln\!\left(\frac{\mu_*}{\mu_0}\right) = 0,\quad \mu_* 
= 
\mu_0 
\exp\!\left(
\frac{10(4\pi)^2}{133\,f_2(\mu_0)}
\right)
\end{equation}
Beyond this scale, we cannot trust the perturbation theory, and one needs a non-perturbative understanding of the theory. 

We assume the reference scale $\mu_0$ to be the scale of inflation $\sim 10^{-5}M_p$ which is the scale associated with the scalaron mass. We set the massive spin-2 IHO scale to be $\mu_2\gtrsim 10^{-4}M_p$ so that the IHO instability drives the Universe to start with inflation in a flat FLRW patch. With these numbers, we get 
\begin{equation}
\mu_* 
= 
\mu_0 \exp\!\left(
\frac{10(4\pi)^2}{133\,f_2(\mu_0)}
\right) M_p \approx 10^{5.17\times 10^{8}} M_p ,
\qquad 
\mu_0 = 10^{-5}M_p, 
\quad 
f_2(\mu_0)=10^{-8} 
\end{equation}
Thus, the Landau pole is at an absurdly enormous scale, exponentially far beyond any physical scale (even beyond any conceivable UV completion scale). This means that the running of $f_2$ is extraordinarily slow\footnote{We can verify this, given $f_2(\mu_0)= 10^{-8}$, at $\mu=M_p$, we get $f_2(M_p)
\approx
\left(1 + 9.7\times 10^{-9}\right)
\times 10^{-8}$.} over any physically relevant energy range, and the coupling remains perturbatively small up to scales vastly exceeding even the Planck scale. The apparent Landau pole arises only at an exponentially large energy, where the effective description itself would already have been superseded by whatever ultraviolet completion governs the theory. Consequently, for phenomenologically small initial values of $f_2$, the one-loop Landau pole is a purely formal artifact with no practical physical consequence. We can easily solve the second equation in \eqref{betaf} and conclude that for the above choice of couplings, the Landau pole for $f_0$ coupling would also be at astronomical large scale (far beyond the Planck scale). 

Although the one-loop running of the dimensionless quadratic couplings exhibits a formal Landau pole (no asymptotic freedom/safety at this order), for phenomenologically small couplings, the pole lies at exponentially large scales, far beyond any physical regime of interest. In this respect, the situation parallels quantum electrodynamics (QED) which also has a Landau pole at way beyond the Planck scale $\mu^e_L =
\mu^e_0 \exp\!\left(
\frac{3\pi}{2\,\alpha(\mu^e_0)}
\right) \sim 10^{262}\,M_p$ ($\alpha(m_e) \approx 1/137$ is the fine structure constant) \cite{Landau1954,Landau1956,PeskinSchroeder,WeinbergQFT2}, yet QED is a consistent, predictive QFT over all accessible energies, while the far-UV completion beyond the formal pole is expected to be governed by nonperturbative physics or new degrees of freedom long before the pole is reached. Therefore, for all practical purposes, quadratic gravity would remain perturbatively renormalizable even until several orders of magnitude beyond the Planck scale. Thus, what we find here is that unitary QQG is more like QED rather than QCD. The QCD analogy is only true is the case of quadratic gravity with spin--2 ghost \cite{Holdom:2016xfn,Holdom:2015kbf} which spoils the unitarity anyway, unless it is saved with contrived quantum prescriptions (See the discussion in Sec.~\ref{sec:prescriptions}). We may not have to worry about Landau pole in unitary QQG, since 
the IHO spin--2 triggered instability would naturally avoid reaching the regime of the Landau pole and the Universe would end up expanding into the infrared scales through cosmic expansion.

\section{Early Universe predictions of unitary quadratic quantum gravity}
\label{sec:QG-early universe}

A decisive test of any quantum theory of gravity lies in its observational implications for the early Universe. In this section, we investigate the cosmological predictions of unitary QQG, with particular emphasis on primordial perturbations generated during the high curvature regime. Our central aim is to determine how the tensor-to-scalar ratio and related observables are modified when the massive spin--2 sector is quantized as an IHO-like field within the direct-sum framework. Although this mode does not introduce additional asymptotic graviton polarizations, its presence can affect the vacuum structure and amplification dynamics of tensor modes in the early Universe. We analyze how this mechanism influences the primordial tensor spectrum and explore possible signatures, including parity asymmetries in the cosmic microwave background fluctuations, that may provide observational access to the underlying IHO spin--2 field dynamics. Despite the presence of a massive spin--2 excitation in the quadratic action, the theory predicts no additional asymptotic gravitational wave polarizations beyond the two transverse-traceless modes of general relativity, in agreement with current experimental bounds.

As discussed in Sec.~\ref {sec:IHO-SM}, inflationary perturbations are IHO fields; thus, they have to be quantized with DQFT (See Sec.~\ref {sec:DQFT}), which involves geometric superselection sectors defined by the discrete spacetime transformations. In this section, we provide two distinct predictions of unitary QQG. i) The effect of IHO spin--2 on the value of tensor-to-scalar ratio and spectral tilts, ii) The DQFT treatment of inflationary quantum fluctuations that leads to parity asymmetry of primordial power spectra (i.e., oscillating even-odd angular power spectra).

\subsection{Tensor-to-scalar ratio and spectral tilts}

Below, we present the scalar- and tensor-perturbed actions of QQG in the Jordan frame. 


\subsubsection{Scalar perturbations:}
To study the scalar perturbations, we start with the general metric perturbation 
\begin{equation}
ds^2=-(1+2\phi)\,dt^2 +2a\,\partial_i B\,dt\,dx^i
+a^2\Big[(1+2\psi)\delta_{ij}+2\partial_i\partial_j E\Big]dx^i dx^j.
\label{eq:scalar_metric}
\end{equation} 
In terms of the gauge-invariant curvature perturbation 
\begin{equation}
\mathcal R \equiv \psi - \frac{H}{\dot F}\,\delta F
\label{eq:R_def}
\end{equation}
where $F= 1 + 2\alpha R$, the second-order action of \eqref{StelleQG} for the curvature perturbation can be obtained as 
\begin{equation}
S^{(2)}_S
=\frac12\int d\tau\,d^3x\; z_s^2(\tau)\,
\Big[(\mathcal R')^2-(\nabla \mathcal R)^2\Big]
\label{eq:scalar_action}
\end{equation} 
The Weyl square term in \eqref{StelleQG} produces no additional propagating scalar degree of freedom and does not modify the reduced quadratic action for the single scalar mode after constraints are eliminated\footnote{Since $\bar W_{\mu\nu\rho\sigma}=0$ on FLRW, the quadratic contribution from the
Weyl-squared term
is
\[
S^{(2)}_{W^2}
=\frac{\beta}{2}\int dt\,d^3x\,a^3 (\delta W)^2 .
\]
For scalar perturbations, we find
\[
(\delta W)^2
=\frac{16}{3}\frac{k^4}{a^8}\,|\Sigma_k|^2 ,
\]
where 
\begin{equation}
    \Sigma
\equiv
(\Phi+\Psi)
+
a\,\frac{d}{dt}\!\left(a\,\sigma\right),
\qquad
\sigma\equiv \dot E-\frac{B}{a}.
\end{equation}
and therefore
\[
S^{(2)}_{W^2,\text{scalar}}
=
\frac{8\beta}{3}
\int dt\,\frac{d^3k}{(2\pi)^3}\,
\frac{k^4}{a^5}\,
|\Sigma_k|^2 ,
\]
which is $k^4$ suppressed, and on superhorizon scales ($k\to 0$), it dies fast and cannot source the conserved curvature perturbation.} \cite{Deruelle:2010kf}. Here
\begin{equation}
z_s^2 \equiv 2a^2 Q_s,
\label{eq:zs_Qs}
\end{equation}
where the exact kinetic coefficient $Q_s$ is
\begin{equation}
Q_s
= 6 M_p^2\, F \,
\frac{\epsilon_2^{\,2}}{(1+\epsilon_2)^2}
\label{eq:Qs_exact}
\end{equation}
Define the canonical variable
\begin{equation}
v \equiv z_s\,\mathcal R,
\label{eq:v_def}
\end{equation}
and Fourier expand $v(\tau,\mathbf x)=\int \frac{d^3k}{(2\pi)^3}v_k(\tau)e^{i\mathbf k\cdot\mathbf x}$.
Varying \eqref{eq:scalar_action} gives the exact Mukhanov–Sasaki (MS) equation
\begin{equation}
v_k''+\left(k^2-\frac{z_s''}{z_s}\right)v_k=0,
\qquad z_s=a\sqrt{2Q_s}.
\label{eq:MS_exact}
\end{equation}
The scalar power spectrum ($\Pc_\Rc$) obtained by 
\begin{equation}
{}_{\rm qdS}\langle 0 \vert \hat v_{\mathbf{k}} \hat v_{\mathbf{k}'} \vert 0 \rangle_{\rm qdS}
=
(2\pi)^3 \delta^{(3)}(\mathbf{k}+\mathbf{k}')
\frac{2\pi^2}{k^3}\,\mathcal{P}_{v}(k),
\qquad
\mathcal P_{\mathcal R}(k)\equiv \frac{k^3}{2\pi^2}\frac{1}{2a^2Q_s}|v_k|^2 
\label{eq:PR_def}
\end{equation}
where $\vert 0\rangle_{\rm qdS}$ is the quasi-de Sitter vacuum where we quantize the MS variable $v$. In \eqref{eq:PR_def}, the factor $\frac{1}{2a^2Q_s}$ comes from rescaling the field operator $\hat v$ by the classical factor $z_s$. 
The quantization procedure we follow is unitary QFTCS that relies on DQFT \cite{Kumar:2022zff,Gaztanaga:2024vtr,Gaztanaga:2024nwn,Gaztanaga:2025awe}, which we present in the next section.
Assuming the Bunch-Davies vacuum solution for the mode function (in de Sitter approximation) 
\begin{equation}
    v_k = \frac{1}{\sqrt{2k}}\LF 1-\frac{i}{k\tau} \RF e^{-ik\tau}
\end{equation}
the scalar power spectrum of Starobinsky inflation in unitary QQG can be computed as\footnote{Note that for $k^2 \ll a^2 H^2$, the curvature perturbation power spectrum depends only on 
the background quantities $H$ and $\epsilon$. Therefore, it is convenient to evaluate 
the spectrum at the moment of horizon crossing, defined by $k = aH$. 
Matching the sub-horizon vacuum normalization to the super-horizon solution 
introduces a factor of $\sqrt{2}$ for each mode crossing the horizon \cite{Powell:2006yg,Kinney:2009vz}. 
Consequently, the observed power spectrum satisfies
\begin{equation}
\mathcal{P}_{\mathcal{R}} \big|_{k=aH}
= 2\, \mathcal{P}_{\mathcal{R}} \big|_{k \ll aH}.
\end{equation}}
\begin{equation}
\begin{aligned}
\mathcal P_{\mathcal R}(k) & = \frac{k^3}{2\pi^2}\frac{1}{2a^2Q_s} \frac{1}{2k}\LF 1+\frac{a^2H^2}{k^2} \RF \\ & \simeq
\left.\frac{H^2}{48\pi^2 M_{\rm Pl}^2}\,
\frac{(1+\epsilon_2)^2}{F\,\epsilon_2^{\,2}}
\right|_{k=aH}.
\end{aligned}
\label{eq:PR_eps3}
\end{equation}
which is exactly the same as $R+R^2$ gravity \cite{DeFelice:2010aj,Nojiri:2010wj} since the Weyl square term does not contribute to the scalar sector at the linearized level. 
Evaluating quantities at horizon crossing $k=aH$, we can relate derivatives with respect to $k$ and the number of e-folds $N$ (from the end of inflation) via
\begin{equation}
d\ln k \simeq -\, d\ln N .
\end{equation}
The scalar spectral index then follows:
\begin{equation}
n_s - 1
= \frac{d\ln \mathcal{P}_\mathcal{R}}{d\ln k}
\simeq -\,\frac{d\ln \mathcal{P}_\mathcal{R}}{dN}
= -\,\frac{2}{N}.
\end{equation}
For $N \sim 50$--$60$, this agrees with the Planck 2018 result \cite{Planck:2018jri}
\begin{equation}
n_s = 0.9649 \pm 0.0042.
\end{equation}
CMB analyses parameterize the spectrum as
\begin{equation}
\mathcal{P}_\mathcal{R}(k)
= A_s \left(\frac{k}{k_*}\right)^{n_s-1},
\end{equation}
with $A_s = 2.2 \times 10^{-9}$ at $k_* = 0.05\,\mathrm{Mpc}^{-1}$.
This normalization fixes $m_0 \sim 1.3 \times 10^{-5} M_{\rm Pl}$ (for $N=55$ with assumptions on instant reheating \cite{Jeong:2023zrv}) and
\begin{equation}
H \sim 5.6 \times 10^{-5} M_{\rm Pl}
\sim 1.3 \times 10^{14}\,\mathrm{GeV},
\end{equation}
so the inflationary scale is $\mathcal{O}(10^{14}\,\mathrm{GeV})$.

It is important to mention the interpretation of ACT+DESI constraints on $n_s$. Recent CMB data from the Atacama Cosmology Telescope (ACT) combined with the Dark Energy Spectroscopic Instrument (DESI) reported slightly higher values of $n_s$. To be precise, the combined
Planck–ACT (P-ACT) analysis gives $n_s = 0.9709 \pm  0.0038$, while including CMB lensing and BAO
information from DESI (P-ACT-LB) raises it further to $n_s = 0.9743 \pm 0.0034$ \cite{AtacamaCosmologyTelescope:2025nti}. Presuming the number of e-foldings to be between $N = 50-60$, some investigations \cite{Kallosh:2025ijd,Kallosh:2025rni} perceived it as evidence against Starobinsky-like inflationary scenarios and started to tweak the shape of inflaton potentials in order to fit with these new data interpretations. In recent months, there has been a surge in the ad-hoc inflationary model building from numerous claims of UV-completions to match with ACT-DESI-Planck results. 

First of all it is worth noticing that the number of e-foldings during inflation does depend on the physics of reheating \cite{Kofman:1997yn,Ellis:2015pla,Iacconi:2023mnw,Jeong:2023zrv} which we cannot directly probe with the current observations \cite{Martin:2014nya}. 
Secondly, ACT does not directly measure the scalar spectral index $n_s$; rather, the reported preference for a higher $n_s$ is a derived, model-dependent inference obtained within a specific CMB+foreground+$\Lambda$CDM likelihood framework and typically in combination with external datasets. The extraction of $n_s$ relies on assumptions regarding foreground modeling, lensing reconstruction, calibration uncertainties, optical-depth priors, and the adopted cosmological parameter space. When DESI information is incorporated, the inference further inherits assumptions about late-time cosmology phenomenological extensions of spatially flat $\Lambda$CDM such as $w_0$–$w_a$ dark energy parameterizations, together with survey specific modeling choices (e.g. baryon acoustic oscillation (BAO), only versus full-shape analyzes, reconstruction procedures, galaxy bias modeling, covariance estimation, and scale cuts). These late-time assumptions are not orthogonal to early universe parameters: through degeneracies in the CMB likelihood, shifts in $\Omega_m$, $H_0$, and $A_s$ propagate directly into the posterior for $n_s$.
In particular, the scalar tilt is correlated with the matter density and expansion rate because these parameters jointly determine the shape and phase of the acoustic peaks, the damping tail, and the lensing amplitude. Consequently, any dataset combination that produces mild shifts or tensions in $\Omega_m$ or related parameters will generically induce a corresponding shift in the inferred $n_s$.
It is therefore methodologically incorrect to interpret a DESI conditioned shift in $n_s$ as a model independent measurement of the primordial spectrum, or to promote it to a decisive exclusion of inflationary attractor models such as Starobinsky $R^2$ inflation. Such conclusions remain conditional on the assumed late-time cosmological framework, priors, and dataset combinations, and should be presented as such. For more details, we suggest the reader to the critical recent studies cautioning on ACT-DESI-Planck data conclusions on changing $n_s$ upwards \cite{McDonough:2025lzo,DESI:2025gwf,daCosta:2024grm}. These studies indicate that the upward shift in $n_s$ is driven primarily by BAO--CMB tension\footnote{The term ``BAO--CMB tension'' refers to a statistical mismatch between parameters inferred from CMB anisotropies ($z \sim 1100$) and BAO measurements of late-time structure ($z \sim 0.1$--$2$). While the CMB constrains the angular sound horizon and parameters such as $\Omega_m$, $H_0$, $A_s$, and $n_s$, BAO probes the late-time expansion history. If the two datasets prefer slightly different values of $\Omega_m$ or $H_0$, their combination can induce correlated shifts in other parameters, including $n_s$, through degeneracies in the CMB likelihood.} and emphasize that the most significant change in $n_s$ occurs specifically when DESI data are included.

\subsubsection{Tensor perturbations}
The perturbed metric for the transverse--traceless (TT) perturbations is
\begin{equation}
ds^2=a(\tau)^2\left[-d\tau^2+\left(\delta_{ij}+2h_{ij}\right)dx^i dx^j\right],
\qquad
\partial_i h_{ij}=0,\qquad h_{ii}=0.
\end{equation}
The TT quadratic action for the unitary QQG takes the form
\begin{equation}
S_{\rm TT}^{(2)}
=\frac{1}{2}\int d\tau\, d^3x\;
a^2(\tau)\left[
F\left(h_{ij}'h_{ij}'-\partial_q h_{ij}\partial_q h_{ij}\right)
+\beta\,\Big(\mathcal D h_{ij}\Big)\Big(\mathcal D h_{ij}\Big)
\right],
\label{eq:STT_dS_start}
\end{equation}
where $F=M_p^2+2\alpha\bar R$,
prime denotes $d/d\tau$, and the de Sitter-covariant second-order operator acting on TT tensors is
\begin{equation}
\mathcal D h_{ij}\equiv h_{ij}''+2\frac{a'}{a}h_{ij}'-\nabla^2 h_{ij}.
\label{eq:Dop_def}
\end{equation}
(For TT perturbations on a conformally flat background, the Weyl-squared contribution reduces to a perfect square
built from $\mathcal D h_{ij}$ at quadratic order.)

\begin{equation}
S_T^{(2)} \;=\; \frac{1}{2} \int d\tau \,
\Big[
A \big( h'_{ij} h'_{ij} - k^2 h_{ij} h_{ij} \big)
\;+\;
\beta \, (\mathcal{D} h_{ij})(\mathcal{D} h_{ij})
\Big] .
\end{equation}
where $A(\tau) \equiv a^2(\tau) F(\tau)$.

Decompose into helicities and Fourier modes,
\begin{equation}
h_{ij}(\tau,\mathbf x)
=\sum_{s=\pm}\int\frac{d^3k}{(2\pi)^3}\,
e^{i\mathbf k\cdot\mathbf x}\,e^{(s)}_{ij}(\hat{\mathbf k})\,h_s(\tau,k),
\end{equation}
so $S_{\rm TT}^{(2)}=\sum_s\int d\tau\int\!\frac{d^3k}{(2\pi)^3}\,L_k[h_s]$, with
\begin{equation}
L_k[h]
=\frac{a^2}{2}\left[
F\left(h'^2-k^2 h^2\right)
+\beta\left(h''+2\frac{a'}{a}h'+k^2 h\right)^2
\right].
\label{eq:Lk_dS_4th}
\end{equation}
Introduce an auxiliary variable
\begin{equation}
\chi(\tau,k)\equiv h''+2\frac{a'}{a}h'+k^2 h,
\label{eq:q_def_dS}
\end{equation}
which allows us to write an equivalent second-order Lagrangian
\begin{equation}
L_k[h,\chi]
=\frac{a^2}{2}\left[
F\left(h'^2-k^2 h^2\right)
-\beta \chi^2
+2\beta \chi\left(h''+2\frac{a'}{a}h'+k^2 h\right)
\right].
\label{eq:Lk_hq_beforeIBP_dS}
\end{equation}
Integrating by parts to eliminate $h''$ (dropping total derivatives) gives
\begin{equation}
L_k[h,\chi]
=\frac{a^2}{2}\left[
F h'^2
-2\beta \chi' h'
- Fk^2 h^2
+2\beta k^2 \chi h
-\beta \chi^2
-4\beta\frac{a'}{a}\chi h'
\right].
\label{eq:Lk_hq_2nd_dS}
\end{equation}
The last term can be absorbed by completing the square in the kinetic structure after defining
the shifted field
\begin{equation}
\tilde h \equiv h-\frac{\beta}{F}\chi,
\qquad\Rightarrow\qquad
\tilde h' = h' - \frac{\beta}{F}\chi'.
\label{eq:htilde_dS_def}
\end{equation}
After straightforward algebra (the same cancellation mechanism as in Minkowski, now with the $a'/a$ terms),
the Lagrangian becomes diagonal in $(\tilde h,\chi)$:
\begin{equation}
L_k
=\frac{a^2}{2}\left[
F\left(\tilde h'^2-k^2\tilde h^2\right)
-\frac{\beta^2}{F}\left(\chi'^2-k^2 q^2\right)
+\beta \chi^2
\right].
\label{eq:Lk_diag_tildeh_q_dS}
\end{equation}
Introduce canonically normalized variables
\begin{equation}
u \equiv \sqrt{F}\,a\,\tilde h,
\qquad
v \equiv \frac{|\beta|}{\sqrt{F}}\,a\,\chi,
\label{eq:uv_def_dS}
\end{equation}
so that the action becomes a sum of two decoupled second-order systems,
\begin{equation}
S_{\rm TT}^{(2)}
=\frac12\sum_{s=\pm}\int d\tau\int\frac{d^3k}{(2\pi)^3}\left[
u_s'^2-\omega_u^2(\tau)u_s^2
\right]
-\frac12\sum_{s=\pm}\int d\tau\int\frac{d^3k}{(2\pi)^3}\left[
v_s'^2-\omega_v^2(\tau)v_s^2
\right],
\label{eq:STT_uv_dS}
\end{equation}
with
\begin{equation}
\omega_u^2(\tau)=k^2-\frac{a''}{a},
\qquad
\omega_v^2(\tau)=k^2+a^2\tilde m_2^2
+\frac{1}{2}\left(\frac{A'}{A}\right)'
-\frac{1}{4}\left(\frac{A'}{A}\right)^2,
\qquad
\tilde m_2^2\equiv -\frac{F}{\beta}.
\label{eq:omegas_dS}
\end{equation}
Thus, $\beta<0\Rightarrow {\tilde m}_2^2>0$ (massive spin--2 ghost sign in the action),
while $\beta>0\Rightarrow m_2^2<0$ (dual-IHO band).

The physical metric perturbation is related to the diagonal fields by
\begin{equation}
h = \tilde h + \frac{\beta}{F}\chi
\label{eq:h_in_uv_schematic}
\end{equation}
so even if the $v$-sector is not taken as an external late-time state, it contributes off-shell by renormalizing
the normalization of the $u$-sector in the effective action for $h$.
At tree level, integrating out $v$ (equivalently, eliminating $q$ off-shell from the quadratic action)
produces an effective kinetic prefactor for the massless tensor mode (See Appendix.~\ref{sec:ZT-derivation} for the detailed derivation)
\begin{equation}
S^{\rm TT}_{\rm eff}[u]
=\frac12\sum_{s=\pm}\int d\tau\int\frac{d^3k}{(2\pi)^3}\;
Z_T\left[u_s'^2-\left(k^2-\frac{z_t''}{z_t}\right)u_s^2\right],
\qquad
Z_T=1+\frac{2H^2}{m_2^2}
=1-\frac{2\beta H^2}{F},
\label{eq:ZT_def}
\end{equation}
where $z_t \equiv \frac{M_p}{2}\, a \sqrt{F}$ and
\begin{equation}
\frac{z_t''}{z_t}=\mathcal{H}^2\LF  2-\epsilon_1+3\epsilon_3+\epsilon_3^2-\epsilon_1\epsilon_3+\epsilon_3\epsilon_4\RF,\quad \epsilon_4 = \frac{\dot\epsilon_3}{H\epsilon_3}
\end{equation}
Therefore, the tensor power spectrum of unitary QQG would become 
\begin{equation}
\mathcal P_T(k)
=\frac{1}{\pi^2}\frac{H^2}{F}\;Z_T^{-1}\Bigg\vert_{k=aH}
\label{eq:PT_final}
\end{equation}
From \eqref{eq:PR_eps3} and \eqref{eq:PT_final} we can obtain the tensor-to-scalar ratio 
\begin{equation}
    r= 48\epsilon_2^2 \LF \frac{12\alpha}{12\alpha-\beta} \RF \approx \frac{12}{N^2} \frac{12\alpha}{12\alpha-\beta}
    \label{tensR}
\end{equation}
where we applied the approximation $F\approx 24\alpha H^2$. Structurally the result in \eqref{tensR} matches with the derivations in \cite{Deruelle:2010kf,Anselmi:2020lpp,Salvio:2022mld,Bianchi:2025tyl} when applied to our case of $\beta>0$. 
From the above tensor-to-scalar ratio expression, we can deduce that $12\alpha >\beta>0$. For the scalaron mass $m_0\sim 1.3\times 10^{-5} M_p$ (taking $N=55$), we get \begin{equation}
12\alpha \simeq 1.18343\times10^{10}.
\end{equation}
Using this, in Table.~\ref{tab:r_beta_correction} we can estimate the effect of $\beta$ on the tensor-to-scalar ratio 
\begin{table}[t]
\centering
\begin{tabular}{c c c c}
\hline
$\beta$ & $\dfrac{12\alpha}{12\alpha-\beta}$ & $r$ & $\dfrac{\Delta r}{r_0}$ \\
\hline
$0$        & $1.00000$ & $0.0039669$ & $0\%$ \\
$10^7$     & $1.00085$ & $0.0039703$ & $0.085\%$ \\
$10^8$     & $1.00852$ & $0.0040008$ & $0.852\%$ \\
$5\times10^8$ & $1.04408$ & $0.0041418$ & $4.41\%$ \\
$10^9$     & $1.09230$ & $0.0043331$ & $9.23\%$ \\
$2\times10^9$ & $1.20336$ & $0.0047737$ & $20.34\%$ \\
$5\times10^9$ & $1.73287$ & $0.0068732$ & $73.29\%$ \\
$10^{10}$  & $6.46960$ & $0.0256668$ & $546.96\%$ \\
\hline
\end{tabular}
\caption{Correction to the tensor-to-scalar ratio $r$ from the Weyl-squared parameter $\beta$, for $N=55$ and $1/(6\alpha)=(1.3\times10^{-5})^2$. The uncorrected Starobinsky value is $r_0\simeq 0.0039669$.}
\label{tab:r_beta_correction}
\end{table}
The results in Table.~\ref{tab:r_beta_correction} indicate that the Weyl-squared sector acts as a genuine quantum-gravity correction to the inflationary prediction for $r$. If $\beta$ is small compared to $12\alpha$, the deviation from the Starobinsky result is tiny, at the sub-percent level, and would be challenging to detect observationally. However, once $\beta$ approaches $10^9$ for the present parameter choice, the correction becomes of order $10\%$, suggesting that sufficiently precise future measurements of primordial $B$-modes could in principle probe the Weyl-squared contribution and thereby test this aspect of quantum gravity.

\subsection{Parity asymmetry of primordial power spectra} 

In this section, we present the detailed (unitary) quantization procedure of inflationary quantum fluctuations within the context of DQFT. This is shortly called the "direct-sum inflation (DSI)" (See \cite{Gaztanaga:2024vtr,Gaztanaga:2024nwn,Gaztanaga:2025awe} for more details) 
\begin{equation}
S^{(2)}  = \frac{1}{2}
\int d\tau\, d^3x \,
\left[
(v')^2
- (\nabla v)^2
+ \frac{z_s''}{z_s} v^2
\right],
\end{equation}
We quantize the MS variable $v$ \eqref{eq:v_def}, which means promoting it to an operator. According to DQFT, split a single quantum field operator ($\hat v$) as a direct-sum of two components ($\hat v_\pm$), which is similar to how it is done in Minkowski \eqref{minQuant}.
\begin{equation}
		\begin{aligned}
			\hat{v}  & = \frac{1}{\sqrt{2}} \hat{v}_{(+)}\LF \tau,\, \textbf{x} \RF \oplus  \frac{1}{\sqrt{2}} \hat{v}_{(-)}\LF -\tau,\, -\textbf{x} \RF \\ 
   & = \frac{1}{\sqrt{2}} \begin{pmatrix}
				\hat{v}_{+} \LF \tau,\,\textbf{x} \RF & 0 \\ 
				0 & \hat{v}_- \LF -\tau,\, -\textbf{x} \RF
			\end{pmatrix}
		\end{aligned}
		\label{fieldmat}
	\end{equation}
which can be expanded as 
\begin{equation}
	\begin{aligned}
& \hat{v}_{\pm }=   
\int \frac{ d^3k}{\LF 2\pi \RF^{3/2}} \Bigg[ c_{\LF\pm \RF\textbf{k}} {v}_{\pm,\,k} e^{\pm i\textbf{k}\cdot \textbf{x}} + c_{\LF\pm \RF\textbf{k}}^\dagger {v}_{\pm,\,k}^\ast e^{\mp i\textbf{k}\cdot \textbf{x}} \Bigg]
\end{aligned}
\label{vid}
\end{equation} 
where $c_{\pm},\,c_{\pm}^\dagger$ are canonical creation and annihilation operators 
\begin{equation}
\bigl[ \hat c_{(\pm)\mathbf{k}},\, \hat c^{\dagger}_{(\pm)\mathbf{k}'} \bigr]
= \delta^{(3)}\!\bigl(\mathbf{k}-\mathbf{k}'\bigr),
\qquad
\bigl[ \hat c^{\dagger}_{(\pm)\mathbf{k}'},\, \hat c^{\dagger}_{(\mp)\mathbf{k}'} \bigr] = \bigl[ \hat c_{(\pm)\mathbf{k}'},\, \hat c^{\dagger}_{(\mp)\mathbf{k}'} \bigr]
= 0.
\end{equation}
The inflationary (quasi-de Sitter) vacuum, according to DQFT, is 
\begin{equation}
    \vert 0\rangle_{\rm qdS} = \begin{pmatrix}
         \vert 0_+\rangle_{\rm qdS} & \\  \vert 0_-\rangle_{\rm qdS}
    \end{pmatrix},\quad c_{(\pm)\textbf{k}}\vert 0_{\pm}\rangle_{\rm qdS} = 0\,.
    \label{qdSvac}
\end{equation}
which are defined by the set of mode functions 
\begin{equation}
	v_{\pm,\,k}^{\prime\prime}+ \LF k^2-\frac{{\nu}_s^{\LF \pm\RF 2}-\frac{1}{4}}{\tau^2} \RF v_{\pm,\,k}^2 =0,\quad \nu_s^{\pm} \approx \frac{3}{2}\pm 2\epsilon_1
	\label{MS-equation}
\end{equation}
where we can see that the time-dependent masses of the IHO fields $v_{\pm}$ are asymmetric around the de Sitter case, which is $\nu_s^\pm = \frac{3}{2}$. This is because the de Sitter metric is $\Pc\Tc$ symmetric (See \eqref{FLRWdS} and \eqref{tdSsym}) whereas the inflationary quasi-de Sitter spacetime breaks the time reversal symmetry \eqref{tdSsym} by the slow-roll parameters \eqref{slrdef}. Recall that time is a parameter in quantum theory; in the context of gravity, the dynamics of spacetime further complicates the non-linear nature of time and its role in defining the behavior of quantum fields. In DSI, we deform the quantum vacuum for the parity conjugate fields $v_\pm$ by choosing a vacuum that explicitly geometrically breaks the degeneracy exactly by the amount of slow-roll parameters. In other words, the MS field components in DSI are deformed-$\Pc\Tc$ mirror of each other 
\begin{equation}
S_{\pm}^{(2)} = \frac{1}{2}
\int d\tau\, d^3x \,
v_\pm \Kc_\pm v_\pm 
\end{equation}
such that the average of their kernels is exactly de Sitter
\begin{equation}
    \Kc_{\rm dS} = \frac{\Kc_++\Kc_- }{2} = -\pd_{\tau}^2+ \nabla^2+ \frac{2}{\tau^2}
\end{equation}
where 
\begin{equation}
    \Kc_\pm = -\pd_{\tau}^2+ \nabla^2+ \frac{2\pm \Delta(\tau)}{\tau^2},\quad \nu_s^2-\frac{1}{4}= \Delta(\tau)
\end{equation}
It is important to note that in the DSI framework, there is no duplication of degrees of freedom. It is a geometric-consistency construction: slow-roll deforms the PT-mirror sectors with opposite signs so that the average kernel remains exactly de Sitter. This preserves the global Bunch–Davies analytic structure (and hence the standard averaged power spectrum). This configuration of the vacuum has led to the parity asymmetry observed in the CMB \cite{Gaztanaga:2024vtr}.

More explicitly, parity conjugate MS components in DSI are given by 
\begin{equation}
v_{\pm, k}  =  v^{\rm dS}_{\pm, k} \LF 1\pm \Delta v_k\RF
\end{equation}
where
\begin{equation}
    v_{\pm, k}^{\rm dS} = \sqrt{\frac{1}{2k}} e^{\mp ik\tau}\LF 1\mp\frac{i}{k\tau} \RF,\quad \Delta v_k = \LF 2\epsilon_1 \RF \LT \frac{1}{H_{3/2}^{(1)} \LF  \frac{k}{k_\ast} \RF} \frac{\pd H^{(1)}_{\nu_s}\LF \frac{k}{k_\ast} \RF}{\pd\nu_s} \Bigg\vert_{\nu_s=\frac{3}{2}} \RT
    \label{modeqds}
\end{equation}
where $\epsilon_1 \approx \frac{1}{2N}$. Converting this into the mode functions of curvature perturbation, we get
\begin{equation}
    \Rc_{\pm \textbf{k}} = \frac{1}{a\sqrt{2Q_s}} v_{\pm k} = \Rc_{\pm\,\textbf{k}}^{\rm dS}\LF 1\pm \Delta v\RF   \quad ; \quad \Rc_{\pm\,\textbf{k}}^{\rm dS} \equiv \frac{1}{a\sqrt{2Q_s}} v_{\pm,\,\textbf{k}}^{\rm dS} 
    \label{eq:vpm}
\end{equation}
If we compute two-point correlations of curvature perturbation in the DSI vacuum 
\begin{equation}
\begin{aligned}
{}_{\rm qdS}\langle 0 \vert \hat \Rc_{\mathbf{k}} \hat \Rc_{\mathbf{k}'} \vert 0 \rangle_{\rm qdS} & = \frac{1}{a\sqrt{2Q_s}} \Bigg[{}_{\rm qdS}\langle 0_+ \vert \hat v_{(+)\mathbf{k}} \hat v_{(+)\mathbf{k}'} \vert 0_+ \rangle_{\rm qdS}+{}_{\rm qdS}\langle 0_- \vert \hat v_{(-)\mathbf{k}} \hat v_{(-)\mathbf{k}'} \vert 0_- \rangle_{\rm qdS}\Bigg]
 \\ & =
(2\pi)^3 \delta^{(3)}(\mathbf{k}+\mathbf{k}')
\frac{2\pi^2}{k^3}\,\mathcal{P}_{\Rc}(k),
\end{aligned}
\label{PRDSI}
\end{equation}
we recover exactly what we have in \eqref{eq:PR_def} and \eqref{eq:PR_eps3}. 
The key DSI prediction is that the {\it two} parity-conjugate sectors have unequal
two-point correlations. Equivalently, the curvature spectrum splits into
\begin{equation}
P_{\Rc}^{\pm}(k)\;\simeq\;P_{\Rc}(k)\,\bigl(1\pm \Delta P_v(k)\bigr),
\label{eq:PRpm_177}
\end{equation}
Here, the fractional asymmetry is
\begin{equation}
\Delta P_v(k)
=(1-n_s)\,\text{Re}\!\left[
\frac{2}{H^{(1)}_{3/2}\!\left(\tfrac{k}{k_*}\right)}
\left.\frac{\partial H^{(1)}_{\nu_s}\!\left(\tfrac{k}{k_*}\right)}{\partial \nu_s}\right|_{\nu_s=3/2}
\right],
\label{eq:DeltaPv_179}
\end{equation}
The average of the $\Pc\Tc$-conjugate sectors reproduces the standard inflationary power spectrum, ensuring consistency with the success of conventional inflationary predictions. The observable deviation appears only in parity-sensitive statistics of large-scale CMB anisotropies.

If our vacuum choice is not parity asymmetric, then we get $\Delta P_v=0$, but this is not what we observed in the data \cite{Gaztanaga:2024vtr,Gaztanaga:2024nwn}.
Because $Y_{\ell m}(-\hat n)=(-1)^\ell Y_{\ell m}(\hat n)$, a parity-odd modulation
$\delta T(\hat n)=-\delta T(-\hat n)$ feeds {\it opposite signs} into even vs.\ odd multipoles,
leading to the angular-spectrum prediction (See \cite{Gaztanaga:2024nwn} or Section.~7 in \cite{Gaztanaga:2025awe})
\begin{equation}
C_\ell^{\rm DSI}
=
C_\ell^{\rm SI}\Bigl[1+(-1)^{\ell+1}\Delta C_\ell\Bigr],
\qquad
\Delta C_\ell=\frac{1}{C_\ell^{\rm SI}}\int_0^{k_c}\frac{dk}{k}\,
A_s\Bigl(\frac{k}{k_*}\Bigr)^{n_s-1} j_\ell^2\!\Bigl(\frac{k}{k_s}\Bigr)\,\Delta P_v(k), 
\label{eq:Cell_odd_even}
\end{equation}
where $C_\ell^{\rm SI}$ is the angular power spectrum of standard inflation (SI) with near scale-invariant power-law power spectrum 
\begin{equation}
    C_\ell^{\rm SI} = \int_0^{\infty} \frac{dk}{k} A_s\left( \frac{k}{k_\ast} \right)^{n_s - 1} j_\ell^2\left( \frac{k}{k_s} \right)
    \label{eq:SICl}
\end{equation} 
Here $k_c= 0.02 k_\ast$ is the cut-off scale that corresponds to $\ell_{\rm max}$ where CMB low-ell anomalies are observed \cite{Schwarz:2015cma,Planck:2019evm}. Note that the quantum fluctuations in DSI are inherently non-Markovian. Consequently, the influence of these early modes on smaller scales requires further investigation (see Section~5.4 of \cite{Gaztanaga:2024vtr} for a detailed discussion).
Hence, DSI predicts suppression of even $\ell$ and enhancement of odd $\ell$ at low multipoles.
A standard parity indicator is
\begin{equation}
R_{TT}(\ell_{\max})
=
\frac{\sum_{\ell=2}^{\ell_{\max}}\ell(\ell+1)C_{\ell}^{\rm even}}
{\sum_{\ell=3}^{\ell_{\max}}\ell(\ell+1)C_{\ell}^{\rm odd}}
\simeq 0.79 \quad (\ell_{\max}\lesssim 20\text{--}30),
\label{eq:RTT}
\end{equation}
i.e.\ about $\sim 20\%$ excess power in odd multipoles.
The statistical evidence for DSI under Planck 2018 data is shown in \cite{Gaztanaga:2024vtr,Gaztanaga:2024nwn,Gaztanaga:2025awe} by comparing posterior model probabilities $p(M|D)$ (not only likelihoods $p(D|M)$). Using Monte-Carlo sky realizations (See Table~2 \cite{Gaztanaga:2025awe}), the parity asymmetry on large scales predicted by DSI is found to be 650 times more probable than SI with combined indicators of low quadrupole $C_2$ and $R_{TT}$.

Up to this point, we have shown that the direct-sum formulation of
inflationary quantum fluctuations naturally leads to two $\Pc\Tc$-conjugate sectors. It is crucial, however, to clarify the origin of the resulting asymmetry.
First, the slow-roll parameter $\epsilon_1 = -\dot H/H^2$
is a scalar characterizing the background solution and does not change sign under reversal of the arrow of time within the same cosmological
history. The $\Pc\Tc$-related sectors therefore do not correspond to backgrounds with $\epsilon_1$ and $-\epsilon_1$. The asymmetry does not arise from flipping the slow-roll dynamics. The origin of the asymmetry lies instead in horizon thermality and in the analytic structure of quantum two-point functions.

In the exact de Sitter spacetime, the Bunch–Davies vacuum can be defined
by analytic continuation from the Euclidean four-sphere \cite{Allen:1985ux}.
The regularity of the Euclidean correlator enforces periodicity in
imaginary conformal time with period
\begin{equation}
\beta_T = \frac{2\pi}{\vert H\vert }.
\end{equation}
This implies that the Wightman functions
The Wightman functions are therefore
\begin{equation}
G^{>}(x,x')
=
\langle 0_{\rm BD}|
\hat v(x)\hat v(x')
|0_{\rm BD}\rangle,
\qquad
G^{<}(x,x')
=
\langle 0_{\rm BD}|
\hat v(x')\hat v(x)
|0_{\rm BD}\rangle,
\end{equation}
of the Mukhanov–Sasaki
variable. With the momentum space representation 
\begin{equation}
G^{>}(x,x')
=
\int\frac{d^3k}{(2\pi)^3}
e^{i\mathbf{k}\cdot(\mathbf{x}-\mathbf{x}')}
v_k(\tau)v_k^*(\tau'),
\end{equation}
\begin{equation}
G^{<}(x,x')
=
\int\frac{d^3k}{(2\pi)^3}
e^{i\mathbf{k}\cdot(\mathbf{x}-\mathbf{x}')}
v_k(\tau')v_k^*(\tau).
\end{equation}
of the Wightman functions, we can easily check the Kubo–Martin–Schwinger (KMS) condition \cite{Kubo1957,MartinSchwinger1959,Haag1967,KapustaGale2006}
\begin{equation}
G^{>}(t-i\beta_T,\mathbf{k})
=
G^{<}(t,\mathbf{k})\quad {\rm or}\quad G^{>}\!\left(\tau e^{i\vert H\vert \beta_T}, \tau'; k\right)
=
G^{<}\!\left(\tau, \tau'; k\right)
\end{equation}
which characterizes thermal behavior at temperature $T=\vert H\vert /(2\pi)$. It is important to note that the KMS condition tells us about an important symmetry in de Sitter space, which is that near $\tau=0$ the de Sitter mode behaves as a fractional power
\begin{equation}
\begin{aligned}
v_k(\tau)=\frac{1}{\sqrt{2k}}\,
\sqrt{\frac{\pi}{2}}\,
e^{i\left(\nu+\frac12\right)\frac{\pi}{2}}\,
\sqrt{-k\tau}\,H^{(1)}_{\nu}(-k\tau) \Big\vert_{k\tau\to 0} 
& \;\sim\;\frac{1}{\sqrt{k}}\,
(-k\tau)^{\frac12-\nu} \\ 
& =\frac{1}{\sqrt{k}}\exp\!\left[\Big(\frac12-\nu\Big)\log(-k\tau)\right].
\end{aligned}
\end{equation}
The point $\tau=0$ is a branch point in the complex $\tau$-plane.
A $2\pi$ analytic continuation corresponds to a closed loop
\begin{equation}
\tau \;\to\; e^{\pm 2\pi i}\tau ,
\end{equation}
where the sign distinguishes counter-clockwise and clockwise orientation.
Since
\begin{equation}
\log(-\tau)\;\to\;\log(-\tau)\pm 2\pi i,
\end{equation}
The mode acquires the monodromy phase
\begin{equation}
v_k(\tau)\;\to\;e^{\mp 2\pi i\nu}\,v_k(\tau).
\end{equation}
Thus, the monodromy is a geometric phase associated with a closed loop in the
complexified time plane. 
The monodromy phase directly controls the KMS condition.
In flat de Sitter,
\begin{equation}
\tau = -\frac{1}{H}e^{-Ht}
\quad \Rightarrow \quad
t \to t - i\beta_T,
\qquad
\beta_T = \frac{2\pi}{\vert H\vert }
\quad \Longleftrightarrow \quad
\tau \to e^{2\pi i}\tau.
\end{equation}
Thus the KMS shift corresponds to a $2\pi$ rotation around the branch point at $\tau=0$. For a single sector with index $\nu$,
\begin{equation}
v_k(e^{2\pi i}\tau) = e^{-2\pi i\nu} v_k(\tau),
\end{equation}
and hence mode-by-mode
\begin{equation}
G_k^{>}(t-i\beta_T,\tau')
=
e^{-2\pi i\nu}\, G_k^{>}(\tau,\tau').
\end{equation}
In exact de Sitter ($\nu=\tfrac32$), the BD branch choice ensures that
the $2\pi$ rotation exchanges the Wightman orderings,
\begin{equation}
G_k^{>}(t-i\beta_T,\tau')
=
G_k^{<}(t,\tau').
\end{equation}
In DSI we have sectorial Wightman functions 
\begin{equation}
G^{>,\pm}(t,t';k)
=
\langle 0_\pm|
\hat v_\pm(t)\hat v_\pm(t')
|0_\pm\rangle,
\qquad
G^{<,\pm}(t,t';k)
=
\langle 0_\pm|
\hat v_\pm(t')\hat v_\pm(t)
|0_\pm\rangle 
\end{equation}
that follow from \eqref{fieldmat} and \eqref{qdSvac}. 

In quasi-de Sitter, in the DSI, we impose a deformed vacuum for the parity conjugate sectors by the choice of Hankel functions with 
\begin{equation}
\nu_\pm=\frac32\pm 2\epsilon_1,
\end{equation}
So the monodromy becomes
\begin{equation}
e^{-2\pi i\nu_\pm}
=
e^{-3\pi i}\,e^{\mp 4\pi i\epsilon_1}.
\end{equation}
Thus, the mode functions in DSI pick up a geometric phase under complex rotation as 
\begin{equation}
    \hat v_{k}\LF \tau e^{2\pi i} \RF  = \begin{pmatrix}
\hat v_{+,k}\!\left(\tau e^{2\pi i}\right) & 0 \\
0 & \hat v_{-,k}\!\left(-\tau e^{2\pi i}\right)
\end{pmatrix}
=
\frac{1}{\sqrt{2}}
\begin{pmatrix}
e^{4\pi\epsilon_1}\,\hat v_{+,k}(\tau) &  0 \\
0 & e^{-4\pi\epsilon_1}\,\hat v_{-,k}(-\tau)
\end{pmatrix}
\label{vgeomph}
\end{equation}
where $\hat v_{
\pm,\,k} = c_{(\pm),\,\textbf{k}} v_{\pm,\,k}$.
This phase effect is similar in spirit to a Berry phase \cite{Berry1984,ShapereWilczek1989}. In the Berry phase, when the physical conditions of a quantum system (such as a magnetic field) are slowly changed along a closed cycle and brought back to their original values, the wavefunction returns to the same state but acquires an additional phase that depends only on the path taken. A closely related phenomenon occurs here: when the mode functions are analytically continued around the special point 
$\tau=0$
in the complex time plane, they also return to themselves with an extra phase. In this sense the phase arises from completing a loop. The difference is that in the present case the loop occurs in the analytic structure of the wave equation (complex time) rather than from slowly varying physical parameters, so the phase originates from the branch structure of the mode solutions rather than from adiabatic evolution of the physical parameters. We can notice that the parity conjugate sectors picks up phase factors with opposite signs (i.e., $e^{\pm 4\pi\epsilon_1}$) reflecting the fact that the fields $v_\pm$ are $\Pc\Tc$ mirror states.

The slow-roll deformation, therefore, modifies the KMS relation sectorwise,
\begin{equation}
G_k^{>,\pm}(t-i\beta_T,\tau')
=
e^{\mp 4\pi i\epsilon_1}\,
G_k^{<,\pm}(t,\tau')
+
\mathcal{O}(\epsilon_1^2).
\end{equation}
Thus, each $\Pc\Tc$ sector individually violates exact KMS at linear order.
However, combining the mirror sectors,
\begin{equation}
G_k^{>,\text{tot}}
=
\frac12\left(G_k^{>,+}+G_k^{>,-}\right),
\end{equation}
one obtains
\begin{equation}
G_k^{>,\text{tot}}(t-i\beta_T)
=
\cos(4\pi\epsilon_1)\,
G_k^{<,\text{tot}}(t)
+
\mathcal{O}(\epsilon_1^2).
\end{equation}
Since
\begin{equation}
\cos(4\pi\epsilon_1)=1+\mathcal{O}(\epsilon_1^2),
\end{equation}
the leading $\mathcal{O}(\epsilon_1)$ KMS deformation cancels in the mirror-symmetric combination.
Hence, while slow-roll evolution softly breaks thermality in each sector,
the $\Pc\Tc$ mirror structure preserves the KMS condition
to first order in $\epsilon_1$ for the combined state.

Because $\Pc\Tc$ exchanges the two analytic branches, these slow-roll
corrections induce a small mismatch between the $\Pc\Tc$-conjugate sectors.
Therefore, in quasi–de Sitter inflation,
$\dot H \neq 0$ softly breaks this symmetry, leading to
$\mathcal{O}(\epsilon)$ corrections in the analytic structure of the
propagators and, consequently, to a sector dependent realization of
the vacuum. Thus, the parity asymmetric vacuum arises through the natural slow-roll deformation of the exact KMS symmetric de Sitter state once time-translation invariance is softly broken.

In summary, in quasi--de Sitter spacetime, $\Pc\Tc$ is no longer an exact symmetry of the background geometry since $\dot H \neq 0$. 
However, $\Pc\Tc$ remains the geometric map relating the two superselection sectors (parity conjugate regions) in DSI. 
Therefore, the slow-roll deformation must enter antisymmetrically in the sector kernels,
\begin{equation}
\Kc_-  -\Kc_{\mathrm{dS}}
=
-\left( \Kc_+ - \Kc_{\mathrm{dS}} \right),
\end{equation}
so that the two sectors remain $\Pc\Tc$ mirror images of one another. 
Equivalently, in terms of the Hankel indices,
\begin{equation}
\nu^{(+)}_s - \frac{3}{2}
=
-\left( \nu_s^{(-)} - \frac{3}{2} \right),
\end{equation}
ensuring that the deformation away from exact de Sitter lifts the $\Pc\Tc$ degeneracy while preserving the mirror structure of the quantum states.

Applying the same DSI quantization procedure with geometric superselection sectors to the inflationary tensor modes
\eqref{hijexp}
\begin{equation}
\begin{aligned}
\hat{u}  & = \frac{1}{\sqrt{2}} \hat{u}_{(+)}\LF \tau,\, \textbf{x} \RF \oplus  
\frac{1}{\sqrt{2}} \hat{u}_{(-)}\LF -\tau,\, -\textbf{x} \RF \\ 
& = \frac{1}{\sqrt{2}} 
\begin{pmatrix}
\hat{u}_{+} \LF \tau,\,\textbf{x} \RF & 0 \\ 
0 & \hat{u}_- \LF -\tau,\, -\textbf{x} \RF
\end{pmatrix}
\end{aligned}
\label{fieldmat}
\end{equation}
with the Fourier expansion 
\begin{equation}
\begin{aligned}
\hat{u}_{\pm }=
\int \frac{ d^3k}{\LF 2\pi \RF^{3/2}}
\Bigg[
d_{\LF\pm \RF\mathbf{k}}\, {u}_{\pm,\,k} e^{\pm i\mathbf{k}\cdot \mathbf{x}}
+
d_{\LF\pm \RF\mathbf{k}}^\dagger {u}_{\pm,\,k}^\ast e^{\mp i\mathbf{k}\cdot \mathbf{x}}
\Bigg]
\end{aligned}
\label{uid}
\end{equation}
where the creation and annihilation operators satisfy 
\begin{equation}
\bigl[ \hat d_{(\pm)\mathbf{k}},\, \hat d^{\dagger}_{(\pm)\mathbf{k}'} \bigr]
= \delta^{(3)}\!\bigl(\mathbf{k}-\mathbf{k}'\bigr),
\qquad
\bigl[ \hat d^{\dagger}_{(\pm)\mathbf{k}'},\, \hat d^{\dagger}_{(\mp)\mathbf{k}'} \bigr]
=
\bigl[ \hat d_{(\pm)\mathbf{k}'},\, \hat d^{\dagger}_{(\mp)\mathbf{k}'} \bigr]
= 0 .
\end{equation}
The $\Pc\Tc$ mirror mode functions in the DSI quasi-de Sitter vacuum  can be obtained through 
\begin{equation}
u_{\pm,k}'' + \left(k^2 - \frac{\nu_t^{(\pm)2} - \tfrac{1}{4}}{\tau^2}\right) u_{\pm,k} = 0,
\qquad
\nu_t^{\pm} \approx \frac{3}{2} \pm 3\epsilon_1^2 .
\end{equation}
Similar to \eqref{vgeomph} these tensor modes pick up geometric phase in the quasi-de Sitter vacuum as 
\begin{equation}
    \hat u_{k}\LF \tau e^{2\pi i} \RF  = \begin{pmatrix}
\hat u_{+,k}\!\left(\tau e^{2\pi i}\right) & 0 \\
0 & \hat u_{-,k}\!\left(-\tau e^{2\pi i}\right)
\end{pmatrix}
=
\frac{1}{\sqrt{2}}
\begin{pmatrix}
e^{6\pi\epsilon_1^2}\,\hat u_{+,k}(\tau) &  0 \\
0 & e^{-6\pi\epsilon_1^2}\,\hat u_{-,k}(-\tau)
\end{pmatrix}
\end{equation}
This 
yields parity asymmetric gravitational power spectra \cite{Kumar:2022zff,Gaztanaga:2025awe}
\begin{equation}
\Pc_{h\pm}
= \int d^3r \, e^{-i\mathbf{k}\cdot\mathbf{r}}\,
{}_{\rm qdS}\langle 0_{\pm}\vert
\hat h_{ij\pm}\!\left(\pm\tau,\pm\mathbf{x}\right)
\hat h_{ij\pm}\!\left(\pm\tau,\pm\mathbf{x}'\right)
\vert 0_\pm\rangle_{\rm qdS}
\;\approx\;
P_h\left(1\pm \Delta \Pc_u\right),
\label{eq:pTdsi}
\end{equation}
where
\begin{equation}
P_h
= A_t\!\left(\frac{k}{k_\ast}\right)^{n_t},
\qquad
r=-8n_t,
\qquad
\Delta \Pc_u
=
\left(\frac{r}{8}\right)
\operatorname{Re}\!\left[
\frac{2}{H_{3/2}^{(1)}\!\left(\frac{k}{k_\ast}\right)}
\frac{\partial H^{(1)}_{\nu_s}\!\left(\frac{k}{k_\ast}\right)}
{\partial\nu_s}
\Bigg|_{\nu_s=\frac{3}{2}}
\right].
\label{eq:delpPT}
\end{equation}
Here $r=\frac{A_t}{A_s}<0.032$ denotes the tensor-to-scalar ratio, which is
constrained from above by the latest BICEP2/Planck/Keck Array measurements of
CMB $B$-mode polarization
\cite{BICEP:2021xfz,Tristram:2021tvh}.

Equation \eqref{eq:pTdsi} shows that in DSI, with $\Pc\Tc$ based geometric quantization the tensor
power spectrum acquires an asymmetric amplitude for two-point correlations of
tensor modes evaluated at parity conjugate points of physical space. This
effect is conceptually distinct from the tensor signatures predicted by
parity-violating theories of gravity, where the asymmetry arises through
chiral gravitational wave polarizations and can lead to phenomena such as
cosmological birefringence
\cite{Lue:1998mq,Gluscevic:2010vv,Contaldi:2003zv,Komatsu:2022nvu}.

The tensor power spectrum \eqref{eq:pTdsi} leads to a parity-asymmetric angular power spectrum for the CMB $B$-mode polarization, which can be written as
\begin{equation}
C_{\ell,\,BB}^{\rm DSI}
=
C_{\ell,\,BB}^{\rm SI}
\left[
1 + (-1)^{\ell+1}\,\Delta \mathcal{C}_{\ell,\,BB}
\right],
\label{clBB}
\end{equation}
where the statistically isotropic contribution is
\begin{equation}
C_{\ell,\,BB}^{\rm SI}
=
\int_0^{\infty} \frac{dk}{k}\,
A_t
\left(
\frac{k}{k_\ast}
\right)^{n_t}
T_{\ell,\,BB}^{2}
\left(
\frac{k}{k_s}
\right),
\label{eq:SIClB}
\end{equation}
and the parity-asymmetric correction is given by
\begin{equation}
\Delta \mathcal{C}_{\ell,\,BB}
=
\frac{1}{C_{\ell,\,BB}^{\rm SI}}
\int_0^{k_c}
\frac{dk}{k}\,
A_t
\left(
\frac{k}{k_\ast}
\right)^{n_t}
T_{\ell,\,BB}^{2}
\left(
\frac{k}{k_s}
\right)
\Delta \mathcal{P}_u(k).
\label{eq:RDclBB}
\end{equation}
Here $T_{\ell,\,BB}$ denotes the radiation transfer function associated with the $B$-mode polarization \cite{Zaldarriaga:1996xe}.

\section{Conclusions}
\label{sec:conclusions}

Quadratic gravity has long been regarded as the unique perturbatively renormalizable modification of Einstein's gravity in four dimensions. Its apparent drawback has always been the presence of an additional massive spin--2 degree of freedom, traditionally understood as a ghost that spoils the unitarity of the theory. Thus, quadratic gravity has been thought to be a pathological theory by the majority of physicists. Alternative frameworks of quantum gravity which have complicated the concept of Planck scale physics with less and less connection to the criterion of renormalizability and to the core principles of QFT.

In this work we revisited the longstanding problem of unitarity in quadratic gravity from a new but a foundational perspective, motivated by the physics of IHOs which is omnipresent across various scientific domains including the standard model of particle physics and QFT at the gravitational horizons. We find a remarkable revelation that the IHO-like instability is what is needed for a consistent UV complete quantum gravity. Our formulation presents a novel understanding that the IHO-like instability carried by off-shell quantum spin-2 degree of freedom achieves unitarity and UV completion for quadratic gravity (or Stelle gravity).

To be more explicit, we showed that the spin--2 ghost of quadratic gravity can be replaced by an IHO-like (called "dual-IHO") spin--2 degree of freedom rather than a pathological negative norm excitation (i.e, ghost). This occurs by a change in sign of the coefficient of Weyl square term in the quadratic gravity action \eqref{StelleQG}.
This approach replaces the conventional ghost instability by a healthy dynamical instability associated with saddle point dynamics, which is also prevalent in the contexts of the SM Higgs mechanism and quantum fields near gravitational horizons.

Building on this insight, the direct-sum quantization framework  we developed earlier brings here a consistent understanding of quantum IHO(-like) degrees of freedom. 
Within this formulation, the spin--2 transverse traceless sector of quadratic gravity admits a unitary quantization, restoring probabilistic consistency while preserving renormalizability. We demonstrated explicitly that scattering amplitudes satisfy the optical theorem. We clarified how imaginary parts do not arise through standard cutting identities (i.e., Cutkowsky rules) because of the spacelike support of the IHO-like modes. Our analysis further suggests that the massive spin--2 sector, when treated as a dual-IHO system, effectively decouples from asymptotic observables while remaining dynamically relevant in the deep ultraviolet virtual quantum processes. This provides a concrete mechanism by which quadratic gravity can remain unitary without introducing additional propagating polarizations in physical graviton states. A particularly sharp result of our construction is the derivation of the dual-IHO propagator from first principles via the Källén–Lehmann representation. The spectral density $\rho$ vanishes identically for two independent and self-consistent reasons: first, the dual-IHO ground state is not square-integrable, so no normalizable vacuum exists from which to define the KL matrix elements; second, the propagator pole sits at spacelike momentum $k^2 =\mu_2^2 > 0$, which lies strictly outside the timelike integration domain of the KL representation. Together, these enforce the principal value propagator as the unique physically consistent prescription, not as an external input, but as a theorem derived from the Hilbert space structure of the theory. This is the precise mechanism by which the dual-IHO spin–2 mode contributes to UV renormalization through real dispersive loop effects, while contributing zero imaginary part to any amplitude at any loop order, thereby leaving the optical theorem intact.

Embracing the finite action principle, unitary QQG might be free from cosmological and BH singularities.
From a cosmological standpoint, the dual-IHO dynamics offers a natural route toward singularity avoidance and a nonsingular origin of the Universe, supplying a driving instability that replaces classical divergences by quantum controlled evolution.
Importantly, the IHO-like instability in quadratic gravity has a universal appeal. The same IHO structure (which is nothing but the Berry--Keating $H = xp$ Hamiltonian \cite{Berry1999}), appears in the descriptions of near-horizon QFTCS, in tachyonic nature of $\mathbb Z_2$ symmetry breaking of the SM, in inflationary (quasi-de Sitter) perturbations. Interestingly Berry-Keating IHO is also widely discussed in other fields such as condensed matter physics. Thus, 
the recurrence of this structure across such disparate areas strongly suggests that the inverted harmonic oscillator is not an anomaly of higher-derivative gravity, but a universal dynamical template of unstable or horizon-dominated systems in fundamental physics.

More broadly, our results highlight a unifying role played by IHOs across quantum field theory, gravitational horizons, and UV behavior of higher-derivative gravity. Therefore, rather than signaling inconsistency, inverted oscillator dynamics encodes horizon physics, tachyonic phases, and ultraviolet gravitational behavior within a single conceptual framework. This perspective opens a viable path toward a unitary, renormalizable theory of gravity and motivates further investigation into its observational consequences, including early Universe dynamics and tensor sector signatures. Interestingly, with dual-IHO spin--2 field, the renormalization group structure of QQG is analogous to that of QED i.e., QQG remains perturbatively well defined and predictive across all physically relevant scales because the Landau pole (as per 1-loop beta function) in QQG appears parametrically far beyond the Planck regime, where new physics or nonperturbative effects are expected to intervene. The presence of the spin–2 IHO degree of freedom does not destabilize this structure; rather, it participates in the UV running while remaining absent from asymptotic spectra. On the observational side, the same sector leaves a controlled imprint in the early Universe. In particular, the tensor-to-scalar ratio receives a correction
\begin{equation}
    r\approx \frac{12}{N^2} \frac{12\alpha}{12\alpha-\beta}
\end{equation}
so that the Starobinsky value is recovered as a lower bound in the limit 
$\beta\rightarrow 0$ (See Table.~\ref{tab:r_beta_correction}). The inclusion of the spin–2 IHO sector therefore, predicts a mild enhancement of primordial tensor modes without introducing extra propagating polarizations. This provides a concrete and testable link between ultraviolet consistency, horizon-scale dynamics, and cosmological observables.
In addition, we argue that a parity asymmetry on large scales (i.e., more power in the odd multipoles compared to even multipoles) becomes natural in the scalar and tensor sectors of the power spectra. Strikingly, they are of the same type and reflect the $\Pc\Tc$ breaking nature of quasi-de Sitter vacuum.
We conclude that quadratic gravity, when reformulated through direct-sum quantum theory and IHO quantization, becomes a consistent UV-complete theory of quantum gravity which is unitary and renormalizable. 

Thus, quadratic gravity is not merely a higher-derivative correction to Einstein theory, but a consistent and predictive framework in which ultraviolet completion, cosmology, and quantum field theory in curved spacetime are naturally unified.

\acknowledgments

This paper is dedicated to the memory of K. S. Stelle and A. A. Starobinsky. KSK acknowledges the support from the Royal Society in the name of the Newton International Fellowship. This research was funded by
Fundação para a Ciência e a Tecnologia grant number UIDB/MAT/00212/2025 and COST
action 23130. The authors thank Gerard 't Hooft, Alexey S. Koshelev, Luca Buoninfante, Damiano Anselmi, John Donoghue, David Wands, Enrique Gaztañaga, Mariam Bouhmadi L\'opez, Masahide Yamaguchi, Obinna Umeh, and Mathew Hull for useful discussions on various occasions.

\appendix

\section{BRST projection and the physical Hilbert space}
\label{app:BRST}

We briefly recall the role of BRST symmetry in the covariant quantization of
higher-derivative gravity. This is the same structure used by Stelle in his proof of
renormalizability of quadratic gravity \cite{Stelle:1976gc}. The purpose of this appendix is
not to rederive the optical theorem, already stated in Eq.~(4.33), but to clarify which
states belong to the physical Hilbert space of the gauge-fixed theory.

In a gravitational theory, the metric contains gauge redundancy because two metrics
related by an infinitesimal diffeomorphism describe the same physical geometry. If
$g_{\mu\nu}=\bar g_{\mu\nu}+h_{\mu\nu}$, then schematically
\begin{equation}
\delta_\xi h_{\mu\nu}
=
\nabla_\mu \xi_\nu+\nabla_\nu \xi_\mu
+{\cal O}(h\xi),
\end{equation}
where $\xi^\mu(x)$ is the infinitesimal diffeomorphism parameter. To define a propagator
one must impose a gauge condition,
\begin{equation}
F_\mu[h]=0.
\end{equation}
In the path integral this gauge condition is inserted together with the corresponding
Faddeev--Popov determinant,
\begin{equation}
1
=
\int {\cal D}\xi\,
\delta\!\left(F_\mu[h^\xi]\right)
\det{\cal M},
\qquad
{\cal M}_{\mu\nu}
=
\frac{\delta F_\mu[h^\xi]}{\delta \xi^\nu}\bigg|_{\xi=0}.
\end{equation}
The determinant $\det{\cal M}$ is represented by introducing anticommuting
Faddeev--Popov ghost and anti-ghost fields, $c^\mu$ and $\bar c_\mu$:
\begin{equation}
\det{\cal M}
=
\int {\cal D}\bar c\,{\cal D}c\,
\exp\left[
i\int d^4x\,\sqrt{-\bar g}\,
\bar c_\mu {\cal M}^{\mu}{}_{\nu}c^\nu
\right].
\end{equation}
These fields are not physical matter fields. They appear only because the gauge-fixing
determinant must be represented inside the covariant path integral. This is why
anticommuting ghost variables occur even in a theory of gravity.

After gauge fixing, the original diffeomorphism invariance is no longer manifest in the
same form. What remains is a residual symmetry of the gauge-fixed action: the
Becchi--Rouet--Stora symmetry, usually called BRS or BRST symmetry. In this symmetry,
the ordinary diffeomorphism parameter $\xi^\mu$ is replaced by the ghost field $c^\mu$.
A convenient schematic form of the BRST transformations is
\begin{align}
\delta_B g_{\mu\nu} &= {\cal L}_{c} g_{\mu\nu}, \\
\delta_B c^\mu &= -c^\nu \nabla_\nu c^\mu, \\
\delta_B \bar c_\mu &= B_\mu, \\
\delta_B B_\mu &=0,
\end{align}
where $B_\mu$ is an auxiliary field. The essential algebraic property is nilpotency,
\begin{equation}
\delta_B^2=0.
\end{equation}
Nilpotency expresses the closure of the gauge symmetry in the gauge-fixed language. The gauge-fixed action can then be written schematically as
\begin{equation}
S_{\rm tot}
=
S_{\rm QG}
+
\delta_B\Psi_{\rm gf},
\label{eq:BRST_total_action}
\end{equation}
where $\Psi_{\rm gf}$ is a Grassmann-odd gauge-fixing functional. A standard choice is
\begin{equation}
\Psi_{\rm gf}
=
\int d^4x\,\sqrt{-\bar g}\,
\bar c_\mu
\left(
Y^{\mu\nu}F_\nu[h]
+
\frac{\xi}{2}B^\mu
\right).
\end{equation}
Here $Y^{\mu\nu}$ denotes the operator used to weight the gauge condition. In Stelle's
covariant quantization of higher-derivative gravity, the gauge-fixing term is chosen so that
the quadratic kinetic operator can be inverted while preserving the required large-momentum
behaviour of the graviton propagator \cite{Stelle:1976gc}. Acting with $\delta_B$ gives
\begin{align}
\delta_B\Psi_{\rm gf}
=
\int d^4x\,\sqrt{-\bar g}\,
\left[
B_\mu Y^{\mu\nu}F_\nu[h]
+
\frac{\xi}{2}B_\mu B^\mu
-
\bar c_\mu Y^{\mu\nu}{\cal M}_{\nu\rho}c^\rho
\right],
\end{align}
which is the gauge-fixing term together with the Faddeev--Popov ghost action.
Since $S_{\rm QG}$ is diffeomorphism invariant and $\delta_B^2=0$, the gauge-fixed
action satisfies
\begin{equation}
\delta_B S_{\rm tot}=0.
\end{equation}
This BRST invariance is the symmetry behind the Slavnov identities used in Stelle's
renormalization proof \cite{Stelle:1976gc}. These identities ensure that the gauge-fixed
calculation still respects the original gauge symmetry and that the allowed divergences
have the correct gauge-invariant structure.
The conserved charge associated with BRST symmetry is denoted $Q_B$ and satisfies
\begin{equation}
Q_B^2=0.
\end{equation}
The physical Hilbert space is then defined by BRST cohomology:
\begin{equation}
{\cal H}_{\rm phys}
=
\frac{\ker Q_B}{\operatorname{im}Q_B}.
\label{eq:BRST_cohomology}
\end{equation}
This means that physical states obey
\begin{equation}
Q_B|\psi\rangle=0,
\end{equation}
while states of the form
\begin{equation}
|\psi\rangle=Q_B|\chi\rangle
\end{equation}
are pure gauge and are identified with zero. In this way, the covariant formalism contains
longitudinal metric components and Faddeev--Popov ghosts in intermediate calculations,
but they do not appear as physical asymptotic states.

This BRST projection removes gauge redundancy; it does not remove genuine
gauge-invariant poles of the propagator. Therefore the conventional massive spin--2 ghost
of quadratic gravity is not removed by BRST. For the usual $\beta<0$ sign, this pole is
timelike,
\begin{equation}
k^2=-m_2^2<0,
\end{equation}
and belongs to the transverse-traceless spin--2 sector. It is not a Faddeev--Popov ghost and not a longitudinal gauge mode. The $\beta>0$ theory studied in this paper is different. The additional spin--2 sector is a
dual-IHO sector with spacelike pole
\begin{equation}
k^2=\mu_2^2>0.
\end{equation}
As shown in Sec.~\ref{sec:KL-PV}, this pole lies outside the Källén--Lehmann spectral domain of
physical asymptotic states. Equivalently, the dual-IHO spin--2 sector carries no physical
spectral weight and its internal Green function is the principal-value Green distribution \eqref{eq:PV-propagator}.
It contributes only virtually, through real dispersive amplitudes and ultraviolet
renormalization, but it does not supply an on-shell physical state.

Thus the role of BRST in the present argument is limited and precise. It defines the
physical state space of the covariantly gauge-fixed gravitational theory. The
Källén--Lehmann/principal-value analysis then shows that the dual-IHO spin--2 sector does
not furnish an element of that physical asymptotic state space. This is why the covariant
BRST formulation and the transverse-traceless analysis used in the main text are consistent
with each other.

\section{Effective quadratic action for the massless graviton during inflation}
\label{sec:ZT-derivation}

We start from the quadratic TT action (one Fourier mode and one helicity; the sum over polarizations is implicit)
\begin{equation}
S^{(2)}_{TT}
=\frac12\int d\tau\,\frac{d^3k}{(2\pi)^3}\; L[h] ,
\qquad
L[h]=\frac{a^2}{2}\Big\{F\big(h'^2-k^2h^2\big)+\beta\,\chi^2\Big\},
\label{eq:TT-start}
\end{equation}
where
\begin{equation}
\chi \equiv Dh \equiv h''+2\frac{a'}{a}h' + k^2 h .
\label{eq:D-operator}
\end{equation}
This is a four-derivative Gaussian theory. A convenient second-order representation is obtained by introducing
an auxiliary field $\Xi$,
\begin{equation}
\beta\,\chi^2 \equiv -\beta\,\Xi^2+2\beta\,\Xi\,\chi ,
\label{eq:aux-identity}
\end{equation}
so that
\begin{equation}
L[h,\Xi]=\frac{a^2}{2}\Big\{F\big(h'^2-k^2h^2\big)-\beta \Xi^2+2\beta \Xi\,Dh\Big\}.
\label{eq:Lhs}
\end{equation}
Integrating out $\Xi$ reproduces the original four-derivative action since the $\Xi$ equation of motion gives
$\Xi=Dh$.

\paragraph{Diagonal second-order form and the physical tensor variable.}
The same quadratic theory can be written in a manifestly two-pole second-order form in terms of a massless
mode $\tilde h$ and a massive mode $\xi$,
\begin{equation}
S^{(2)}_{TT}
=\frac12\int d\tau\,\frac{d^3k}{(2\pi)^3}\Big[
\tilde h\,\mathcal K_0\,\tilde h \;-\; \xi\,\mathcal K_2\,\xi
\Big].
\label{eq:diag-form}
\end{equation}
Here $\mathcal K_0$ and $\mathcal K_2$ are the (self-adjoint) quadratic kernels for the massless and massive
spin--2 sectors in conformal time. For TT modes on exact de Sitter they take the operator form
\begin{align}
\mathcal K_0
&=Fa^2\Big(\partial_\tau^2+2\frac{a'}{a}\partial_\tau+k^2\Big),
\label{eq:K0}
\\
\mathcal K_2
&=Fa^2\Big(\partial_\tau^2+2\frac{a'}{a}\partial_\tau+k^2+a^2 \tilde m_2^2\Big),
\label{eq:K2}
\end{align}
with 
\begin{equation}
    \tilde m_2^2\equiv -\frac{F}{\beta}.
    \label{eq:medef}
\end{equation}
The overall relative sign between the two sectors in \eqref{eq:diag-form} encodes whether the second pole
corresponds to a ghost ($\tilde m_2^2>0$) or to a dual-IHO mode ($\tilde m_2^2<0$), but the derivation below is
purely Gaussian and does not depend on this interpretation.

The key point is that in the Jordan frame, the matter sources couple to the {metric} perturbation $h$.
Hence, the observable correlator is $\langle h h\rangle$, not $\langle \tilde h\tilde h\rangle$. In the
diagonal basis, the source-coupled field is a linear combination. We parametrize this by
\begin{equation}
h=\tilde h+\xi,
\qquad\Rightarrow\qquad
\tilde h=h-\xi.
\label{eq:h-tilde-chi}
\end{equation}

\paragraph{Gaussian elimination of the heavy mode.}
Substituting \eqref{eq:h-tilde-chi} into \eqref{eq:diag-form} yields
\begin{align}
S^{(2)}_{TT}[h,\xi]
&=\frac12\int d\tau\,\frac{d^3k}{(2\pi)^3}\Big[
(h-\xi)\mathcal K_0(h-\xi)-\xi\,\mathcal K_2\,\xi
\Big]
\nonumber\\
&=\frac12\int d\tau\,\frac{d^3k}{(2\pi)^3}\Big[
h\,\mathcal K_0\,h
-2\xi\,\mathcal K_0\,h
+\xi(\mathcal K_0-\mathcal K_2)\xi
\Big].
\label{eq:Shchi}
\end{align}
Varying with respect to $\xi$ gives the saddle equation
\begin{equation}
\frac{\delta S}{\delta\xi}=0
\quad\Rightarrow\quad
-\mathcal K_0 h +(\mathcal K_0-\mathcal K_2)\xi=0
\quad\Rightarrow\quad
\xi=(\mathcal K_0-\mathcal K_2)^{-1}\mathcal K_0\,h.
\label{eq:chi-solution}
\end{equation}
Substituting \eqref{eq:chi-solution} back into \eqref{eq:Shchi} gives the exact tree-level effective action
for the source-coupled tensor mode $h$:
\begin{equation}
S^{(2)}_{\rm eff}[h]
=\frac12\int d\tau\,\frac{d^3k}{(2\pi)^3}\;
h\left[
\mathcal K_0
-\mathcal K_0(\mathcal K_0-\mathcal K_2)^{-1}\mathcal K_0
\right]h.
\label{eq:Seff-exact}
\end{equation}
This is a nonlocal (operator-valued) form factor multiplying the light kernel.

\paragraph{Low-energy (superhorizon) reduction and $Z_T$.}
Using \eqref{eq:K0}--\eqref{eq:K2}, the difference of kernels is purely the mass term,
\begin{equation}
\mathcal K_0-\mathcal K_2
=-\frac{F}{4}a^4 \tilde m_2^2,
\qquad\Rightarrow\qquad
(\mathcal K_0-\mathcal K_2)^{-1}
=-\frac{4}{F}\frac{1}{a^4 \tilde m_2^2}.
\label{eq:Kdiff}
\end{equation}
Inserting \eqref{eq:Kdiff} into \eqref{eq:Seff-exact} yields
\begin{equation}
S^{(2)}_{\rm eff}[h]
=\frac12\int d\tau\,\frac{d^3k}{(2\pi)^3}\;
h\,\mathcal K_0\left[
1+\frac{\mathcal K_0}{\frac{F}{4}a^4 \tilde m_2^2}
\right]h.
\label{eq:Seff-factor}
\end{equation}
At horizon exit/superhorizon scales, the relevant massless dynamics are governed by the Mukhanov variable
$u=(\sqrt{F}/2)\,a\,h$, which obeys
\begin{equation}
u''+\left(k^2-\frac{z_t''}{z_t}\right)u=0.
\label{eq:u-eom}
\end{equation}
On exact de Sitter $z_t''/z_t=2a^2H^2$, and near freeze-out $k^2\sim a^2H^2$, so the operator ratio appearing
in \eqref{eq:Seff-factor} reduces to a number,
\begin{equation}
\frac{\mathcal K_0}{\frac{F}{4}a^4 m_2^2}\;\longrightarrow\;\frac{2H^2}{\tilde m_2^2}.
\label{eq:ratio-to-number}
\end{equation}
Therefore, the tree-level effective action for the source-coupled tensor mode can be written as
\begin{equation}
S^{(2)}_{\rm eff}[h]
\simeq \frac12\int d\tau\,\frac{d^3k}{(2\pi)^3}\;
Z_T\;h\,\mathcal K_0\,h,
\qquad
Z_T=1+\frac{2H^2}{\tilde m_2^2}.
\label{eq:Seff-ZT}
\end{equation}
Using \eqref{eq:medef}, we equivalently obtain
\begin{equation}
Z_T=1+\frac{2H^2}{\tilde m_2^2}
=1-\frac{2\beta H^2}{F}.
\label{eq:ZT-final}
\end{equation}
By the rescaling the field $h\to \frac{1}{a\sqrt{F}}h$, we get the effective action in \eqref{eq:ZT_def}.

\bibliographystyle{utphys}
\bibliography{ref.bib}

\end{document}